\documentclass[12pt]{article}
\usepackage[nosort]{cite}
\usepackage{amsmath,amsthm,amsfonts,amssymb,amscd,mathrsfs,slashed,graphicx,braket}
\usepackage{epsfig, xcolor, cancel, wasysym, float, subfig, hyperref}
\usepackage[british]{babel}

\newlength{\xtrawidth}
\newlength{\xtraheight}
\numberwithin{equation}{section}
\numberwithin{table}{section}
\numberwithin{figure}{section}
\begin{document}
%%%%%%%%%%%%%%%%%%%%%%%%%%%%
%%%%%%%%%%%%%%%%%%%%%%%%%%%%

\pagenumbering{gobble} 
\begin{center}
\hfill BONN--TH--2018--07\\
\vskip 0.6in
%%
%{\Large\bf{Dark matter in the KL moduli stabilzation scenario}}\\[2ex]
%{\Large\bf{with SUSY breaking sector dual to $\mathcal{N} = 1$~SQCD}}
{\Large\bf{Dark matter in the KL moduli stabilization scenario}}\\[2ex]
{\Large\bf{with SUSY breaking sector from $\mathcal{N} = 1$~SQCD}}
\vskip 0.5in
{\bf
Thaisa C.~da~C.~Guio$^{1}$ and 
Ernany R. Schmitz$^{1}$}
\vskip 0.15in
{\it $^{1}\,$Bethe Center for Theoretical Physics, Physikalisches Institut der Universit\"at Bonn,\\
Nussallee 12, D-53115 Bonn, Germany}
\\

\vskip 0.15in
{\tt tguio@th.physik.uni-bonn.de}\\
{\tt ernany@th.physik.uni-bonn.de}
\end{center}
\vskip 0.15in
\begin{center} {\bf Abstract} \end{center}
 
We investigate neutralino dark matter from a string/M-theory perspective. Using the Kallosh-Linde (KL) scenario to stabilize the string moduli requires supersymmetry breaking for uplifting to a de Sitter vacuum. We consider the free magnetic dual description of $\mathcal{N}=1$ SUSY QCD with massive flavours, the Intriligator-Seiberg-Shih model (ISS), as an F-term dynamical SUSY breaking sector. This framework allows for a gravitino mass in the TeV range. Moreover, due to the plethora of particles from the ISS sector, we investigate the consequences of coupling the MSSM with the KL-ISS setup to obtain constraints from both late entropy production and the dark matter relic density. In addition to thermal neutralino production, we consider neutralino production via the decays of gravitinos and ISS fields.
 
\newpage
\pagenumbering{roman}  
\setcounter{page}{1}
\tableofcontents

\newpage
\pagenumbering{arabic}
\setcounter{page}{1}

%%%%%%%%%%%%%%%%%%%%%%%%%%%%
%%%%%%%%%%%%%%%%%%%%%%%%%%%%
\section{Introduction}
%%%%%%%%%%%%%%%%%%%%%%%%%%%%
%%%%%%%%%%%%%%%%%%%%%%%%%%%%

The existence of dark matter has been widely corroborated by a number of observational evidences, such as rotation curves of galaxies \cite{Rubin:1980zd}, gravitational lensing in galaxy clusters \cite{GravitationalLensing} and the observed angular power spectrum of the cosmic microwave background (CMB) \cite{Planck}, which is well fitted within the $\Lambda$CDM model. 

The most prominent candidates for a dark matter particle are QCD axions \cite{PecceiQuinn01, PecceiQuinn02} and supersymmetric particles, such as sneutrinos, gravitinos and neutralinos \cite{Nilles:1983ge, Martin:1997ns, Drees_book}. Other examples are SIMPS \cite{Bernal:2015bla}, complex scalars \cite{Sanchez-Vega:2014rka} and heavy neutrinos \cite{Sanchez-Vega:2015qva}. In supersymmetric scenarios, natural candidates for dark matter are neutralinos, which are LSPs (lightest supersymmetric particles) guaranteed to be stable in conserved R-parity models \cite{Drees_book, Dimopoulos:1981zb, Farrar:1978xj, Fayet:1974pd, Fayet:1977yc, Salam:1974xa, Pagels:1981ke}. Supersymmetric extensions of the Standard Model can, themselves, be low-energy effective theories of String Theory/M-theory, more fundamental scenarios with strings and branes. Therefore, we include neutralinos that appear as by-products of decays of gravitinos, which in turn have their masses constrained by SUSY breaking F-terms as well as by properties of the so-called moduli.

Moduli --- together with stringy axions --- are the only remnants in four-dimensional low-energy effective theories resulting from compactifications of the non-observed extra dimensions in string theories/M-theory. In geometrical/topological terms, the moduli parametrize the size and shape of the internal manifold formed by the extra dimensions \cite{Candelas:1990pi, Greene:1996cy}. For example, in Calabi-Yau compactifications, three types of moduli appear. The first are the so-called K\"ahler moduli, which are scalar fields corresponding to deformations of Ricci-flat metrics on the internal manifold. The second are the so-called complex structure moduli, which are scalar fields corresponding to deformations of the complex structure of the internal manifold. And the third are additional scalar fields that arise both from expanding RR and NS-NS fluxes in basis of harmonic forms as well as from the dilaton in type I, type II and heterotic string theories.

In such scenarios of four-dimensional theories arising from more fundamental theories such as string theories/M-theory, the moduli dynamics is extremely relevant for our understanding of the cosmological evolution of the Universe both during and after inflation \cite{Baumann:2014nda}. However, their existence can also pose serious problems for our four-dimensional world. In fact, one of the biggest tasks for string/M-theory cosmology is to solve the so-called moduli stabilization problem, i.e., to find vacua in which all the moduli have positive squared-masses. Furthermore, moduli are required to be quite heavy with masses of $\mathcal{O}(30$~TeV) or higher in order not to affect Big-Bang nucleosynthesis (BBN), thus avoiding the so-called moduli problem \cite{Polonyi_original, Polonyi_problem, Note_polonyi}. 

Many models of moduli stabilization have been constructed in the literature. The two leading ones are the Large Volume Scenario (LVS) \cite{Balasubramanian:2005zx} and the KKLT scenario \cite{Kachru:2003aw}. An extension of the KKLT has been proposed in the so-called KL scenario \cite{Kallosh:2004yh}, which solves its cosmological and phenomenological issues, such as an extremely large gravitino mass of $\mathcal{O}(10^{10}$~GeV) or low-scale inflation. The moduli that are stabilized via the KL scenario are the ones --- together with parameters from SUSY breaking F-terms --- that will set the mass of the gravitino, which then decays to the neutralino dark matter candidate. 

For the moduli stabilization scenarios mentioned in the previous paragraph, the moduli are fully stabilized in the sense that there are no instabilities and no flat directions in the scalar potential. Nevertheless, these scenarios can only provide anti-de Sitter (AdS) or Minkowski vacua. Achieving realistic de Sitter (dS) vacua from stabilized AdS/Minkowski vacua is known as uplifting. This mechanism requires the identification of a supersymmetry breaking sector that introduces a positive contribution to raise the potential from negative/zero to positive values. Still this uplifting has to be implemented without de-stabilizing the moduli.

In this work, we perform F-term uplifting with a class of models of dynamical supersymmetric breaking, namely, models that are dual to strongly-coupled $\mathcal{N}=1$ supersymmetric $SU(N_c)$ QCD with $N_f$ massive chiral multiplets (flavours) in the free magnetic dual range $N_c + 1 \leq N_f < 3N_c/2$. These models are known as ISS models after the work of Intriligator, Seiberg and Shih \cite{ISS_model}. In this range of $N_f$, they are weakly coupled in the IR and, therefore, the low-energy analysis in their original work was performed using Seiberg duality \cite{Seiberg:1994pq}. The ISS models are highly motivated as natural embeddings into higher dimensional fundamental theories, namely type IIA string theory. In fact, they appear as low-energy effective theories on intersecting NS 5-branes and D-branes \cite{Giveon:1998sr, Giveon:2009ur, Ooguri:2006bg}. 

We then combine the ISS model with the stabilized AdS vacuum in the KL moduli stabilization scenario and, furthermore, merge this uplifted KL--ISS scenario with the Minimal Supersymmetric Standard Model (MSSM). We obtain constraints for the moduli, the ISS fields, and the gravitino to be in agreement with the cosmological evolution of the Universe in such a way that entropy production is not problematic, i.e., in a way that there is no entropy dilution of products from Big-Bang Nucleosynthesis (BBN). Furthermore, we analyze decays of the ISS fields to MSSM fields and lighter ISS fields, and obtain constraints for the neutralino dark matter production based on the observed dark matter relic density \cite{Planck}. 

Our paper is organized in the following way: in section \ref{Sec:F-term_KL-ISS}, we give a short review of both the KL moduli stabilization scenario and the ISS setup for F-term dynamical supersymmetry breaking. We also briefly review how F-term uplifting via an ISS sector can achieve dS vacua solutions in the KL moduli stabilization scenario, and how this provides the masses for the relevant fields in our cosmological/phenomenological analysis, namely, the moduli, the ISS fields, and the gravitino. In section \ref{Sec:Interactions_DecayModes}, we analyse decay channels of the ISS fields to MSSM fields and combinations of lighter ISS fields. In section \ref{Sec:Post-inflation_Entropy}, we analyze the importance of the ISS fields together with the inflaton and the modulus, the last one via its relation to the gravitino mass, for the cosmological evolution of the Universe. Here, we are particularly interested in obtaining constraints on the superpotential coupling $h$ and the energy scale $M$ of the ISS model. We check how to avoid possible and problematic late entropy production, which would dilute products of BBN. In section \ref{Sec:DarkMatter}, we determine the resulting thermal and non-thermal neutralino dark matter relic density abundance from decays of both relativistic/non-relativistic gravitinos and products from decays of the ISS fields. We present our conclusions in section \ref{Sec:Conclusions}.

%%%%%%%%%%%%%%%%%%%%%%%%%%%%
%%%%%%%%%%%%%%%%%%%%%%%%%%%%
\section{The KL--ISS scenario} \label{Sec:F-term_KL-ISS}
%%%%%%%%%%%%%%%%%%%%%%%%%%%%
%%%%%%%%%%%%%%%%%%%%%%%%%%%%

In this section, we give a brief review of both the KL moduli stabilization scenario and the ISS model. We work here with the AdS vacuum in the KL scenario and dynamically break supersymmetry on it, explaining how it leads to vacuum uplifting. We present the superfield content and the four-dimensional relevant quantities needed for the rest of the work, namely the K\"ahler potential, the superpotential, and the scalar potential.

%%%%%%%%%%%%%%%%%%%%%%%%%%%%
\subsection{The KL moduli stabilization scenario}
\label{Subsec:KL}
%%%%%%%%%%%%%%%%%%%%%%%%%%%%

Before we start with the proper KL moduli stabilization scenario \cite{Kallosh:2004yh}, we give a brief review of the original KKLT moduli stabilization scenario \cite{Kachru:2003aw}, the cosmological and phenomenological issues arising from it.

In the original work of KKLT moduli stabilization scenario, it was argued that all moduli are stabilized in a controlable way in the context of compactifications of type IIB string theory on Calabi-Yau manifolds with the presence of fluxes. 

To be more precise, the starting point is F-theory compactified on an elliptically fibered Calabi-Yau fourfold $X$ --- see, e.g., Calabi-Yau fourfold constructions for F-theory in \cite{Klemm:1996ts}. The manifold $M$ of the fibration encodes the data from the type IIB geometry whereas the type IIB axiodilaton $\tau$ is to be associated with deformations in the complex structure of the elliptic fiber. 

In the absence of fluxes, this setup has a certain locus in the moduli space which is of type IIB compactified on an orientifolded Calabi-Yau threefold $M$ \cite{Sen:1997gv}. Moreover, the focus is on a setup with only one\footnote{The generalization to several K\"ahler moduli $\rho_i$ is also discussed in \cite{Giddings:2001yu}, but does not introduce new aspects compared to what is presented here.} chiral superfield, the K\"ahler modulus $\rho$, from $h^{1,1}(M)=1$ as in \cite{Giddings:2001yu}. In the presence of fluxes, there appears the familiar Gukov-Vafa-Witten superpotential \cite{Gukov:1999ya} 
\begin{equation} \label{eq:KKLT_W}
	W = \int_M G_3 \wedge \Omega~,
\end{equation} 
where $\Omega$ is the holomorphic $(3,0)$-form on $M$, and $G_3 = F_3 - \tau H_3$ depends on the three-forms from RR and NS fluxes in type IIB --- $F_3$ and $H_3$, respectively --- and on the axiodilaton~$\tau$. 

Furthermore, following conventions in \cite{Giddings:2001yu}, the tree-level K\"ahler potential reads
\begin{equation} \label{eq:KKLT_K}
K = -3 \text{ln}(\rho + \bar{\rho}) - \text{ln}(\tau + \bar{\tau}) - \text{ln}\left(-i\int_M \Omega \wedge \bar{\Omega}\right)~.
\end{equation}

Using the covariant derivative $D_a W = \partial_a W + (\partial_a K) W$ and the moduli space K\"{a}hler metric $K_{a\bar{b}} = \partial_a \partial_{\bar{b}} K$, it follows that the four-dimensional effective $\mathcal{N}=1$ supergravity scalar potential for chiral superfields labelled by $a$ and $b$,
\begin{equation} \label{eq:KKLT_V_1}
	V  = e^K (K^{a\bar{b}} D_a W \overline{D}_{\bar{b}} \overline{W} - 3|W|^2)~,
\end{equation}
leads to a positive semi-definite contribution, characteristic of the so-called no-scale models since the superpotential (\ref{eq:KKLT_W}) is independent of the modulus $\rho$. 

For generic solutions, all complex structure moduli from F-theory on an elliptically fibered Calabi-Yau fourfold are fixed at the very large mass scale $m \sim \alpha'/R^3$, where $\alpha' \sim T^{-1}$ with the string tension $T$, and $R$ is the radius of the orientifolded Calabi-Yau threefold $M$. However, the volume modulus $\rho$ is left unfixed and we, therefore, study the effective theory associated to it. Notice that there can be non-perturbative corrections to the superpotential \footnote{Corrections to the K\"ahler potential have been neglected due to stabilization of $\rho$ at large values compared with the string scale, but there would still be an AdS supersymmetric vacuum \cite{Kachru:2003aw}.}, either from Euclidean D3-branes or from gaugino condensation from stacks of D7-branes wrapping 4-cycles in $M$. For both cases, the corrections follow a similar behaviour, namely, there are exponential corrections to the superpotential for $\rho$ at large volume \cite{Witten:1996bn} .

The analysis of the vacuum structure follows from the tree-level K\"ahler potential --- only for the volume modulus $\rho$ --- and the superpotential with non-perturbative corrections, respectively given by
\begin{equation} \label{eq:KKLT_Final}
	\begin{split}
K_{\text{KKLT}} & = -3 \text{ln}(\rho + \bar{\rho})~, \\
	W_{\text{KKLT}} & = W_0 + A e^{-a\rho}~,
	\end{split}
\end{equation}
where $W_0<0$ is a tree-level constant contribution from fluxes and $A,a>0$ are coefficients determined by either of the two non-perturbative corrections mentioned in the previous paragraph. The result is that, since supersymmetry is preserved due to $D_{\rho}W_{\text{KKLT}}=0$, and the scalar potential has a negative value, the volume modulus $\rho$ is stabilized in an AdS supersymmetric vacuum.  

The KKLT moduli stabilization scenario was very successful in its attempt to construct the first semi-realistic models for string cosmology. However, it had some issues \cite{Kallosh:2004yh}. Firstly, the gravitino mass is extremely large, namely $m_{3/2} \sim  6 \times 10^{10}$ GeV; secondly, the Hubble parameter during inflation has to satisfy $H \lesssim m_{3/2}$, since the modulus mass is stabilized at a mass $\sim m_{3/2}$. In other words, we either have large scale inflation with a large gravitino mass or a very low-scale inflation with a gravitino mass of $\mathcal{O}(\text{TeV})$. The problem with the first case is that we would need phenomenology from large scale supersymmetry and the problem with the second case is that there are so far no string theory inspired models with stable extra dimensions for such a low-scale inflation --- in Planck units $H \lesssim 10^{-15}$.

These issues were later dealt with in the new KL scenario for moduli stabilization \cite{Kallosh:2004yh}. It is a consistent framework where the gravitino mass is small  --- in other words, supersymmetry breaking happens at low-scales --- and inflation takes place at high-scales. Compared with the original KKLT, the modifications that were introduced in the KL scenario to solve these issues are: firstly, to allow for stabilization of the volume modulus in a supersymmetric Minkowski vacuum; secondly, to modify the superpotential to a racetrack expression, namely
\begin{equation} \label{eq:KL_W}
	W_{\text{KL}} =  W_0 + Ae^{-a\rho} - B e^{-b\rho}~,
\end{equation}
where $W_0<0$ is a tree-level constant contribution from fluxes and $A,B,a,b>0$ are coefficients determined by non-perturbative corrections. The K\"ahler potential is still the same as in the KKLT scenario, namely 
\begin{equation} \label{eq:KL_K}
K_{\text{KL}} = -3 \text{ln}(\rho + \bar{\rho})~.
\end{equation}

At a supersymmetric Minkowski vacuum, both $W_{\text{KL}}=0$ and $D_{\rho}W_{\text{KL}}=0$ must be satisfied. This implies that the scalar potential vanishes, and we have a supersymmetric Minkowski minimum with a vanishing gravitino mass. Moreover, the gravitino mass does not depend on the height of the potential, which in the KKLT was related to the Hubble parameter of inflation, $H\lesssim m_{3/2}$. Therefore, any possible uplifting would not give a contribution depending on the Hubble parameter and, therefore, it would still be possible to have high-scale inflation without necessarily constraining the gravitino mass to be also large, i.e., having in fact low-scale supersymmetry breaking. Notice that it is also possible to find AdS vacuum with $W_{\text{KL}} \neq 0$ and $D_{\rho}W_{\text{KL}}=0$. In this case it is also found that there is no dependence on the Hubble parameter for the gravitino mass. 

In both moduli stabilization scenarios, uplifting of the AdS/Minkowski vacuum can be performed by explicitly breaking supersymmetry with non-perturbative terms from the addition of several anti-D3-branes that, however, do not add further moduli to the discussion. For a sufficient warped background, such anti-D3-branes give a small contribution to the AdS/Minkowski vacuum turning it to a small positive value without compromising the mechanisms that stabilized the moduli in the first place. 

%%%%%%%%%%%%%%%%%%%%%%%%%%%%
\subsection{The ISS model} 
\label{Subsec:ISS}
%%%%%%%%%%%%%%%%%%%%%%%%%%%%

In contrast to the uplifting via explicit supersymmetry breaking from anti-D3-branes, in this work we consider the uplifting of the AdS vacuum in the KL scenario via dynamical supersymmetry breaking. For this purpose, we use the so-called ISS model \cite{ISS_model}. Before we review the uplifting and its consequences for the KL-ISS scenario we will deal with in this work, we first present the details of the ISS model itself.

The use of strong gauge dynamics in dynamical supersymmetry breaking that can explain the hierarchy between the Planck scale and the weak scale is known since the works by Witten \cite{Witten:1981nf, Witten:1982df}. Sufficient conditions for the occurrence of dynamical supersymmetry breaking were suggested in \cite{Affleck:1983vc, Affleck:1983mk, Affleck:1984mf, Affleck:1984xz}.

There have been some models discussing dynamical supersymmetry breaking in a stable ground state but they are not only complicated but also pose various issues for phenomenology --- see reviews \cite{Shadmi:1999jy, Giudice:1998bp, Terning:2003th}. In the original work of the ISS model \cite{ISS_model}, much simpler and phenomenologically viable models were constructed by allowing dynamical supersymmetry breaking to happen in metastable vacua. 

The ISS model we deal with here consists of a theory which arises in the so-called free magnetic dual range $N_c + 1\leq N_f < 3N_c/2$ of $SU(N_c)$ $\mathcal{N}=1$ SUSY QCD with confinement scale $\Lambda$ coupled to $N_f$ chiral multiplets (flavours) $Q^i$ in the $N_c$ representation and $\bar{N}_f \left(=N_f\right)$ chiral multiplets $\tilde{Q}_{\tilde{i}}$ in the $\bar{N_c}$ representation, where $i,\tilde{i}=1,\ldots,N_f$ \cite{Intriligator:1995au}. The anomaly free global symmetry of SUSY QCD is 
\begin{equation}
	SU(N_f)_L \times SU(N_f)_R \times U(1)_B \times U(1)_R~.
\end{equation}
The transformations for the quarks $Q$ and $\tilde{Q}$ are given by
\begin{eqnarray}
	\begin{split}
	Q~~~~~ &\left(N_f,1,1, 1\right)~,\\
	\tilde{Q}~~~~~ &\left(1,\bar{N}_f,-1, 1\right)~.
	\end{split}
\end{eqnarray}

By studying the non-Abelian Coulomb phase of SUSY QCD at the origin of the moduli space \cite{Seiberg:1994bz}, one arrives at two dual descriptions apart from the conformal field theory of interacting quarks and gluons in the regime $3N_c/2 < N_f < 3N_c$. The original description is the free non-Abelian electric phase in terms of ``electric" variables, which is an $SU(N_c)$ gauge theory with $N_f$ chiral multiplets (flavours). The dual description is the free non-Abelian magnetic phase in terms of ``magnetic" variables, which is an $SU(N)$, where $N=N_f-N_c$, theory with $N_f$ flavours and an extra gauge invariant massless field. As expected, when one theory is weakly coupled, the other theory is strongly coupled in the sense of Seiberg duality \cite{Seiberg:1994pq}. In the first phase, for $N_f \geq 3N_c$, the original electric theory is not asymptotically free, the electric variables are free in the IR and the magnetic ones are infinitely strongly coupled. In the second phase, for $N_c + 1 \leq N_f < 3N_c/2$, the magnetic theory is not asymptotically free, the magnetic variables are free in the IR and the electric ones are infinitely strongly coupled. 

Therefore, when the ISS model is mentioned, it is meant that one deals with the IR free, low-energy effective theory of the magnetic dual of $SU(N_c)$ $\mathcal{N}=1$ SUSY QCD in the range $N_c + 1 \leq N_f < 3N_c/2$ with $N=N_f-N_c$. In other words, as mentioned in the previous paragraph, it is given by the IR free, low-energy effective theory from $SU(N_f-N_c)$ with $N_f$ flavours and $N_{f}^2$ extra gauge invariant massless fields. More precisely, the ISS model consists of ISS fields $\phi_{\text{ISS}}$, which collectively denote the chiral superfields\footnote{Here $S^i_j$ are the $N_f^2$ extra gauge invariant massless fields.} $q_i^a, \tilde{q}^j_b, S^i_j$, where $i,j=1,\ldots,N_f$ are flavour indices, $a,b=1,\ldots,N$, and $N_f>N=N_f-N_c$ \cite{ISS_model}. With $U(1)_R$ standing for R-parity symmetry, the global symmetry group is
\begin{equation}
	SU(N) \times SU(N_f)_L \times SU(N_f)_R \times U(1)_B \times U(1)' \times U(1)_R~.
\end{equation} 

The transformations for the fields $q$, $\tilde{q}$ and $S$ are given by
\begin{eqnarray}
\begin{split}
q~~~~~& \left(N,\bar{N}_f,1,1,1,0\right)~,\\
\tilde{q}~~~~~& \left(\bar{N},1,N_f,-1, 1,0\right)~,\\
S~~~~~& \left(1,N_f,\bar{N}_f,0, -2,2\right)~.
\end{split}
\end{eqnarray}

Following notation in \cite{Strong_stabilization_Dudas}, the K\"ahler potential and the tree-level superpotential ---~without gauging $SU(N)$~--- are respectively given by
\begin{align} 
	 K_{\text{ISS}} ={}& |q|^2 + |\tilde{q}|^2 + |S|^2 = q_i^a \bar{q}_a^i + \tilde{q}^i_a \bar{\tilde{q}}^a_i + S^i_j \bar{S}^j_i ~, \label{eq:ISS_K}\\
	W_{\text{ISS}} ={}& h(\text{Tr}\tilde{q}Sq - M^2\text{Tr}S) = h(\tilde{q}^i_a S^j_i q_j^a - M^2 S^i_j \delta_i^j)~, \label{eq:ISS_W}
\end{align}
where $h$ is a dimensionless coupling and $M \ll M_{\rm P}$ is the energy scale of the ISS model.  Both parameters $h$ and $M$ will be constrained in our phenomenological and cosmological analysis in sections \ref{Sec:Post-inflation_Entropy} and \ref{Sec:DarkMatter}. The second term in (\ref{eq:ISS_W}) explicitly breaks the global symmetry group to $SU(N) \times SU(N_f)_V \times U(1)_B \times U(1)_R$. Moreover, if the supergravity embedding is taken, $U(1)_R$ is explicitly broken.

%%%%%%%%%%%%%%%%%%%%%%%%%%%%
\subsection{Consequences of F-term uplifting in the KL-ISS scenario}
\label{Subsec:F-term_Consequences}
%%%%%%%%%%%%%%%%%%%%%%%%%%%%

After reviewing both the KL moduli stabilization scenario and the ISS model, we now present more details of the relevant four-dimensional low-energy effective expressions and particle content in our scenario, focusing on the consequences of F-term uplifting to the gravitino mass and to soft terms such as $A$-terms.

In this work, we consider the following particle content: the modulus field $\rho$, the ISS fields $\phi_{\text{ISS}}$, the inflaton field $\eta$, and the MSSM fields which we collectively denote by $\phi$. 

The K\"ahler potential and superpotential for the inflaton are the same as in \cite{Inflaton_decay}, namely $K(\eta-\bar{\eta})^2,S\bar{S})$ and $W(\eta)=Sf(\eta)$ where  $S$ is a stabilizer field. From these functions, it is possible to conclude that $\eta$ does not decay into gravitinos (condition which simplifies our study and is discussed in section \ref{Sec:DarkMatter}).

The canonical K\"ahler potential for the MSSM fields is given by
\begin{equation} \label{eq:MSSM_K}
	K_{\text{MSSM}} = \phi \bar{\phi}~.
\end{equation}
Furthermore, we call the tree level superpotential for the MSSM fields $W_{\text{MSSM}} = W(\phi)$. 

We will consider additionally a Giudice-Masiero
term \cite{Giudice_Masiero}
\begin{equation} \label{eq:GiudiceMasiero}
	 K_{\text{GM}} = c_{H}H_{1}H_{2}+\textrm{h.c.}~,
\end{equation}
where $H_{1}$ and $H_{2}$ are the Higgs superfields of the MSSM and $c_{H}$ is a constant with no mass dimension. This term is often phenomenologically required due to the fixing of $\textrm{tan}\beta$ within the four-dimensional low-energy effective theory context \cite{Strong_stabilization_Dudas,GM_pheno1,GM_pheno2}. 
As it is clear, we use canonical K\"ahler potentials for all the fields except for the volume modulus $\rho$ and the inflaton $\eta$. 
 
Within the context of inflaton plus MSSM fields, we add the KL scenario to have cosmology and phenomenology motivated by string theories/M-theory and use the ISS model to uplift the, so far, non-positive vacuum. 

The uplifting of the vacuum energy in the KL scenario can be done in the following way. We start with a combination of the KL scenario with the ISS model, reviewed in sections \ref{Subsec:KL} and \ref{Subsec:ISS}, respectively. This means that we consider the following combination of K\"ahler potential and superpotential, respectively,
\begin{align} 
	K_{\text{KL-ISS}} ={}& -3 \text{ln}(\rho + \bar{\rho}) + |q|^2 + |\tilde{q}|^2 + |S|^2,\label{eq:KL+ISS_K}\\
	W_{\text{KL-ISS}} ={}& W_0 + Ae^{-a\rho} - B e^{-b\rho} + h(\text{Tr}\tilde{q}Sq - M^2\text{Tr}S),\label{eq:KL+ISS_W}
\end{align} 
where $W_0<0$ and $A,B,a,b>0$ are constants that have been introduced in the previous sections. 

Now we review the vacuum structure of this combination \cite{Strong_stabilization_Dudas}. Let Im$\rho = 0$ and Re$\rho = \sigma$ for simplicity. Furthermore, let $\sigma_0$ be the value of $\rho$ at its minimum. The supersymmetric Minkowski vacuum $V_{\text{KL}}(\sigma_0) = 0$ in the KL scenario must satisfy 
\begin{equation} \label{eq:SUSY_Minkowski}
	\begin{split}
	D_{\rho}W_{\text{KL}}|_{\sigma=\sigma_0} & =\partial_{\rho}W_{\text{KL}} + \partial_{\rho} K_{\text{KL}}W_{\text{KL}}|_{\sigma=\sigma_0}=0~, \\
	W_{\text{KL}}(\sigma_0) & =0~.
	\end{split}
\end{equation}
Allowing $W_{\text{KL}}(\sigma_0) \neq 0$, namely $W_{\text{KL}}(\sigma_0) \equiv \Delta$, shifts the minimum to a supersymmetric AdS minimum
\begin{equation} \label{eq:SUSY_AdS}
	V_{\text{KL}}\left(\sigma_{0}\right) \simeq -3m_{3/2}^{2} \simeq -\frac{3\Delta^{2}}{8\sigma_{0}^{3}}~. 
\end{equation}
We still have $D_{\rho}W_{\text{KL}}|_{\sigma=\sigma_0+\delta\sigma}=0$ for it to be supersymmetric, where $\delta\sigma~\ll~\sigma_0$ \cite{Linde:2011ja}. Therefore, an upflifting contribution is required to turn the vacuum to positive values (dS), and to obtain the correct cosmological constant $\frac{\Lambda}{M_{P}^{2}}\sim10^{-120}$. Note that one can obtain $m_{3/2}\neq 0$ while local SUSY is not broken, i.e., $D_iW=0$, however the vacuum can only be AdS, $\Lambda<0$.

The uplifting is performed via F-term with the ISS model. By working out the first derivative of the four-dimensional effective $\mathcal{N}=1$ supergravity scalar potential (\ref{eq:KKLT_V_1}), namely $\partial_{\phi_\textrm{ISS}}V_{\text{KL-ISS}}=0$, the metastable ISS vacuum $(S_0,q_0,\tilde{q}_0)$ is given by
\begin{eqnarray}
(S_{0})^i_j & = & 0~, \label{eq:VEV_S}\\
(q_{0})^a_i & = & M\delta^a_i~,\label{eq:VEV_q}\\
(\tilde{q}_{0})^j_b & = & M\delta^j_b~.\label{eq:VEV_qtilde}
\end{eqnarray}
From a matrix viewpoint, $q_{0}$ and $\tilde{q}_{0}$ can be written as
\begin{equation} \label{eq:VEV_q_matrix}
	\begin{split}
q_0 & = \left(\begin{array}{c}
M\mathbb{I}_{N\times N}\\
0_{\left(N_{f}-N\right)\times N}
\end{array}\right)~,\\
\tilde{q}_0 & = \left(\begin{array}{cc}
M\mathbb{I}_{N\times N} & 0_{N\times\left(N_{f}-N\right)}\end{array}\right)~.
	\end{split}
\end{equation}
These are the VEVs responsible for spontaneous symmetry breaking in the ISS model that allow for a possible uplifting of the AdS vacuum in the KL scenario --- given in equation (\ref{eq:SUSY_AdS}). In fact, from terms $e^{K_{\text{KL-ISS}}}\partial_{\{q,\tilde{q},S\}}W_{\text{KL-ISS}} \partial_{\{\bar{q},\bar{\tilde{q}},\bar{S}\}}\overline{W}_{\text{KL-ISS}}$ in the four-dimensional scalar potential, these VEVs imply that the minimum of the KL-ISS potential is given by --- with $M_{\rm P}=1$ ---
\begin{equation} \label{eq:KL-ISS_ScalarPotential}
V_{\textrm{min}}=\frac{e^{2NM^{2}}}{(2\sigma_{0})^{3}}\{\Delta^{2}(-3+2NM^{2})+h^{2}M^{4}(N_{f}-N)\}~. 
\end{equation}
Since $M \ll M_{\rm P}$, we neglect the term $2NM^2$ compared with -3 in the first parenthesis of (\ref{eq:KL-ISS_ScalarPotential}) unless a huge $N = N_f - N_c$ of $\mathcal{O}(10^{10})$ is taken for it to be relevant, which is unlikely. Using this fact and that the scalar potential minimum must equal the small but positive contribution from the cosmological constant $M_{P}^{4}\left(\frac{\Lambda}{M_{P}^{2}}\right)\sim10^{-120}M_{P}^{4} \simeq 0$, we obtain a constraint for the parameter $\Delta$, namely\footnote{The cosmological constant is set to approximately zero by positive values which, therefore, is in agreement with the fact that we have uplifted the KL vacua to positive values. We choose here the extreme case of a vanishing cosmological constant only to get a lower bound constraint for the superpotential $\Delta$ that is necessary to perform the uplifting.}
\begin{equation} \label{eq:Delta}
|\Delta| \simeq \sqrt{\frac{N_{f}-N}{3}}hM^{2}~.
\end{equation}

The expected gravitino mass due to the supersymmetry breaking sector --- the ISS sector with parameters $h$ and $M$ --- and the value $\sigma_0 = \text{Re}\rho$ of the modulus at the minimum, is given by
\begin{equation}
\label{eq:Mgravitino}
m_{3/2}=\left\langle e^{K/2}W\right\rangle \simeq \frac{|\Delta|}{\left(2\sigma_{0}\right)^{3/2}}e^{NM^{2}} \simeq \frac{e^{NM^{2}}}{\left(2\sigma_{0}\right)^{3/2}}\sqrt{\frac{N_{f}-N}{3}}hM^{2}~.
\end{equation}
Restoring the reduced Planck mass $M_{\rm P}$, it is given by
\begin{equation} \label{eq:Mass_Gravitino_NonNaturalUnits}
m_{3/2} \simeq \frac{e^{NM^{2}/M_{\rm P}^2}}{\left(2\sigma_{0}/M_{\rm P}\right)^{3/2}}\sqrt{\frac{N_{f}-N}{3}}h\left(\frac{M}{M_{\rm P}}\right)^{2}M_{\rm P} ~.
\end{equation}
Since $M$ is assumed to be well below the Planck scale $M_{P}$ --- recall smallness of $M$ due to the dynamical nature of the supersymmetry breaking sector ---, we set $e^{NM^{2}/M_{\rm P}^2} = 1$ in (\ref{eq:Mass_Gravitino_NonNaturalUnits}). Throughout this work, we use the gravitino mass with unity reduced Planck mass ($M_{\rm P}=1$), i.e.,  
\begin{equation} \label{eq:Mass_Gravitino}
m_{3/2} \simeq \frac{1}{\left(2\sigma_{0}\right)^{3/2}}\sqrt{\frac{N_{f}-N}{3}}hM^{2}~.
\end{equation}
Since $m_{3/2}$ depends on parameters, such as $N_f$ and $N$, that do not cover much parameter space, we must now comment on the rigidity of $m_{3/2}$. We recall that $N_{f}$ and $N$ reflect the size of the global symmetry groups $\textrm{SU}\left(N_{f}\right)$ and $\textrm{SU}\left(N\right)$, respectively. Therefore, even if they differ quite sizeably, let us say $\left(N_{f}-N\right)\sim20$, the variation on the gravitino mass would not be large. Secondly, we will see in the next section that the parameters of the KL superpotential (\ref{eq:KL_W}) fix $\sigma_{0} \simeq 13.86 M_{\rm P}$. Therefore, the relevant free parameters in determining the gravitino mass are $M$ and $h$. For the rest of the work, we shall assume $m_{3/2} \apprge 10^{-13} M_{P}$ for its decay to occur before BBN such that there is no gravitino problem. With the use of (\ref{eq:Mass_Gravitino}), this leads to the following range for $M$
\begin{equation}\label{eq:Constraint_GravM}
	M \apprge 3.82 \times 10^{-6}h^{-1/2}M_{P}~.
\end{equation} 
Since the parameter $M$ will be used as a control parameter for the entropy production by decays of the ISS fields and for the dark matter production, we should obtain a consistent model if our analysis is in agreement with this range. Equation (\ref{eq:Constraint_GravM}) then gives the first constraint we must consider for the phenomenological/cosmological analysis that appears in sections \ref{Sec:Post-inflation_Entropy} and \ref{Sec:DarkMatter} later in this work, culminating in the analysis of dark matter production. 

We should mention some words in respect to the generation of soft terms via F-term SUSY
breaking from the ISS sector and the resulting particle spectrum. This has been analyzed extensively in \cite{Strong_stabilization_Dudas}. In general, for models with strong moduli stabilization, the generated $A$-terms and the gaugino masses are very small at tree level. This happens because, for example, $\bar{S}\ \overline{D}_{\bar{S}}\overline{W}_{\text{KL-ISS}} \propto M^{2}W_{\text{ISS}}\ll W_{\text{ISS}}$ --- and similar terms with respect to $q$ and $\tilde{q}$ --- and the strong condition $D_{\rho}W_{\text{KL-ISS}} \ll W_{\text{KL-ISS}}$ after SUSY breaking. Indeed the tree-level value for $A$ is $A\propto\frac{m_{3/2}}{m_{\rho}}m_{3/2} \ll m_{3/2}$ since $m_{\rho} \gg m_{3/2}$. Furthermore, the condition $D_{\rho}W_{\text{KL-ISS}} \ll W_{\text{KL-ISS}}$ and the modulus mass of $\mathcal{O}(10^{-3}M_{\rm P})$ also imply that the gaugino masses are undesirably small, namely $m_{1/2}\propto\frac{m_{3/2}}{m_{\rho}}m_{3/2}\partial_{\rho}\textrm{ln}h_{A} \ll m_{3/2}$, where $h_A$ denotes the gauge kinetic functions. Therefore, one should apply a one-loop level calculation to generate both $A$-terms and gaugino masses, resembling anomaly mediated models. For acceptable gaugino masses, $m_{3/2}$ is forced to assume high values $\mathcal{O}\left(10-1000\textrm{ TeV}\right)$ in order to compensate for suppressed loop pre-factors. The resulting spectrum then resembles that of split supersymmetry, with light gaugino masses and soft scalar masses of $\mathcal{O}(m_{3/2})$.

%%%%%%%%%%%%%%%%%%%%%%%%%%%%
\subsection{The modulus and the ISS fields}
\label{Subsec:Modulus_ISS}
%%%%%%%%%%%%%%%%%%%%%%%%%%%%

In the previous section we focused on the direct consequences of F-term uplifting within the KL-ISS scenario. Now, we discuss the properties of the modulus $\rho$ and the ISS fields $\phi_{\text{ISS}}$ that will be relevant to the cosmological analysis of the primordial Universe. From this subsection on, we write the ISS fields only with lowered indices for an easy display.

To obtain the masses for the ISS and the modulus fields, we first compute the $8 \times 8$ non-diagonal mass matrix for $\rho, \bar{\rho}, S, \bar{S}, q, \bar{q}, \tilde{q}, \bar{\tilde{q}}$ directly from the scalar potential (\ref{eq:KKLT_V_1}) in our scenario and then diagonalize it. Namely, we use the scalar potential
\begin{equation} \label{eq:KKLT_V_2}
	V  = e^K (K^{a\bar{b}} D_a W \overline{D}_{\bar{b}} \overline{W} - 3|W|^2)~,
\end{equation}
with
\begin{equation} \label{eq:K_W}
	\begin{split}
	K & \equiv K_{\text{KL-ISS}} + K_{\text{MSSM}} + K_{\text{GM}} + K((\eta-\bar{\eta})^2) ~,\\
	W & \equiv W_{\text{KL-ISS}} + W_{\text{MSSM}} + W(\eta)~.	
	\end{split}
\end{equation} 

\noindent $\bullet$ The modulus field $\rho$
\vspace*{0.15cm}

We start the analysis with the modulus field $\rho$. After diagonalization of the mass matrix, we obtain the mass of the modulus $\rho$ to be
\begin{equation} \label{eq:Mass_Modulus}
m_{\rho}^{2}=\frac{2}{9}AaBb\left(a-b\right)\left[\frac{Aa}{Bb}\right]^{\frac{-a-b}{a-b}}\textrm{ln}\left(\frac{aA}{bB}\right)+\mathcal{O}\left(M^{2}\right)~.
\end{equation}
Thus, both real and imaginary components obtain the same mass. If we set, for example, $a=0.1,\,b=0.05,\,A=1,\,B=1$, we obtain $m_{\rho}\simeq2.19\times10^{-3}M_{P}$. This value is much heavier than the inflaton reference mass we will use in this work, namely $m_{\eta}=10^{-5}M_{P}$ --- in the simplest chaotic inflation models, $m_{\eta}\sim6\times10^{-6}M_{P}$. As shown in \cite{Inflaton_decay}, one of the conditions to ignore the dynamics of the modulus field during inflation is that its mass must be much heavier than the inflaton mass. In this way, the modulus and the inflaton decouple and they can be studied separately. This fact implies that the modulus field does not receive contributions from the inflaton potential during inflation. Therefore, its VEV has the same value both during and after inflation, which in turn leads to a vanishing post-inflationary oscillation amplitude, i.e., $\left\langle \rho\right\rangle _{\textrm{amp.}}=\left|\left\langle \rho\right\rangle _{\textrm{ins.}}-\left\langle \rho\right\rangle _{\textrm{min.}}\right|=0$, where $\left\langle \rho\right\rangle _{\textrm{ins.}}$ is the VEV
of $\rho$ during inflation and $\left\langle \rho\right\rangle _{\textrm{min.}}$
is the VEV of $\rho$ at the offset of inflation. For these reasons, we neglect the evolution of the modulus field $\rho$ after inflation. For completeness, the $\rho$ value at the minimum of the potential is given by the VEV of its real part $\textrm{Re}\rho$, namely $\sigma_{0}$, which we compute to be $\sigma_{0}=\frac{1}{a-b}\textrm{ln}\left(\frac{aA}{bB}\right)+\mathcal{O}\left(M^{2}\right)$. Using the same parameters as we used to obtain the mass for the modulus field $\rho$ after equation (\ref{eq:Mass_Modulus}), we obtain $\sigma_{0}\simeq 13.86\,M_{P}$.\\

\noindent  $\bullet$ The ISS scalar fields $\phi_{\text{ISS}}$\\

We continue the analysis for the ISS fields. We write the following linear combinations $Q_{1},Q_{2},Q_{3},Q_{4}$ for $q, \bar{q}, \tilde{q}, \bar{\tilde{q}}$,\footnote{Note that we denote the scalar parts of the ISS superfields also by $q_{ai}$, $\tilde{q}_{ia}$ and $S_{ij}$.} firstly with $i=1,\ldots,N$ and $a=1,\ldots,N$,
\begin{eqnarray} \label{eq:Lc_1}
\textrm{Re, Im}\left[Q_{1}\right] & = & \frac{1}{2} (q_{ai}\pm\bar{q}_{ai}+\tilde{q}_{ia}\pm\bar{\tilde{q}}_{ia})~,\\ \label{eq:Lc_2}
\textrm{Re, Im}\left[Q_{2}\right] & = & \frac{1}{2} \left[q_{ai}\pm\bar{q}_{ai}-\left(\tilde{q}_{ia}\pm\bar{\tilde{q}}_{ia}\right)\right]~,
\end{eqnarray}
and, secondly with $i=N+1,\ldots,N_{f}$
and $a=1,\ldots,N$, 
\begin{eqnarray} \label{eq:Lc_3}
\textrm{Re, Im}\left[Q_{3}\right] & = & \frac{1}{2}(q_{ai}\pm\bar{q}_{ai}\pm\tilde{q}_{ia}+\bar{\tilde{q}}_{ia})~,\\ \label{eq:Lc_4}
\textrm{Re, Im}\left[Q_{4}\right] & = & \frac{1}{2}\left[q_{ai}\pm\bar{q}_{ai}-\left(\pm\tilde{q}_{ia}+\bar{\tilde{q}}_{ia}\right)\right]~.
\end{eqnarray}
For comparison, these combinations were given in \cite{ISS_model}
by the following parametrizations, for $i=1,\ldots,N$ and $a=1,\ldots,N$,
\begin{eqnarray}
	q_{ai}\rightarrow2^{-1/2}\left(\chi_{+}+\chi_{-}\right)_{ai}~,\\
	\left(\tilde{q}_{ia}\right)^{\textrm{T}}\rightarrow2^{-1/2}\left(\chi_{+}-\chi_{-}\right)_{ai}~,
\end{eqnarray}
and, for $i=N+1,\ldots,N_{f}$
and $a=1,\ldots,N$,
\begin{eqnarray}
	q_{ai}\rightarrow2^{-1/2}\left(\rho_{+}+\rho_{-}\right)_{ai}~,\\
	\left(\tilde{q}_{ia}\right)^{\textrm{T}}\rightarrow2^{-1/2}\left(\rho_{+}-\rho_{-}\right)_{ai}~.
\end{eqnarray}
Following it, we would obtain 
\begin{eqnarray}
	\textrm{Re, Im}\left[Q_{1}\right] & = 2^{-1/2}\left(\chi_{+}\pm\chi_{+}^{*}\right)_{ai}~,\\
	\textrm{Re, Im}\left[Q_{2}\right] & = 2^{-1/2}\left(\chi_{-}\pm\chi_{-}^{*}\right)_{ai}~,\\
	\textrm{Re, Im}\left[Q_{3}\right] & = 2^{-1/2}\left(\rho_{\pm}\pm\rho_{\pm}^{*}\right)_{ai}~,\\
	\textrm{Re, Im}\left[Q_{4}\right] & = 2^{-1/2}\left(\rho_{\mp}\pm\rho_{\mp}^{*}\right)_{ai}~.
\end{eqnarray}

We show in table \ref{tab:Masses_ISS} the number of real or imaginary components as well as the mass eigenvalues for each of the 6 mass eigenstates constructed from the ISS fields $S, \bar{S}, q, \bar{q}, \tilde{q}, \bar{\tilde{q}}$ after diagonalization of the mass matrix.

In \cite{ISS_model}, massless Goldstone modes were predicted to exist. In our analysis, this would correspond to both the imaginary and the real components of $Q_{4}$ as well as to the imaginary part of $Q_{2}$. At tree-level, one is unable to notice that $\textrm{Re}[Q_2]$ is actually a pseudo-Goldstone whose mass is given by higher order corrections, as we analyze below. The reason why the scalar mass spectrum yields $N^2+2\left(N_f-N\right)N=2N_fN-N^2$ massless bosons is that the VEVs of $(q,\tilde{q})$ break the original symmetry $SU(N)\times SU(N_f)_V \times U(1)_B$ with $N^2 + N_f^2 -1$ generators into $SU(N)_V\times SU(N_f-N)_V \times U(1)_{B^{'}}$ with $(N^2-1)+(N_f-N)^2$ generators, where $SU(N_f)_V$ breaks into $SU(N)_V\times SU(N_f-N)_V\times U(1)_{B^{'}}$, whereas the original $SU(N)\times U(1)_B$ is completely broken. We also take into account that the group $SU(N)_V$ is not spontaneously broken since, in the region $i\otimes a=N^2$, there exists $Q_2$ with null VEV. Furthermore, the $S$ field transforming as a singlet under $SU(N_f)_V$ gives no contribution to the massless Goldstone mode analysis.

\begin{table}[H] 
\begin{center}
\begin{tabular}{|l|c|c|}
\hline 
ISS scalar mass eingenstate & Number$_{\textrm{Re/Im}}$ &  Mass\tabularnewline
\hline 
\hline 
$S_{1}\equiv S_{ij}\left(i\otimes j\in N^{2}\right)$ & $N^{2}$ & $\sqrt{\frac{6}{N_{f}-N}}\left(\frac{M_{P}}{M}\right)m_{3/2}$\tabularnewline
\hline 
$S_{2}\equiv S_{ij} \left(i\otimes j\in N_{f}^{2}-N^{2}\right)$ & $\left(N_{f}+N\right)\left(N_{f}-N\right)$ & $\mathcal{O}\left(m_{3/2}\right)$\tabularnewline
\hline 
$Q_{1}\equiv\textrm{Lc}\left[q_{ai},\tilde{q}_{ia} \right]\left(i\otimes a\in N^{2}\right)$ & $N^{2}$ & $\sqrt{\frac{6}{N_{f}-N}}\left(\frac{M_{P}}{M}\right)m_{3/2}$\tabularnewline
\hline 
$Q_{2}\equiv\textrm{Lc}\left[q_{ai},\tilde{q}_{ia} \right]\left(i\otimes a\in N^{2}\right)$ & $N^{2}$ & $0$ \tabularnewline
\hline 
$Q_{3}\equiv\textrm{Lc}\left[q_{ai},\tilde{q}_{ia} \right]\left(i\otimes a\in N_{f}N-N^{2}\right)$ & $\left(N_{f}-N\right)N$ & $\sqrt{\frac{6}{N_{f}-N}}\left(\frac{M_{P}}{M}\right)m_{3/2}$\tabularnewline
\hline 
$Q_{4}\equiv\textrm{Lc}\left[q_{ai},\tilde{q}_{ia} \right]\left(i\otimes a\in N_{f}N-N^{2}\right)$ & $\left(N_{f}-N\right)N$ & $0$ \tabularnewline
\hline
\end{tabular}
\caption{Tree-level largest contributions to the masses of the ISS scalar mass eigenstates. The notation is such that $N^{2}$ is the cartesian product $N\otimes N$ with
$N=\{1,\ldots,N\}$, $N_{f}^{2}$ is the cartesian product $N_{f}\otimes N_{f}$, with $N_f=\{1,\ldots,N_f\}$, and $N_{f}N$ is the cartesian product $N_{f}\otimes N$ with $N=\{1,\ldots,N\}$ and $N_f=\{1,\ldots,N_f\}$. Furthermore, Lc stands for the linear combinations given from equations (\ref{eq:Lc_1}) to (\ref{eq:Lc_4}), and the second column gives the number of real or imaginary components for each type of ISS scalar mass eigenstates.}
\label{tab:Masses_ISS}
\end{center}
\end{table}
\vspace*{-0.5cm}
As to quantum corrections to the masses, we recall that one-loop calculations in \cite{ISS_model} generate an additional $\mathcal{O}\left(m_{3/2}\frac{M_{P}}{M}\right)$ mass to the real and imaginary parts of $S_{2}^{'}$ --- a subset of $S_{2}$, namely $S_{ij}$ with indices $i,j>N$ --- as well as to the real part of $Q_{2}$. More precisely,
\begin{equation}
	\begin{split}
	e^{K/2}m_{S_{2}^{'}}^{\textrm{1-loop}}&= e^{K/2}\left(\frac{\textrm{ln}\left(4\right)-1}{8\pi^{2}}\right)^{1/2}\sqrt{N}h^{2}M\\
	&= \left(\frac{3\left(\textrm{ln}\left(4\right)-1\right)}{8\pi^{2}}\right)^{1/2}\sqrt{\frac{N}{N_{f}-N}}h\left(\frac{M_{P}}{M}\right)m_{3/2}~,\\
	e^{K/2}m_{Q_{2}}^{\textrm{1-loop}} &= e^{K/2}\left(\frac{\textrm{ln}\left(4\right)-1}{8\pi^{2}}\right)^{1/2}\sqrt{N_{f}-N}h^{2}M\\ 
	&= \left(\frac{3\left(\textrm{ln}\left(4\right)-1\right)}{8\pi^{2}}\right)^{1/2}h\left(\frac{M_{P}}{M}\right)m_{3/2}~.
	\end{split}
\end{equation}
Since $M\ll M_{P}$, we can safely consider the one-loop contribution
to yield the mass of the real and imaginary components of $S_{2}^{'}$,
as well as of the real component of $Q_{2}$.

For a better understanding of the mass matrices for the ISS fields $S$ and $Q$ after diagonalization, we display them below in a diagramatic form. They are given by, respectively,
\begin{equation} \label{eq:matrix_S}
S = \left(
\begin{array}{c|c}
(S_1)_{N \times N} & (S_2^{\text{nd}})_{N \times (N_f-N)} \\
\hline
(S_2^{\text{nd}})_{(N_f-N) \times N} & (S'_2)_{(N_f-N) \times (N_f-N)}
\end{array}
\right)~,
\end{equation}
\begin{equation} \label{eq:matrix_Q}
Q = \left(
\begin{array}{c|c}
(Q_1~\&~Q_2)_{N \times N} & (Q_3~\&~Q_4)_{N \times (N_f-N)}
\end{array}
\right)~.
\end{equation}
Here we see the splitting of $S_2$ into non-diagonal pieces $(S_2^{\text{nd}})_{N \times (N_f-N)}$, \ $(S_2^{\text{nd}})_{(N_f-N) \times N}$, and the subset $S'_2$ which receives mass through one-loop calculations beyond its tree-level mass of $\mathcal{O}(m_{3/2})$ as presented above. Furthermore we see that $Q_1$ and $Q_2$ as well as $Q_3$ and $Q_4$ mix in the block forms sketched above. The indices here refer to the dimensionality of their corresponding blocks.

A further comment is in order. When one considers the ISS model alone, the VEVs (\ref{eq:VEV_S}), (\ref{eq:VEV_q})
and (\ref{eq:VEV_qtilde}) render $\partial V_{\textrm{ISS}}/\partial\phi_{\text{ISS}}=0$. However, when the modulus field $\rho$ contribution is included, i.e., $\left.W_{\rho}\right|_{\textrm{min}}=\Delta$,
the first derivatives of the ISS fields change according to
\begin{equation}
	\begin{split}
	\partial V_{\textrm{KL-ISS}}/\partial\left(q,\tilde{q}\right)_{ia} &= \mathcal{O}\left(m_{3/2}^{2}M_{P}\left\langle q,\tilde{q}\right\rangle_{ia} \right)~,\\
	\partial V_{\textrm{KL-ISS}}/\partial S_{ij} &= \mathcal{O}\left(m_{3/2}^{2}M_{P}\right)\delta_{ij}~.
	\end{split}
\end{equation} 
To cancel these effects, the VEVs of $q_{ia}$, $\tilde{q}_{ai}$, and $S_{ij}$ --- assuming the VEVs to be the diagonal ones --- should obtain corrections, i.e.,
\begin{eqnarray}
	\begin{split}
	\left\langle q_{ia},\tilde{q}_{ai}\right\rangle & = \left(M-\mathcal{O}\left(M^{3}M_{\rm P}^{-2}\right)\right)\delta_{ia}~,\label{eq:VEV_q_2}\\
	\left\langle S_{ij}\right\rangle & =\left(\left(\frac{M}{M_{P}}\right)^{2}M_{P}+\mathcal{O}\left(M^4M_{\rm P}^{-3}\right)\right)\delta_{ij}~~~\text{for}~i,j\leq N~, \label{eq:VEV_S_2}\\
	\left\langle S_{ij}\right\rangle & =\left(\frac{16\pi^2\left(N_f-N\right)}{3\left(\textrm{ln}4-1\right)Nh^2}\left(\frac{M}{M_{P}}\right)^{2}M_{P}+\mathcal{O}\left(M^4M_{\rm P}^{-3}\right)\right)\delta_{ij}\\
	&
	\simeq\left(\frac{N_f-N}{3N}\frac{408.79}{h^2}\left(\frac{M}{M_{P}}\right)^{2}M_{P}+\mathcal{O}\left(M^4M_{\rm P}^{-3}\right)\right)\delta_{ij} ~~~\text{for}~i,j>N~.\label{eq:VEV_S_3}
	\end{split}
\end{eqnarray} 
The last correction is dominated by one-loop contributions. Notice that these corrections cancel the term $2NM^2$ in equation (\ref{eq:KL-ISS_ScalarPotential}).

An observation about the sign of $\Delta$ must be made. When we calculate the modified VEVs, written in (\ref{eq:VEV_q_2}), one must assume either that $h<0$ or $\Delta<0$ for the equation $\partial V_{\text{KL-ISS}}/\partial \phi_\textrm{ISS}$ to vanish. We decide to restrain the sign freedom of $\Delta$ and take it to be negative, while opting for $h>0$.\\

\noindent  $\bullet$ The ISS fermion fields $\chi_{\phi_\textrm{ISS}}$\\

We now present the masses and eigenstates of the fermionic parts ($\chi_{ai}$, $\tilde{\chi}_{ia}$ and $\chi_{ij}$) of the ISS superfields. For $i\otimes a\in N^{2}$ and $i\otimes j\in N^{2}$,
$i=a$ and $i=j$, the eigenstates are

\begin{eqnarray}
\chi_{B1} & = & \frac{1}{\sqrt{5}}\left(\sqrt{2}\chi_{ai}+\sqrt{2}\tilde{\chi}_{ia}+\chi_{ij}\right)+\mathcal{O}\left(M\right)~,\label{eq:chi_B1}\\
\chi_{B2} & = & \frac{1}{\sqrt{5}}\left(\sqrt{2}\chi_{ai}+\sqrt{2}\tilde{\chi}_{ia}-\chi_{ij}\right)+\mathcal{O}\left(M\right)~,\label{eq:chi_B2}\\
\chi_{S1} & = & \frac{1}{\sqrt{2}}\left(\chi_{ai}-\tilde{\chi}_{ia}\right)~. \label{eq:chi_S1}
\end{eqnarray}
For $i\neq a$ and $i\neq j$, one finds
\begin{eqnarray}
\chi_{S3} & = & \frac{1}{2}\left(\chi_{i>a}+\tilde{\chi}_{a<i}-\chi_{i<a}-\tilde{\chi}_{a>i}\right)~,\label{eq:chi_S3}\\
\chi_{S4} & = & \frac{1}{2}\left(\chi_{i>a}-\tilde{\chi}_{a<i}+\chi_{i<a}-\tilde{\chi}_{a>i}\right)~,\label{eq:chi_S4}\\
\chi_{B3} & = & \frac{1}{2\sqrt{2}}\left(\chi_{i>a}-\tilde{\chi}_{a<i}-\chi_{i<a}+\tilde{\chi}_{a>i}-\sqrt{2}\chi_{i<j}+\sqrt{2}\chi_{i>j}\right)+\mathcal{O}\left(M\right)~,\label{eq:chi_B3}\\
\chi_{B4} & = & \frac{1}{2\sqrt{2}}\left(\chi_{i>a}+\tilde{\chi}_{a<i}+\chi_{i<a}+\tilde{\chi}_{a>i}-\sqrt{2}\chi_{i<j}-\sqrt{2}\chi_{i>j}\right)+\mathcal{O}\left(M\right)~,\label{eq:chi_B4}\\
\chi_{B5} & = & \frac{1}{2\sqrt{2}}\left(\chi_{i>a}-\tilde{\chi}_{a<i}-\chi_{i<a}+\tilde{\chi}_{a>i}+\sqrt{2}\chi_{i<j}-\sqrt{2}\chi_{i>j}\right)+\mathcal{O}\left(M\right)~,\label{eq:chi_B5}\\
\chi_{B6} & = & \frac{1}{2\sqrt{2}}\left(\chi_{i>a}+\tilde{\chi}_{a<i}+\chi_{i<a}+\tilde{\chi}_{a>i}+\sqrt{2}\chi_{i<j}+\sqrt{2}\chi_{i>j}\right)+\mathcal{O}\left(M\right)~.\label{eq:chi_B6}
\end{eqnarray}
For example, if $N=2$, we have $\left(\chi_{S3}\right)_{N=2}=\frac{1}{2}\left(\chi_{21}+\tilde{\chi}_{12}-\chi_{12}-\tilde{\chi}_{21}\right)$, and similarly for the other states. For $i\otimes a\in N_{f}N-N^{2}$ and $i\otimes j\in N_{f}^{2}-N^{2}$,
the mass eigenstates are 
\begin{eqnarray}
\chi_{M1} & = & \frac{1}{\sqrt{2}}\left(\chi_{ia}-\tilde{\chi}_{ai}\right)~,\label{eq:chi_M1}\\
\chi_{M2} & = & \frac{1}{\sqrt{2}}\left(\chi_{ia}+\tilde{\chi}_{ai}\right)~,\label{eq:chi_M2}\\
\chi_{S2} & = & \chi_{ij}\quad,\textrm{ for }i\neq j\label{eq:chi_S2}\\
& = & \frac{1}{\sqrt{2}}\left(\chi_{qq}-\chi_{ij}\right)\quad,\textrm{ for }i=j~,\label{eq:chi_S2extra}
\end{eqnarray}
where $q\equiv N+1$. 

As an example, for $N=2$ and $N_{f}=5$ and
$i=j$, we obtain the possibilities $\chi_{S2}=\{\frac{1}{\sqrt{2}}\left(\chi_{33}-\chi_{44}\right),\,\frac{1}{\sqrt{2}}\left(\chi_{33}-\chi_{55}\right)\}$.
The Goldstino has already been subtracted out and is given by
\begin{equation}
\chi_{\textrm{Goldstino}}=\frac{1}{\sqrt{N_{f}-N}}\sum_{i=N+1}^{N_{f}}\chi_{ii}+\mathcal{O}\left(M^{2}\right)~.\label{eq:chi_goldstino}
\end{equation}
Regarding the masses of the above eigenstates, the main contribution
comes from the term $\left(e^{G/2}\frac{W_{ij}}{W}\right)\bar{\chi}_{R}\chi_{L}+\textrm{h.c.}$~. Their values are given in tables \ref{tab:iaN2_table} and \ref{tab:iaNfN_N2}.
\begin{table}[H]
\begin{center}
	\begin{tabular}{|l|c|c|}
		\hline 
		ISS fermion mass eigenstate & Number & Mass\tabularnewline
		\hline 
		\hline 
		$\chi_{B1}\equiv\textrm{Lc}\left[\chi_{ij},\chi_{ai},\tilde{\chi}_{ia}\right]\left(i\otimes a\in N^{2}\right)$ & $N$ & $\sqrt{\frac{6}{N_{f}-N}}\left(\frac{M_{P}}{M}\right)m_{3/2}$\tabularnewline
		\hline 
		$\chi_{B2}\equiv\textrm{Lc}\left[\chi_{ij},\chi_{ai},\tilde{\chi}_{ia}\right]\left(i\otimes a\in N^{2},i\otimes j\in N^{2}\right)$ & $N$ & $\sqrt{\frac{6}{N_{f}-N}}\left(\frac{M_{P}}{M}\right)m_{3/2}$\tabularnewline
		\hline 
		$\chi_{S1}\equiv\textrm{Lc}\left[\chi_{ai},\tilde{\chi}_{ia}\right]\left(i\otimes a\in N^{2},i\otimes j\in N^{2}\right)$ & $N$ & $m_{3/2}$\tabularnewline
		\hline 
		$\chi_{S3}\equiv\textrm{Lc}\left[\chi_{ai},\tilde{\chi}_{ia}\right]\left(i\otimes a\in N^{2}\right)$ & $N\left(N-1\right)/2$ & $m_{3/2}$\tabularnewline
		\hline 
		$\chi_{S4}\equiv\textrm{Lc}\left[\chi_{ai},\tilde{\chi}_{ia}\right]\left(i\otimes a\in N^{2}\right)$ & $N\left(N-1\right)/2$ & $m_{3/2}$\tabularnewline
		\hline 
		$\chi_{B3}\equiv\textrm{Lc}\left[\chi_{ij},\chi_{ai},\tilde{\chi}_{ia}\right]\left(i\otimes a\in N^{2},i\otimes j\in N^{2}\right)$ & $N\left(N-1\right)/2$ & $\sqrt{\frac{6}{N_{f}-N}}\left(\frac{M_{P}}{M}\right)m_{3/2}$\tabularnewline
		\hline 
		$\chi_{B4}\equiv\textrm{Lc}\left[\chi_{ij},\chi_{ai},\tilde{\chi}_{ia}\right]\left(i\otimes a\in N^{2},i\otimes j\in N^{2}\right)$ & $N\left(N-1\right)/2$ & $\sqrt{\frac{6}{N_{f}-N}}\left(\frac{M_{P}}{M}\right)m_{3/2}$\tabularnewline
		\hline 
		$\chi_{B5}\equiv\textrm{Lc}\left[\chi_{ij},\chi_{ai},\tilde{\chi}_{ia}\right]\left(i\otimes a\in N^{2},i\otimes j\in N^{2}\right)$ & $N\left(N-1\right)/2$ & $\sqrt{\frac{6}{N_{f}-N}}\left(\frac{M_{P}}{M}\right)m_{3/2}$\tabularnewline
		\hline 
		$\chi_{B6}\equiv\textrm{Lc}\left[\chi_{ij},\chi_{ai},\tilde{\chi}_{ia}\right]\left(i\otimes a\in N^{2},i\otimes j\in N^{2}\right)$ & $N\left(N-1\right)/2$ & $\sqrt{\frac{6}{N_{f}-N}}\left(\frac{M_{P}}{M}\right)m_{3/2}$\tabularnewline
		\hline 
	\end{tabular}
	
	\caption{Tree-level largest contributions to the masses of the ISS fermion
		eigenstates within the index space $i\otimes a\in N^{2}$ and $i\otimes j\in N^{2}$.
		The notations $i\otimes a$ and $i\otimes j$ are of the same type as the ones for the	ISS scalar masses table \ref{tab:Masses_ISS}. Furthermore, Lc stands for the linear combinations given from equations (\ref{eq:chi_B1}) to (\ref{eq:chi_B6}), and the second column gives the number of ISS fermions mass eigenstates in the considered index space.\label{tab:iaN2_table}}
		\end{center}
\end{table}
\begin{table}[H]
	\begin{center}
	\begin{tabular}{|l|c|c|}
		\hline 
		ISS fermion mass eigenstate & Number & Mass\tabularnewline
		\hline 
		\hline 
		$\chi_{M1}\equiv\textrm{Lc}\left[\chi_{ai},\tilde{\chi}_{ia}\right]\left(i\otimes a\in N_{f}N-N^{2}\right)$ & $N\left(N_{f}-N\right)$ & $\frac{16\pi^{2}}{\left(\textrm{ln}[4]-1\right)h^{2}}m_{3/2}$\tabularnewline
		\hline 
		$\chi_{M2}\equiv\textrm{Lc}\left[\chi_{ai},\tilde{\chi}_{ia}\right]\left(i\otimes a\in N_{f}N-N^{2}\right)$ & $N\left(N_{f}-N\right)$ & $\frac{16\pi^{2}}{\left(\textrm{ln}[4]-1\right)h^{2}}m_{3/2}$\tabularnewline
		\hline 
		$\chi_{S2}\equiv\textrm{Lc}\left[\chi_{ij}\right]\left(i\otimes j\in N_{f}^{2}-N^{2}\right)$ & $N_{f}^{2}-N^{2}-1$ & $0$\tabularnewline
		\hline 
	\end{tabular}
	
	\caption{Leading order contributions to the masses of the ISS fermion
		eigenstates within the index space $i\otimes a\in N_{f}N-N^{2}$ and
		$i\otimes j\in N_{f}^{2}-N^{2}$. The notations $i\otimes a$
		and $i\otimes j$ are of the same type as the ones for the ISS scalar masses table \ref{tab:Masses_ISS}. Furthermore, Lc stands for the linear combinations given from equations (\ref{eq:chi_M1}) to (\ref{eq:chi_S2extra}), and the second column gives the number of ISS fermions mass eigenstates in the considered index space.\label{tab:iaNfN_N2}}
		\end{center}
\end{table}

%%%%%%%%%%%%%%%%%%%%%%%%%%%%
%%%%%%%%%%%%%%%%%%%%%%%%%%%%
\section{Interactions and Decay Modes} \label{Sec:Interactions_DecayModes}
%%%%%%%%%%%%%%%%%%%%%%%%%%%%
%%%%%%%%%%%%%%%%%%%%%%%%%%%%

In the previous section, we have introduced the particle content of the KL-ISS scenario, discussed the uplifting of the KL vacuum, and coupled this scenario to the inflaton and MSSM fields. In the present section, we give the interaction terms between the ISS and the MSSM fields --- besides the ISS sector self-interactions ---, and compute the corresponding decay rates, which play a very important role in the discussion of the early universe and dark matter production in the next sections \ref{Sec:Post-inflation_Entropy} and \ref{Sec:DarkMatter}. 

If the total decay rate of the ISS fields is sufficiently small, the ISS fields decay quite late, well after inflaton decays reheat the universe. The ISS fields must decay before the onset of BBN in order not to jeopardize the successful BBN predictions of the Standard Model. Note that not only the ISS fields themselves but also their decay products should decay before the onset of BBN.

ISS fields decaying well before BBN may release a large amount of entropy. In fact, some versions of Affleck-Dine baryogenesis \cite{Olive_strongly_stabilized,Affleck_Dine} require a lot of late entropy production in order to yield a baryon asymmetry that matches the asymmetry observed today, $n_{B}-n_{\bar{B}}/s_{0}\sim10^{-10}$. However, baryogenesis can also be accomplished for a small ISS entropy production through the mechanisms given in e.g. \cite{Konstandin:2013caa,Morrissey:2012db}. In this work, we will consider the latter option. We thus require that the decay of the ISS fields release less entropy than inflaton decays do, and this minimizes the difference between our scenario and standard cosmology.

To compute the decay rates, we consider both two-body decays ($\phi_{\text{ISS}} \rightarrow 1 + 2$) and three-body decays ($\phi_{\text{ISS}} \rightarrow 1 + 2 + 3$), and use the following expressions
\begin{eqnarray} \label{eq:DecayRate_Two}
\frac{d\overline{\Gamma}^{12}_{\phi_{\text{ISS}}}}{d\Omega_{\textrm{CM}}} & = & \frac{\left|\mathcal{\overline{M}}^{12}_{\phi_{\text{ISS}}}\right|^{2}}{64\pi^{2}}\frac{S_{12}}{m_{\phi_{\text{ISS}}}^{3}}s~,\\
\overline{\Gamma}^{123}_{\phi_{\text{ISS}}} & = & \frac{1}{m_{\phi_{\text{ISS}}}64\pi^{3}}\int_{0}^{\frac{m_{\phi_{\text{ISS}}}}{2}}dE_{2}\int_{\frac{m_{\phi_{\text{ISS}}}}{2}-E_{2}}^{\frac{m_{\phi_{\text{ISS}}}}{2}}dE_{1}\:\left|\mathcal{\overline{M}}^{123}_{\phi_{\text{ISS}}}\right|^{2}~,
\label{eq:DecayRate_Three}
\end{eqnarray}
where $\text{d}\Omega_{\text{CM}}$ is the phase space differential element, $\mathcal{\overline{M}}^{12}_{\phi_{\text{ISS}}}$ ($\mathcal{\overline{M}}^{123}_{\phi_{\text{ISS}}}$) is the amplitude of the two(three)-body decay, $S_{12}=\left[m_{\phi_{\text{ISS}}}^{2}-\left(m_{1}-m_{2}\right)^{2}\right]^{1/2}\left[m_{\phi_{\text{ISS}}}^{2}-\left(m_{1}+m_{2}\right)^{2}\right]^{1/2}$, $m_{\phi_{\text{ISS}}}$ is the mass of the decaying ISS particle, $s$ is the symmetry factor for indistinguishable final states. Here we denote decay rates where helicities were summed over by the bar line above $\Gamma$. Since the ISS fields are much heavier\footnote{Exceptions are the fields Im$Q_2$  and $Q_4$. However we will see later that these fields do not oscillate after inflation, thus they do not yield important contributions to the energy content of the Universe and would be neglected after all.} than their decay products, we consider all final particles to be massless for simplicity.

In the following, we use the indices
\begin{equation}
	\begin{split}
	i,j,a,b~~~~&\text{for ISS fields}~,\\
	m,n,p,q,r,s~~~~&\text{for MSSM fields}~,\\
	\alpha,\beta~~~~&\text{for both ISS and MSSM fields without distinction}~,\\
	\mu,\nu, \rho, \sigma~~~~&\text{for spacetime indices}~,
	\end{split}
\end{equation}
and similarly for the same indices with an overbar for ISS and MSSM fields, e.g., $\bar{i}$, $\bar{m}$ and $\bar{\alpha}$. Recall that $\delta_{ij} = \text{diag}(1,\ldots,1)$ with $N_f$ entries. We further define
\begin{equation}
	\delta'_{ij} = \text{diag}(\underbrace{1,\ldots,1}_{N},\underbrace{0,\ldots,0}_{N_f-N})~.
\end{equation}

We derive the couplings of the ISS fields to the MSSM fields using the simplifications $K_{m},W_{m}\ll1$ for the MSSM fields. This can be done since the VEVs of the MSSM fields are either zero or at most of the weak scale. Furthermore, we have shown in table \ref{tab:Masses_ISS} and the discussion following it that we have, apart from the null masses of Im$Q_2$ and $Q_4$, 4 types of ISS scalar masses, namely 
\begin{equation} \label{eq:Masses_Types}
	\begin{split}
    ~~~~\sqrt{\frac{6}{N_{f}-N}}\left(\frac{M_{P}}{M}\right)m_{3/2}~, ~~~~
	& \mathcal{O}(m_{3/2})~, \\  \left(\frac{3\left(\textrm{ln}\left(4\right)-1\right)}{8\pi^{2}}\right)^{1/2}\sqrt{\frac{N}{N_{f}-N}}h\left(\frac{M_{P}}{M}\right)m_{3/2}~, ~~~~ & \left(\frac{3\left(\textrm{ln}\left(4\right)-1\right)}{8\pi^{2}}\right)^{1/2}h\left(\frac{M_{P}}{M}\right)m_{3/2}~.
	\end{split}
\end{equation}
The size of their masses is going to directly impact the values of their decay rates --- see equations (\ref{eq:DecayRate_Two}) and (\ref{eq:DecayRate_Three}). 

Within the category of $\left(q,\tilde{q}\right)$, we write the decay rates only for $\textrm{Re}Q_{1}$ and $\textrm{Re}Q_{2}$ using the short notation $Q_1$ and $Q_2$ for the initial particles for an easy display. As we will see, the oscillation amplitude of the former is non-vanishing, and despite the ampitude of Re$Q_2$ being zero, its decay rate is important since it is a decay product of Re$Q_1$ itself. They correspond to masses $\sqrt{\frac{6}{N_{f}-N}}\left(\frac{M_{P}}{M}\right)m_{3/2}$ and $\left(\frac{3\left(\textrm{ln}\left(4\right)-1\right)}{8\pi^{2}}\right)^{1/2}h\left(\frac{M_{P}}{M}\right)m_{3/2}$, respectively. Similarly, within the $S$ set of fields, the ones which will have non-vanishing oscillation amplitudes are the diagonal fields of $\textrm{Re}S_{1}$ and $\textrm{Re}S_{2}$, which possess masses $\sqrt{\frac{6}{N_{f}-N}}\left(\frac{M_{P}}{M}\right)m_{3/2}$ and $\left(\frac{3\left(\textrm{ln}\left(4\right)-1\right)}{8\pi^{2}}\right)^{1/2}\sqrt{\frac{N}{N_{f}-N}}h\left(\frac{M_{P}}{M}\right)m_{3/2}$~, respectively. We then also write decay rates only for Re$S_1$ and Re$S_2$ using the short notation $S_1$ and $S_2$ for the initial particles for an easy display.

Moreover, instead of using the supergravity
Lagrangian density that appears in \cite{Wess_and_Bagger}, we chose to use the form in \cite{Drees_book, Cremmer_etal} --- it can also be found in \cite{Cremmer_Julia_etal} --- because it is expressed only in terms of $G=K+\textrm{ln}\left(W\overline{W}\right)$.

In the following, we present our computations for the decays of the ISS fields to MSSM fields, the gravitino, and ISS scalars and fermions. Again, recall that we are writing all indices for the ISS fields as lower indices for an easy display. 

%%%%%%%%%%%%%%%%%%%%%%%%%%%%
\subsection{MSSM scalars}
\label{Subsec:MSSM_Scalars}
%%%%%%%%%%%%%%%%%%%%%%%%%%%%

The first interaction we consider is the one with MSSM scalars $\phi$. These come from the Lagrangian \cite{Drees_book, Cremmer_etal} --- without
D-term ---
\begin{equation}
\mathcal{L}^{\text{scalars}}_{\phi_{\text{ISS}}}=G_{m\bar{n}}D_{\mu}\phi^{m}D^{\mu}\bar{\phi}^{\bar{n}}-e^{G}\left(G_{\alpha}G^{\alpha\bar{\beta}}G_{\bar{\beta}}-3\right)~.\label{eq:scalarpotential_scalars}
\end{equation}
The first term is the kinetic term for scalars and has no importance in the decay $\phi_{\text{ISS}}\rightarrow\textrm{MSSM}~\text{scalars}$. It is shown here for we must guarantee that we have canonical kinetic terms, which is true for a canonical K\"ahler potential. The interactions of the ISS fields --- $S$, or either the $q$ or the $\tilde{q}$, denoted by $(q_{ia}, \tilde{q}_{ai})$ --- with two MSSM scalars --- $\phi_m$ and $\bar{\phi}_m$ --- are given by
\begin{align}
	\begin{split}\label{eq:Sij_2scalars}
	\mathcal{L}_{S_{ij}}^{\phi_{m}\bar{\phi}_{m}} \hspace*{0.8cm} ={}& \frac{\sqrt{3}m_{3/2} \left(\delta'_{ij}-\delta_{ij}\right)}{\sqrt{N_{f}-N}} \left(\frac{m_{3/2}}{M_{P}}+\sqrt{\frac{3}{N_{f}-N}}\frac{m_{3/2}}{hM^{2}}\frac{W\left(\phi\right)}{M_{P}^{2}}\right)\phi_{m}\bar{\phi}_{m}S_{ij} \\ & \hspace*{10.5cm} + \textrm{h.c.}~,
	\end{split}\\
	\begin{split}\label{eq:q_2scalars}
	\mathcal{L}_{\left(q_{ia},\tilde{q}_{ai}\right)}^{\phi_{m}\bar{\phi}_{m}} ={}& m_{3/2}\delta_{ia}\left(\frac{m_{3/2}M}{M_{\rm P}^2}+\sqrt{\frac{3}{N_{f}-N}}\frac{m_{3/2}}{hM}\frac{W\left(\phi\right)}{M_{P}^3}\right)\phi_{m}\bar{\phi}_{m}\left(q_{ia},\tilde{q}_{ai}\right)+\textrm{h.c.}~,
	\end{split}
\end{align}
where $\phi_{m}\bar{\phi}_{m}$ comes from the first term in the $e^{K\left(\phi,\bar{\phi}\right)}$
expansion. The Giudice-Masiero term and the $\mu$-term of $W\left(\phi\right)$
add the channels $\phi_{\textrm{ISS}}\rightarrow h_{1}^{0,+}+h_{2}^{0,-}$,
where $h_1$ and $h_2$ are the Higgs scalars of the Higgs superfields $H_1$ and $H_2$ in (\ref{eq:GiudiceMasiero}). 

The corresponding decay rates for two MSSM scalars are given by
\begin{eqnarray}
	\overline{\Gamma}^{2\phi}_{S_{1ij}} & \propto & \frac{m_{3/2}^3}{M_{\rm P}^2}\left(\frac{M}{M_{\rm P}}\right)^5\delta^{'}_{ij}~,\label{eq:S1_2scalars}\\
	\overline{\Gamma}^{2\phi}_{S_{2ij}} & = & \frac{1}{2^4\pi}\sqrt{\frac{8\pi^2\left(N_f-N\right)}{3\left(\textrm{ln}(4)-1\right)N}}\frac{3}{N_{f}-N}\frac{m_{3/2}^{3}}{hM_{P}^{2}}\left(\frac{M}{M_{\rm P}}\right)\left(\delta'_{ij}-\delta_{ij}\right)~,\\
	\overline{\Gamma}^{2\phi}_{Q_{1ia}} & = & \frac{1}{2^4\sqrt{2}\pi}\sqrt{\frac{N_f-N}{3}}\frac{m_{3/2}^{3}}{M_{P}^{2}}\left(\frac{M}{M_{\rm P}}\right)^3\delta_{ia}~,\\
	\overline{\Gamma}^{2\phi}_{Q_{2ia}} & = & \frac{1}{2^4\pi }\sqrt{\frac{8\pi^2}{3\left(\textrm{ln}(4)-1\right)}}\frac{m_{3/2}^{3}}{hM_{P}^{2}}\left(\frac{M}{M_{\rm P}}\right)^3\delta_{ia}~.
\end{eqnarray}
We must note that the result (\ref{eq:S1_2scalars}) does not come from (\ref{eq:Sij_2scalars}). Instead, it stems from suppressed terms $\bar{K}_{\bar{S}_{ij}}K_{S_{ij}}$ in (\ref{eq:scalarpotential_scalars}).

The interaction of the ISS fields with three MSSM scalars --- $\phi_{m}$, $\phi_{n}$, and $\phi_{p}$ --- is obtained considering $W\left(\phi\right)\supset W_{mnp}\phi_{m}\phi_{n}\phi_{p}$,
together with the zeroth term in the $e^{K\left(\phi,\bar{\phi}\right)}$ expansion, where $W_{mnp}$ is the derivative with respect to the three MSSM scalars $\phi_{m}\phi_{n}\phi_{p}$. These are given by
\begin{equation}
	\begin{split}
\mathcal{L}_{S_{ij}}^{\phi_{m}\phi_{n}\phi_{p}} & =  \frac{3}{N_{f}-N}\frac{m_{3/2}^{2}}{hM^{2}}W_{mnp}\left(\delta_{ij}-\delta'_{ij}\right)S_{ij}\phi_{m}\phi_{n}\phi_{p}+\textrm{h.c.}~,\\
\mathcal{L}_{\left(q_{ia},\tilde{q}_{ai}\right)}^{\phi_{m}\phi_{n}\phi_{p}} & = \sqrt{\frac{3}{N_{f}-N}}\frac{m_{3/2}^{2}}{hMM_{P}}W_{mnp}\delta_{ia}\phi_{m}\phi_{n}\phi_{p}\left(q_{ia},\tilde{q}_{ai}\right)+\textrm{h.c.}~
	\end{split}
\end{equation}

For three MSSM scalars, the decay rates are given by
\begin{eqnarray}
	\overline{\Gamma}^{3\phi}_{S_{1ij}} & \propto & W_{mnp}^2\frac{M^3m_{3/2}^5}{M_{\rm P}^7}\delta'_{ij}~,\label{eq:S1_3scalars}\\
	\overline{\Gamma}^{3\phi}_{S_{2ij}} & = & \frac{W_{mnp}^{2}}{2^{9}\pi^{3}}\sqrt{\frac{3\left(\textrm{ln}(4)-1\right)N}{8\pi^2\left(N_f-N\right)}}\left(\frac{3}{N_{f}-N}\right)^{2}\frac{m_{3/2}^{5}}{hM^{5}}M_{P}\left(\delta_{ij}-\delta'_{ij}\right)~,\\
	\overline{\Gamma}^{3\phi}_{Q_{1ia}} & = & \frac{\sqrt{2}W_{mnp}^{2}}{2^{9}\pi^{3}}\left(\frac{3}{N_{f}-N}\right)^{3/2}\frac{m_{3/2}^{5}}{h^2M^{5}}\left(\frac{M}{M_{\rm P}}\right)^{2}M_{P}\delta_{ia}~, \label{eq:Q1_3Scalars}\\
	\overline{\Gamma}^{3\phi}_{Q_{2ia}} & = & \frac{W_{mnp}^{2}}{2^{9}\pi^{3}}\sqrt{\frac{3\left(\textrm{ln}\left(4\right)-1\right)}{8\pi^{2}}}\frac{3}{N_{f}-N}\frac{m_{3/2}^{5}}{h M^{5}}\left(\frac{M}{M_{\rm P}}\right)^{2}M_{P}\delta_{ia}~.\label{eq:Q2_3Scalars}
\end{eqnarray}
Here, the rate (\ref{eq:S1_3scalars}) comes from the expansion of the K\"{a}hler exponential, written mathematically as $e^{K}\overline{W}W\left(\frac{\overline{W}_{\bar{S}_{ij}}}{W}\frac{W_{S_{ij}}}{W}\right)\supset \left(\bar{S}_{ij}{S}_{ij}\right)W(\phi)$.

As to the appearance of the $\mu$-term and Giudice-Masiero contributions, the former is suppressed compared with three-scalars decay, since  $\mu\sim\mathcal{O}\left(\textrm{electroweak scale}\right)$. If  $c_{H}\sim\mathcal{O}\left(1\right)$, the latter yields a similar contribution as the one given in $\mathcal{L}_{\phi_{\text{ISS}}}^{\phi_{m}\bar{\phi}_{m}}$ for two-body decays.

%%%%%%%%%%%%%%%%%%%%%%%%%%%%
\subsection{MSSM fermions}
\label{Subsec:MSSM_fermions}
%%%%%%%%%%%%%%%%%%%%%%%%%%%%

The second interaction we explore is the one with MSSM fermions --- $\bar{\chi}^{m}$ and $\chi^n$. The fermionic Lagrangian for Dirac
fermions is given by \cite{Drees_book, Cremmer_etal} 
\begin{eqnarray} \label{eq:Lagrangian_Fermions}
\mathcal{L}^{\text{fermions}}_{\phi_{\text{ISS}}} & = & iG_{m\bar{n}}\bar{\chi}_{R}^{m}\gamma^{\mu}D_{\mu}\chi_{R}^{n}+\left(-G_{mi\bar{n}}+\frac{1}{2}G_{m\bar{n}}G_{i}\right)\bar{\chi}_{R}^{m}\gamma^{\mu}D_{\mu}\phi^{i}_{\text{ISS}}\chi_{R}^{\bar{n}}\nonumber \\
 &  & +e^{G/2}\left(-G_{mn}-G_{m}G_{n} + G_{mn\bar{i}}G^{j\bar{i}}G_{j}\right)\bar{\chi}_{R}^{m}\chi_{L}^{n}+\textrm{h.c.}\\
 & \equiv & A+B+C+\textrm{h.c.}\nonumber 
\end{eqnarray}
We notice that only two-body decays are relevant in this case and also that the $A$ term in equation (\ref{eq:Lagrangian_Fermions}) does not matter for the decay of the ISS fields into fermions. 

Working out the contributions from the $B$ term in equation (\ref{eq:Lagrangian_Fermions}), we obtain
\begin{eqnarray}
	B_{S_{ij}} & = & \frac{1}{2M_{\rm P}}\sqrt{\frac{3}{N_{f}-N}}\left(\delta'_{ij}-\delta_{ij}\right)\bar{\chi}_{R}^{m}\gamma^{\mu}\partial_{\mu}S_{ij}\chi_{R}^{m}~, \label{eq:B_S}\\
	B_{(q_{ia},\tilde{q}_{ai})} & \propto & \frac{1}{2M_{\rm P}}\left(\frac{M}{M_{\rm P}}\right)^3\delta_{ia}\bar{\chi}_{R}^{m}\gamma^{\mu}\partial_{\mu}(q_{ia},\tilde{q}_{ai})\chi_{R}^{m}~, \label{eq:B_q}\label{eq:Bterm_2fermions_q}
\end{eqnarray}
where the term $-G_{mi\bar{n}}=-K_{mi\bar{n}}$ yields zero since the considered K\"ahler terms are
canonic --- we will consider the Giudice-Masiero term below. We did not obtain the result (\ref{eq:Bterm_2fermions_q}) in full form because the order of this term is negligible compared to the $C$ term, as we will see. The smallness of this expression is due to $G_{\left(q_{ia},\tilde{q}_{ai}\right)}=\left(\frac{K_{\left(q_{ia},\tilde{q}_{ai}\right)}}{M_{\rm P}^2}+\frac{W_{\left(q_{ia},\tilde{q}_{ai}\right)}}{W}\right)M_{\rm P}=\mathcal{O}\left(M^3/M_{\rm P}^3\right)\delta_{ia}$.

By computing the decay rates from $B$, we notice that they are suppressed because the masses of the MSSM fermion fields are negligibly small compared to the masses of the ISS fields. For vanishing MSSM fermion masses, we find
\begin{equation}
	\begin{split}
	\Gamma & \propto p_{q}^{\mu}p_{q}^{\nu}\textrm{Tr}\left[\gamma_{\mu}\left(\cancel{p}_{1}+m_{1}\right)\gamma_{\nu}\left(\cancel{p}_{2}-m_{2}\right)\right] \\ 
	& \simeq m_{\phi_{\textrm{ISS}}}^{2}\textrm{Tr}\left[\gamma_{0}\left(\cancel{p}\right)\gamma_{0}\left(\cancel{p}\right)\right] \\
	& = m_{\phi_{\textrm{ISS}}}^{2}\textrm{Tr}\left[\gamma_{0}\left(\gamma_{0}\frac{m_{\phi_{\textrm{ISS}}}}{2}+\gamma_{3}\frac{m_{\phi_{\textrm{ISS}}}}{2}\right)\gamma_{0}\left(\gamma_{0}\frac{m_{\phi_{\textrm{ISS}}}}{2}+\gamma_{3}\frac{-m_{\phi_{\textrm{ISS}}}}{2}\right)\right] \\
	& = 0~.
	\end{split}
\end{equation}

Working out the contributions from the $C$ term in equation (\ref{eq:Lagrangian_Fermions}), we obtain
\begin{eqnarray}
C_{S_{ij}} & = & -\frac{1}{2h}\frac{m_{3/2}}{ M_{\rm P}}\left(\frac{M_{\rm P}}{M}\right)^2 \left(\delta'_{ij}-\delta_{ij}\right)\left(-\frac{W_{mn}}{M_{\rm P}}\right)S_{ij}\bar{\chi}_{R}^{m}\chi_{L}^{n}~,\label{eq:C_S}\\
C_{(q_{ia},\tilde{q}_{ai})} & = & -\frac{1}{2h}\sqrt{\frac{3}{N_f-N}}\frac{m_{3/2}}{M_{\rm P}}\left(\frac{M_{\rm P}}{M}\right)\delta_{ia}\left(\frac{-W_{mn}}{M_{\rm P}}\right)(q_{ia},\tilde{q}_{ai})\bar{\chi}_{R}^{m}\chi_{L}^{n}~.\label{eq:C_q}
\end{eqnarray}

The Giudice-Masiero term $K_{H}=c_{H}H_{1}H_{2}+\textrm{h.c.}$ for Higgs superfields $H_1$ and $H_2$ yields the additional
terms
\begin{eqnarray}
C_{S_{ij}}^{\tilde{h}_{1,2}} & = & \frac{c_H}{2}\sqrt{\frac{3}{N_{f}-N}}\frac{m_{3/2}}{M_{\rm P}}\left(\delta'_{ij}-\delta_{ij}\right)S_{ij}\left(\bar{\tilde{h}}_{R}^{1}\tilde{h}^{2}_L-\bar{\tilde{h}}_{R}^{2}\tilde{h}_{L}^{1}\right)~,\label{eq:GM_S}\\
C_{(q_{ia},\tilde{q}_{ai})}^{\tilde{h}_{1,2}} & = & \frac{c_H}{2}\frac{m_{3/2}}{ M_{\rm P}}\left(\frac{M}{M_{\rm P}}\right)\delta_{ia}(q_{ia},\tilde{q}_{ai})\left(\bar{\tilde{h}}_{R}^{1}\tilde{h}^{2}_L-\bar{\tilde{h}}_{R}^{2}\tilde{h}_{L}^{1}\right)~,\label{eq:GM_q}
\end{eqnarray}
where $\tilde{h}_1$ and $\tilde{h}_2$ are the Higgsino components of the Higgs superfields $H_1$ and $H_2$.

The most important decay rates for $\phi_{\textrm{ISS}}=S_2, Q_1, Q_2$ are given
by the ones with the Higgsinos in the final state, equations (\ref{eq:GM_S}) and (\ref{eq:GM_q}). in fact, we know that the derivative of the MSSM superpotential with respect to the Higgs superfields $H_1$ and $H_2$ satisfies $W_{mn}|_{H_1,H_2} = \mu_0 \ll M_{P}$, since $\mu_0$ is at the weak scale. Therefore, the decay rates are given by
\begin{eqnarray}
\overline{\Gamma}^{2\chi}_{S_{1ij}} & \propto & c_H^2\frac{M^3m_{3/2}^3}{M_{\rm P}^5}\delta'_{ij}~,\\
\overline{\Gamma}^{2\chi}_{S_{2ij}} & = & \frac{c_{H}^{2}}{2^{5}\pi}\sqrt{\frac{3\left(\textrm{ln}(4)-1\right)N}{8\pi^2\left(N_f-N\right)}}\frac{3}{N_{f}-N}h\frac{m_{3/2}^{3}}{MM_{\rm P}}\left(\delta'_{ij}-\delta_{ij}\right)~,\\
\overline{\Gamma}^{2\chi}_{Q_{1ia}} & = & \frac{c^2_H}{2^7\pi}\sqrt{\frac{N_f-N}{6}}\frac{m_{3/2}^3M}{M_{\rm P}^3}\delta_{ia}~,\\
\overline{\Gamma}^{2\chi}_{Q_{2ia}} & = & \frac{c^2_H}{2^7\pi}\sqrt{\frac{3\left(\textrm{ln}(4)-1\right)}{8\pi^2}}h  \frac{m_{3/2}^3M}{M_{\rm P}^3}\delta_{ia}~.
\end{eqnarray}
The decay rate for $S_1$ comes again from higher order terms, namely the expansion of $e^{G/2}$ at the term $C$. For $S_{2ij}$, the interference terms between $B_{S_{ij}}$ and $C_{S_{ij}}$, and between $B_{S_{ij}}$ and $C_{S_{ij}}^{\tilde{h}_{1,2}}$ can be neglected. This happens because $m_{S_{ij}} \gg m_{\tilde{h}_1},m_{\tilde{h}_2}$ and the trace of an odd number of gamma matrices vanishes.

%%%%%%%%%%%%%%%%%%%%%%%%%%%%
\subsection{Two MSSM fermions plus one MSSM Scalar}
\label{Subsec:MSSM_2Fermions_1Scalar}
%%%%%%%%%%%%%%%%%%%%%%%%%%%%

The third interaction we consider is the one with two MSSM fermions $\bar{\chi}_{m}$ and $\chi_{n}$,  and one MSSM scalar $\phi_{p}$, namely $\phi_{\textrm{ISS}}\rightarrow\bar{\chi}_{m}\chi_{n}\phi_{p}$. The effective Lagrangian for this interaction is given by
\begin{equation} \label{eq:L_2f1s}
	\begin{split}
	\mathcal{L}_{\text{ISS}}^{2f+1s} & \simeq \left(\frac{1}{2}W_{mnp}\right)\left(\frac{i\gamma_{\nu}p_{\chi_{m}}^{\nu}+m_{\chi_m}}{\left(p_{\chi_{m}}^{\nu}\right)^{2}-m_{\chi_m}^2}\right)\left[\frac{1}{2}\left(\frac{K_{\phi_{\text{ISS}}}}{M_{\rm P}}+M_{\rm P}\frac{W_{\phi_{\text{ISS}}}}{W}\right)\gamma^{\mu}\right]\partial_{\mu}\phi_{\textrm{ISS}}\bar{\chi}_{m}\chi_{n}\phi_{p}~\\
 	& \simeq \frac{i}{4}\left(\frac{K_{\phi_{\text{ISS}}}}{M_{\rm P}}+M_{\rm P}\frac{W_{\phi_{\text{ISS}}}}{W}\right)\left(W_{mnp}\right)\left(\frac{p_{\nu}^{\chi_{m}}}{\left(p_{\nu}^{\chi_{m}}\right)^{2}}\right)\gamma^{\nu}\gamma^{\mu}\partial_{\mu}\phi_{\textrm{ISS}}\bar{\chi}_{m}\chi_{n}\phi_{p}~,
	\end{split}
\end{equation}
where the process is a two-vertex diagram $\phi_{\textrm{ISS}} \rightarrow \bar{\chi}_{m} \chi_m \rightarrow \bar{\chi}_m\chi_n\phi_p$. One vertex of the diagram yields $\frac{i}{2}\left(K_{\phi_{\text{ISS}}} +\frac{W_{\phi_{\text{ISS}}}}{W}\right)\gamma^{\mu}$
for $\phi_{\text{ISS}} \rightarrow \bar{\chi}_{m} \chi_{m}$, which we obtained from the second term in equation (\ref{eq:Lagrangian_Fermions}). The other vertex given by $\frac{1}{2}W_{mnp}$
corresponds to $\chi_{m}\rightarrow\chi_{n}\phi_{p}$. This is a vertex
within the MSSM coming from a Yukawa coupling term. Here we use the notation that $K_{\phi_{\text{ISS}}}$ and $W_{\phi_{\text{ISS}}}$ are the derivatives of the K\"ahler potential $K$ and the superpotential $W$ with respect to the ISS fields. Finally, the propagator
of the MSSM fermion, $\chi_{m}$ (which is considered massless) is given by $\frac{i\gamma_{\nu}p_{\chi_{m}}^{\nu}}{\left(p_{\chi_{m}}^{\nu}\right)^{2}}$,
where $p_{\chi_{m}}^{\nu}$ is the off-shell momentum of $\chi_{m}$. Hence, the final decay rates for these three-body decay processes are given by 
\begin{equation}
	\overline{\Gamma}^{2\chi+\phi}_{S_{1ij}} \propto  \left|W_{mnp}\right|^{2}\frac{m_{3/2}^3}{M_{\rm P}^2}\left(\frac {M}{M_{\rm P}}\right)\delta'_{ij}~,\label{eq:2ferm_1bos_S1}
\end{equation}
\begin{equation}
	\begin{split}
	\overline{\Gamma}^{2\chi+\phi}_{S_{2ij}} \simeq \left|W_{mnp}\right|^{2}\frac{3.6\times10^{-3}}{2^{12}\pi^{3}} & \left(\frac{3\left(\textrm{ln}(4)-1\right)N}{8\pi^2\left(N_f-N\right)}\right)^{3/2} \\ & \times \frac{3h^3 m_{3/2}^{3}}{M_{\rm P}^{2} (N_{f}-N)}\left(\frac{M_{\rm P}}{M}\right)^3 \left(\delta_{ij}-\delta'_{ij}\right),
	\end{split}
\end{equation}
\begin{equation}
	\overline{\Gamma}^{2\chi+\phi}_{Q_{1ia}} \propto  \left|W_{mnp}\right|^{2}\frac{m_{3/2}^3}{M_{\rm P}^2}\left(\frac {M}{M_{\rm P}}\right)^3\delta_{ia}~,\label{eq:2fermionsplus1scalar_a}
\end{equation}
\begin{equation}
	\overline{\Gamma}^{2\chi+\phi}_{Q_{2ia}} \propto \left|W_{mnp}\right|^{2}\frac{m_{3/2}^3}{M_{\rm P}^2}\left(\frac {M}{M_{\rm P}}\right)^3\delta_{ia}~,\label{eq:2fermionsplus1scalar_b}
\end{equation}
where the pre-factor of $3.6\times10^{-3}$ comes from the integration in equation (\ref{eq:DecayRate_Three}). We notice that equations (\ref{eq:2fermionsplus1scalar_a}) and (\ref{eq:2fermionsplus1scalar_b}) were not obtained in full form for we recall the behaviour of the function $G_{\left(q_{ia},\tilde{q}_{ai}\right)}=\mathcal{O}\left(M^3/M_{\rm P}^3\right)\delta_{ia}$ for $q_{ia}$ and $\tilde{q}_{ai}$ explained in section \ref{Subsec:MSSM_fermions}. Equation (\ref{eq:2ferm_1bos_S1}) is obtained when considering the term $\frac{K_{S_{ij}}}{M_{\rm P}}$ from within the covariant derivative of $S_{ij}$.

%%%%%%%%%%%%%%%%%%%%%%%%%%%%
\subsection{MSSM gauge sector}
\label{Subsec:MSSM_Gluons_Gluinos}
%%%%%%%%%%%%%%%%%%%%%%%%%%%%

The fourth interaction we are interested in is the one with MSSM gauge bosons $g$ and gauginos $\tilde{g}$. Unless otherwise assumed, the gauge kinetic function does not contain any of the ISS fields. Thus, no tree-level coupling to gauge bosons is allowed. However, it is possible for $\phi_{\text{ISS}}$ to decay into two gauge bosons or two gauginos at the quantum level due to anomaly effects \cite{Endo_inflaton_2gravitinos}. The corresponding decay rate is given by
\begin{eqnarray} \label{eq:Decay_Gluon_Gluinos}
\Gamma\left(\phi_{\text{ISS}} \rightarrow gg,\tilde{g}\tilde{g}\right) & \sim & \frac{N_{g}\alpha_g^{2}}{2^8\pi^{3}} \frac{\left|K_{\phi_{\text{ISS}}}\right|^{2}m_{\phi_{\text{ISS}}}^{3}}{M_{\rm P}^4}~,
\end{eqnarray}
where $N_{g}$ is the number of generators of the relevant gauge group and $\alpha_g$ is its corresponding fine structure constant. For each of the ISS fields, we obtain
\begin{eqnarray}
	\Gamma\left(Q_{1ia}\rightarrow gg,\tilde{g}\tilde{g}\right) & \sim & \frac{N_{g}\alpha_g^{2}}{2^8\pi^{3}}\left(\frac{6}{N_{f}-N}\right)^{3/2}\frac{m_{3/2}^{3}}{M M_{\rm P}}\delta_{ia}~,\\
	\Gamma\left(Q_{2ia}\rightarrow gg,\tilde{g}\tilde{g}\right) & \sim & \frac{N_{g}\alpha_g^{2}}{2^8\pi^{3}}\left(\frac{3\left(\textrm{ln}\left(4\right)-1\right)}{8\pi^{2}}\right)^{3/2}h^{3}\frac{m_{3/2}^{3}}{M M_{\rm P}}\delta_{ia}~,\\
	\Gamma\left(S_{1ij}\rightarrow gg,\tilde{g}\tilde{g}\right) & \sim & \frac{N_{g}\alpha_g^{2}}{2^7\pi^{3}}\left(\frac{6}{N_{f}-N}\right)^{1/2}  \frac{m_{3/2}^3 M}{M_{\rm P}^3}\delta^{'}_{ij}~,\\
	\Gamma\left(S_{2ij}\rightarrow gg,\tilde{g}\tilde{g}\right) & \sim & \frac{N_{g}\alpha_g^{2}}{2^6\pi^{3}}\sqrt{\frac{8\pi^2\left(N_f-N\right)}{3\left(\textrm{ln}(4)-1\right)N}} h^{-1}\frac{m_{3/2}^3 M}{M_{\rm P}^3}\left(\delta_{ij}-\delta^{'}_{ij}\right)~.
\end{eqnarray}

%%%%%%%%%%%%%%%%%%%%%%%%%%%%
\subsection{Gravitinos}
\label{Subsec:Gravitinos}
%%%%%%%%%%%%%%%%%%%%%%%%%%%%

The fifth interaction we consider is the one with gravitinos $\psi_\mu \equiv \psi_{3/2}$. The Lagrangian concerning the gravitino is given by the fermionic kinetic and fermionic mass Lagrangians \cite{Drees_book, Cremmer_etal}
\begin{equation} \label{eq:L_Gravitinos}
	\mathcal{L}^{3/2}_{\phi_{\text{ISS}}}=-\frac{i}{8}\epsilon^{\mu\nu\rho\sigma}\bar{\psi}_{\mu}\gamma_{\nu}\psi_{\rho}G_{i}D_{\sigma}\phi^{i}_{\text{ISS}}+\frac{i}{2}e^{G/2}\bar{\psi}_{\mu L}\sigma^{\mu\nu}\psi_{\nu R}+\textrm{h.c.}\equiv D+E~,
\end{equation}
where $\sigma^{\mu\nu}=\frac{i}{2}\left[\gamma^{\mu},\gamma^{\nu}\right]$.

Working out the contributions from the $D$ terms in equation (\ref{eq:L_Gravitinos}), we obtain
\begin{eqnarray}
D_{S_{ij}} & = & \frac{i\left(\delta'_{ij}-\delta_{ij}\right)}{8M_{\rm P}}\sqrt{\frac{3}{N_{f}-N}}\epsilon^{\mu\nu\rho\sigma}\bar{\psi}_{\mu}\gamma_{\nu}\psi_{\rho}\partial_{\sigma}S_{ij}~,\\
D_{(q_{ia},\tilde{q}_{ai})} & \propto & \left(\frac{M}{M_{\rm P}}\right)^3\frac{i\delta_{ia}}{8M_{\rm P}}\epsilon^{\mu\nu\rho\sigma}\bar{\psi}_{\mu}\gamma_{\nu}\psi_{\rho}\partial_{\sigma}\left(q_{ia},\tilde{q}_{ai}\right)~,
\end{eqnarray}
where the term $\mathcal{O}\left(M^3\right)$ stems from the function $G_{\left(q_{ia},\tilde{q}_{ai}\right)}$  which we discussed earlier in section \ref{Subsec:MSSM_fermions}.

Working out the contributions from the $E$ terms in equation (\ref{eq:L_Gravitinos}), we obtain
\begin{eqnarray}
	E_{S_{ij}} & = & \frac{\left(\delta_{ij}-\delta'_{ij}\right)}{8}\sqrt{\frac{3}{N_{f}-N}}\frac{m_{3/2}}{M_{\rm P}}S_{ij}\bar{\psi}_{\mu}\left[\gamma^{\mu},\gamma^{\nu}\right]\psi_{\nu}~,\\
	E_{(q_{ia},\tilde{q}_{ai})} & = & \frac{\delta_{ia}}{8}\frac{m_{3/2}M}{M_{\rm P}^2}\left(q_{ia},\tilde{q}_{ai}\right)\bar{\psi}_\mu\left[\gamma^\mu,\gamma^\nu\right]\psi_\nu~.
\end{eqnarray}

Both $D$ and $E$ terms yield similar contributions, therefore both must be taken into account to obtain the decay rate for each of the particles. The resulting decay rates are given by
\begin{eqnarray}
	\overline{\Gamma}^{2\psi_{3/2}}_{S_{1ij}} & = & \frac{5\sqrt{2}\delta'_{ij}}{18\times 2^5\pi}\left(\frac{3}{N_{f}-N}\right)^{5/2}\frac{m_{3/2}^3}{MM_{\rm P}}~,\label{eq:S1_Gravitinos}\\	
	\overline{\Gamma}^{2\psi_{3/2}}_{S_{2ij}} & = & \frac{5\left(\delta_{ij}-\delta'_{ij}\right)}{18\times 2^6\pi}\frac{3}{N_{f}-N}\left(\frac{3\left(\textrm{ln}\left(4\right)-1\right)N}{8\pi^{2}\left(N_f-N\right)}\right)^{5/2}h^5\frac{m_{3/2}^{3}}{M_{\rm P}^{2}}\left(\frac{M_{\rm P}}{M}\right)^5~,\label{eq:S2_Gravitinos}\\
	\overline{\Gamma}^{2\psi_{3/2}}_{Q_{1ia}} & = & \frac{\sqrt{2}\delta_{ia}}{9\times2^{4}\pi}\left(\frac{3}{N_{f}-N}\right)^{5/2}\frac{m_{3/2}^{3}}{M_{\rm P}^{2}}\left(\frac{M_{\rm P}}{M}\right)^3~, \label{eq:Q1_Gravitinos}\\
	\overline{\Gamma}^{2\psi_{3/2}}_{Q_{2ia}} & = & \frac{\delta_{ia}}{9\times2^{6}\pi}\left(\frac{3\left(\textrm{ln}\left(4\right)-1\right)}{8\pi^{2}}\right)^{5/2}h^5\frac{m_{3/2}^{3}}{M_{\rm P}^{2}}\left(\frac{M_{\rm P}}{M}\right)^3~.\label{eq:Q2_Gravitinos}
\end{eqnarray}
The decay rate for $S_1$ originates from higher order terms due to the K\"ahler potential of the D term and due to the expansion of the exponential of the K\"{a}hler potential in the $E$ term of equation (\ref{eq:L_Gravitinos}).

The remaining task is to study the decays of heavy ISS states into the ISS sector itself, namely lighter ISS scalars and ISS fermions. We do it in the next three subsections below, and then conclude this section on interactions and decay modes with a brief discussion of our results.

%%%%%%%%%%%%%%%%%%%%%%%%%%%%
\subsection{ISS scalars}
\label{Subsec:ISSscalars}
%%%%%%%%%%%%%%%%%%%%%%%%%%%%
We now consider the decay rates from ISS scalars. The decay rates for $S_1$ are calculated from the Lagrangian term $e^G\left(K_{\bar{S}_{ij}}K_{S_{ij}}\right)\supset \frac{m_{3/2}^2}{h}W_{\phi_\textrm{ISS}}S_{ij}\delta^{'}_{ij}$. For $S_2$, they stem from $e^G\left(K_{\bar{S}_{ij}}\frac{W_{S_{ij}}}{W}\right)\supset \frac{3}{N_f-N}\frac{m_{3/2}^2}{hM^2}W_{\phi_\textrm{ISS}}S_{ij}\left(\delta^{'}_{ij}-\delta_{ij}\right)$. And, finally, for $Q_i$, they are obtained from $e^G\left(K_{(\bar{q},\bar{\tilde{q}})_{ia}}K_{(q,\tilde{q})_{ia}}\right)\supset \sqrt{\frac{3}{N_f-N}}\frac{m_{3/2}^2}{hM}W_{\phi_\textrm{ISS}}q_{ia}\delta_{ia}$. 

There are various decays of the ISS fields into two ISS scalars. Here we choose to write only the ones which are most relevant:
\begin{eqnarray}
\overline{\Gamma}\left(S_{1}\rightarrow Q_{4}+Q_{4}^{*}\right) & = & \frac{2\sqrt{2}}{9\sqrt{3}} \frac{\pi^3}{[\text{ln}(4)-1]^2}\frac{(N_f-N)^{7/2}}{N}\frac{m_{3/2}^{3}}{h^{4} M_{\rm P}^2}\left(\frac{M}{M_{\rm P}}\right)^5~,\\
\overline{\Gamma}\left(S_{2}\rightarrow Q_{4}+Q_{4}^{*}\right) & = & \frac{8\sqrt{2}}{\sqrt{3}}\frac{\pi^4}{[\text{ln}(4)-1]^{5/2}}\left(\frac{N_f-N}{N}\right)^{3/2}\frac{m_{3/2}^{3}}{h^{5}M_{\rm P}^2}\left(\frac{M}{M_{\rm P}}\right)~,\\
\overline{\Gamma}\left(Q_{1}\rightarrow Q_{4}+Q_{4}^{*}\right) & = & \frac{1}{\sqrt{6}}\frac{\pi^3}{[\text{ln}(4)-1]^2}\frac{(N_f-N)^{3/2}}{N} \frac{m_{3/2}^{3}}{h^{4}M_{\rm P}^2}\left(\frac{M}{M_{\rm P}}\right)^3~,\\
\overline{\Gamma}\left(Q_{2}\rightarrow Q_{4}+Q_{4}^{*}\right) & = & \frac{8\sqrt{2}}{3\sqrt{3}}\frac{\pi^4}{[\text{ln}(4)-1]^{5/2}}\frac{(N_f-N)^2}{N}\frac{m_{3/2}^{3}}{h^{5}M_{\rm P}^2}\left(\frac{M}{M_{\rm P}}\right)^3~.
\end{eqnarray}
Here the notation is such that, for example, $\overline{\Gamma}(S_1 \rightarrow Q_4 + Q_4^*)$ takes into account two processes, namely $S_1$ to 2Re$Q_4$ and $S_1$ to 2Im$Q_4$. The other decay rates above for two ISS scalars as well as below for three ISS scalars follow the same notation.

For three ISS scalars, we choose again to write only the results for the stronger decay rates, which are given by
\begin{eqnarray}
\overline{\Gamma}\left(S_{1}\rightarrow S_{2}+Q_{4}+Q_{4}^{*}\right) & = & \frac{\sqrt{3}}{2^{10}\sqrt{2}}\frac{1}{\pi^3}N(N_f-N)^{3/2}\frac{m_{3/2}^{5}}{M M_{\rm P}^3}~,\\
\overline{\Gamma}\left(S_{2}\rightarrow S^{\text{nd}}_{2}+Q_{4}+\textrm{Im}Q_{2}\right) & = & \frac{9\sqrt{3}}{2^{12}\sqrt{2}}\frac{\sqrt{\text{ln}(4)-1}}{\pi^4}\frac{N^{5/2}}{(N_f-N)^{3/2}}\frac{m_{3/2}^{5}h}{M^{5}}M_{\rm P}~,\\
\overline{\Gamma}\left(Q_{1}\rightarrow S_{2}+Q_{4}+Q_{4}^{*}\right) & = & \frac{3\sqrt{3}}{2^{10}\sqrt{2}}\frac{1}{\pi^3}N\sqrt{N_f-N}\frac{m_{3/2}^{5}}{M^{3}M_{\rm P}}~,\\
\overline{\Gamma}\left(Q_{2}\rightarrow S^{\text{nd}}_{2}+Q_{4}+\textrm{Im}Q_{2}\right) & = & \frac{3\sqrt{3}}{2^{12}\sqrt{2}}\frac{\sqrt{\text{ln}(4)-1}}{\pi^4}N^2\frac{m_{3/2}^{5}h}{M^{3}M_{\rm P}}~.
\end{eqnarray}
Recall that $S^{\text{nd}}_{2}$ stands for $S_2$ at the subspace $i\otimes j=N(N_f-N)$ with tree-level mass $\mathcal{O}(m_{3/2})$ as explained in section 2.

%%%%%%%%%%%%%%%%%%%%%%%%%%%%
\subsection{ISS fermions}
\label{Subsec:ISSfermions}
%%%%%%%%%%%%%%%%%%%%%%%%%%%%

We now consider the decay rates from ISS fermions. For $S_1$, the decay rate is calculated from $e^{G/2}\left(\frac{W_{ij}}{W}\right)\bar{\chi}^i_R\chi^j_L+\textrm{h.c.}\supset \sqrt{\frac{3}{N_f-N}}\frac{m_{3/2}}{hM^2}W_{ij}\bar{\chi}^i_R\chi^j_L+\textrm{h.c.}$, whereas the other decay rates come fom $e^{G/2}\left(K_iK_j\right)\bar{\chi}^i_R\chi^j_L+\textrm{h.c.}\supset m_{3/2}K_iK_j\bar{\chi}^i_R\chi^j_L+\textrm{h.c.}$~.

The stronger decay rates from decays of the ISS fields into two ISS fermions have the following expressions
\begin{align}
\overline{\Gamma}\left(S_{1}\rightarrow\chi_{S1}+\bar{\chi}_{S1}\right)  ={}& \frac{3\sqrt{3}}{4\sqrt{2}}\frac{1}{\pi}\frac{1}{(N_{f}-N)^{3/2}}\frac{m_{3/2}^{3}}{M_{\rm P}^{2}}\left(\frac{M_{\rm P}}{M}\right)^5~,\\
\overline{\Gamma}\left(S_{1}\rightarrow\chi_{S3}+\bar{\chi}_{S3}~,\chi_{S4}+\bar{\chi}_{S4}\right) ={}& \frac{3\sqrt{3}}{2\sqrt{2}}\frac{1}{\pi}\frac{N-1}{(N_{f}-N)^{3/2}}\frac{m_{3/2}^{3}}{M_{\rm P}^{2}}\left(\frac{M_{\rm P}}{M}\right)^5~,
\label{eq:nondiagonal_S1}\\
\overline{\Gamma}\left(S_{2}\rightarrow\chi_{S2}+\bar{\chi}_{S2}\right) ={}& \frac{8\sqrt{2}}{3\sqrt{3}}\frac{\pi^2}{(\text{ln}(4)-1)^{3/2}}\frac{(N_f-N)^{5/2}}{N^{3/2}} \frac{m_{3/2}^{3}}{h^{3}M_{\rm P}^2}\left(\frac{M}{M_{\rm P}}\right)^3~,\\
\overline{\Gamma}\left(Q_{1}\rightarrow\chi_{S1}+\bar{\chi}_{S1}\right) ={}& \frac{\sqrt{3}}{2^4\sqrt{2}}\frac{1}{\pi}\frac{N}{\sqrt{N_f-N}}\frac{m_{3/2}^{3}}{M_{\rm P}^2}\left(\frac{M}{M_{\rm P}}\right)~,\\
\overline{\Gamma}\left(Q_{2}\rightarrow\chi_{S1}+\bar{\chi}_{S1}\right) ={}& \frac{\sqrt{3}}{2^6\sqrt{2}}\frac{\sqrt{\text{ln}(4)-1}}{\pi^2}N\frac{m_{3/2}^{3} h}{M_{\rm P}^2}\left(\frac{M}{M_{\rm P}}\right)~,
\end{align}
where $\chi_{S1}$, $\chi_{S2}$, $\chi_{S3}$, $\chi_{S4}$ are the small mass fermions given in tables \ref{tab:iaN2_table} and \ref{tab:iaNfN_N2}.

%%%%%%%%%%%%%%%%%%%%%%%%%%%%
\subsection{Two ISS fermions plus one ISS scalar}
\label{Subsec:ISSfermions+ISSscalar}
%%%%%%%%%%%%%%%%%%%%%%%%%%%%

The last decay we deal with is the one of the ISS fields to two ISS fermions and one ISS complex scalar. Two contributions for this kind of process are relevant to discuss, namely the contribution from 4-point vertices and the contribution with an intermediate fermion propagator. The former   normally stem from the higher order term $\frac{1}{3}e^{G/2}\left(\frac{W_iW_j}{W^2}\right)\bar{\chi}^i_R\chi^j_L+\text{h.c.}\supset \frac{1}{3}\frac{3}{N_f-N}\frac{m_{3/2}}{h^2M^4}W_iW_j\bar{\chi}^i_R\chi^j_L+\textrm{h.c.}$~. However, for the relevant four-point process here, namely for the decay of $S_2$ as we see below, the lower order four-point interaction $\frac{1}{3}\sqrt{\frac{3}{N_f-N}}\frac{m_{3/2}}{hM^2}K_iW_j\bar{\chi}^i_R\chi^j_L + \text{h.c.}$ turns out to be more relevant since it does not vanish like the higher order interaction. Regarding the processes with an intermediate fermion propagator, the largest amplitudes contain at each vertex the coupling $\sqrt{\frac{3}{N_f-N}}\frac{m_{3/2}}{hM^2}W_{ij}$ and the fermion propagator itself. The stronger decay rate for $S_2$ happens through a 4-point vertex whereas the stronger decay rates for $S_1$, $Q_1$ and $Q_2$ happen via fermion exchange. They are given by
\begin{align}
	\begin{split}
	\overline{\Gamma}\left(S_{1}\rightarrow\chi_{S1}+\bar{\chi}_{S1}+Q_{2}\right) ={}& \frac{3\sqrt{3}}{2^9\sqrt{2}}\frac{1}{\pi^3}\frac{1}{(N_f-N)^{5/2}}\left(\frac{m_{3/2}}{M}\right)^5 M_{\rm P} \\ & \simeq 3.62\times 10^{-9} \frac{1}{(N_f-N)^{3/2}}\frac{m_{3/2}^{3}h^2}{M_{\rm P}^2}\left(\frac{M_{\rm P}}{M}\right)~,
	\end{split}\\
	\begin{split}
	\overline{\Gamma}\left(S_{2}\rightarrow\chi_{S_2} + \bar{\chi}_{S_2} + \text{Re}Q_4 \right) ={}& \frac{\sqrt{3}}{2^{15}\sqrt{2}}\frac{[\text{ln}(4)-1]^{3/2}}{\pi^6} \frac{N^{5/2}}{(N_f-N)^{3/2}}\left(\frac{m_{3/2}}{M}\right)^5 h^3M_{\rm P}\\ & \simeq 1.46\times 10^{-13} \frac{N^{5/2}}{(N_f-N)^{1/2}}\frac{m_{3/2}^{3}h^5}{M_{\rm P}^2}\left(\frac{M_{\rm P}}{M}\right)~,
		\end{split}\\
	\begin{split}
	\overline{\Gamma}\left(Q_{1}\rightarrow\chi_{S1}+\bar{\chi}_{S1}+Q_{2}\right) ={}& \frac{27\sqrt{3}}{2^{15}\sqrt{2}}\frac{1}{\pi^3}\frac{1}{(N_f-N)^{5/2}}\frac{m_{3/2}^5}{M^9}M_{\rm P}^5\\ & \simeq 5.09\times10^{-10}\frac{1}{(N_f-N)^{3/2}} \frac{m_{3/2}^{3}h^{2}}{M_{\rm P}^{2}}\left(\frac{M_{\rm P}}{M}\right)^5~,
	\end{split}\\
	\begin{split}
\overline{\Gamma}\left(Q_{2}\rightarrow\chi_{S1}+\bar{\chi}_{S1}+\text{Im}Q_{2}\right) ={}& \frac{45\sqrt{3}}{2^{24}\sqrt{2}}\frac{[\text{ln}(4)-1]^{3/2}}{\pi^6}\frac{1}{N_f-N}\frac{m_{3/2}^5 h^3}{M^9} M_{\rm P}^5 \\ & \simeq 2.57\times 10^{-14} \frac{m_{3/2}^{3}h^{5}}{M_{\rm P}^{2}}\left(\frac{M_{\rm P}}{M}\right)^5~.
	\end{split}
\end{align}
Here we use the notation $\overline{\Gamma}\left(S_{1}\rightarrow\chi_{S1}+\bar{\chi}_{S1}+Q_{2}\right)$ to denote the process with an ISS complex scalar $Q_2$ in the products. In other words, this notation denotes two decays, namely~$\overline{\Gamma}\left(S_{1}\rightarrow\chi_{S1}+\bar{\chi}_{S1}+ \text{Re}Q_{2}\right)$ and $\overline{\Gamma}\left(S_{1}\rightarrow\chi_{S1}+\bar{\chi}_{S1}+ \text{Im}Q_{2}\right)$. A similar notation is employed for the decay of $Q_1$.

%%%%%%%%%%%%%%%%%%%%%%%%%%%%
\subsection{Brief discussion on results for decay rates}
\label{Subsec:Results_Decays}
%%%%%%%%%%%%%%%%%%%%%%%%%%%%

Now we briefly comment on our results regarding the decay rates computed above. Assuming $N=1$ and a moderate value for $N_f$ of order $\mathcal{O}(1)$ while still respecting the relation $N_f>3N$, we observe that the largest contributions to the total decay rates of the ISS fields
originate from their decays
\vspace*{0.15cm}

$\bullet$ to gravitinos via $\left(S_2,\,Q_1,\,Q_2\right) \rightarrow \bar{\psi}_{3/2}\psi_{3/2}$~; 

$\bullet$ to two ISS fermions via $S_1 \rightarrow \bar{\chi}_{S1}+\chi_{S1}$~;

$\bullet$ and to two ISS fermions plus one ISS scalar via both $Q_{1}\rightarrow\chi_{S1}+\bar{\chi}_{S1}+Q_{2}$ and \\ \hspace*{0.8cm} $Q_{2}\rightarrow\chi_{S1}+\bar{\chi}_{S1}+\textrm{Im}Q_{2}$~. 
\vspace*{0.15cm}
We summarize these results below,
\begin{eqnarray}
	\Gamma^\textrm{total}_{S_1} & \simeq & \overline{\Gamma}\left(S_1 \rightarrow \bar{\chi}_{S1}+\chi_{S1}\right)~,\label{eq:S_1_decay_total}\\
	\Gamma^\textrm{total}_{S_2} & \simeq & \overline{\Gamma}^{2\psi_{3/2}}_{S_{2ij}}~,\label{eq:S_2_decay_total}\\
	\Gamma^\textrm{total}_{Q_1} & \simeq & \overline{\Gamma}^{2\psi_{3/2}}_{Q_{1ia}} + \overline{\Gamma}\left(Q_{1}\rightarrow\chi_{S1}+\bar{\chi}_{S1}+Q_{2}\right)~,\label{eq:Q_1_decay_total}\\
	\Gamma^\textrm{total}_{Q_2} & \simeq & \overline{\Gamma}^{2\psi_{3/2}}_{Q_{2ia}} + \overline{\Gamma}\left(Q_{2}\rightarrow\chi_{S1}+\bar{\chi}_{S1}+\textrm{Im}Q_{2}\right)~.\label{eq:Q_2_decay_total}
\end{eqnarray}
For the ones relevant for the analysis in sections \ref{Sec:Post-inflation_Entropy} and \ref{Sec:DarkMatter}, we will use the following short notations

$\bullet$ $\overline{\Gamma}\left(S_1 \rightarrow \bar{\chi}_{S1}+\chi_{S1}\right)$ as $\Gamma_{S_1}^{\chi\chi}$~; 

$\bullet$ $\overline{\Gamma}\left(Q_{1}\rightarrow\chi_{S1}+\bar{\chi}_{S1}+Q_{2}\right)$ as $\Gamma_{Q_1}^{\chi\chi Q_2} = \Gamma_{Q_1}^{\chi\chi\text{Re}Q_2} + \Gamma_{Q_1}^{\chi\chi\text{Im}Q_2}$~;

$\bullet$ and $\overline{\Gamma}\left(Q_{2}\rightarrow\chi_{S1}+\bar{\chi}_{S1}+\textrm{Im}Q_{2}\right)$ as $\Gamma_{Q_2}^{\chi\chi\textrm{Im}Q_2}$.\\

In the following sections \ref{Sec:Post-inflation_Entropy} and \ref{Sec:DarkMatter}, we obtain constraints on $h$ and $M$ imposed by entropy dilution and dark matter production. We expect to confirm the smallness of $M$ reminescent of its dynamical origin within the ISS SUSY breaking sector.

%%%%%%%%%%%%%%%%%%%%%%%%%%%%
%%%%%%%%%%%%%%%%%%%%%%%%%%%%
\section{Post-inflation dynamics and entropy production} \label{Sec:Post-inflation_Entropy}
%%%%%%%%%%%%%%%%%%%%%%
%%%%%%%%%%%%%%%%%%%%%%

So far we have obtained the masses for the relevant fields  considered in this work --- the moduli, the gravitino, and the ISS fields --- as well as interactions of the ISS fields with the MSSM fields, the gravitino, and  light ISS scalars and fermions. In this section and in section \ref{Sec:DarkMatter}, we set these masses and interactions against the constraints of small entropy dilution at the primordial Universe and dark matter production. We leave the discussion of the latter for the next section and focus instead on the entropy issue in the present section.

%%%%%%%%%%%%%%%%%%%%%%%%%%%%
\subsection{Oscillations}
\label{subsec:Oscillations}
%%%%%%%%%%%%%%%%%%%%%%%%%%%%

We start with a detailed study of oscillations of the inflaton $\eta$ and the ISS fields. We determine which of the ISS fields are relevant for the subsequent analysis and, furthermore, discuss the epochs of the decays of the inflaton $\eta$ and the relevant ISS fields so that we set these against entropy dilution and dark matter constraints.

The procedure is to analyze the ISS fields after inflation has already occurred together with possible modifications to the ISS model due to the inflaton field itself, which has been neglected in our analysis so far. 

Recall the VEVs of the ISS fields without contributions from the inflaton $\eta$, see equation (\ref{eq:VEV_q_2}):
\begin{eqnarray}
\begin{split}
\left\langle q_{ia},\tilde{q}_{ai}\right\rangle & \simeq M\delta_{ia}~,\\
\left\langle S_{ij}\right\rangle & \simeq\left(\frac{M}{M_{P}}\right)^{2}M_{P}\,\delta_{ij}~~~\text{for}~i,j\leq N~, \\
\left\langle S_{ij}\right\rangle &
\simeq\frac{N_f-N}{3N}\frac{408.79}{h^2}\left(\frac{M}{M_{P}}\right)^{2}M_{P}\,\delta_{ij}~~~\text{for}~i,j>N~.
\end{split}\label{eq:VEVs_ISS}
\end{eqnarray} 

We now add possible contributions from the inflaton $\eta$, which means that we introduce the following term to the ISS scalar potential \cite{Olive_strongly_stabilized, Olive_non-problems, Randall_inflation_flatdirections}
\begin{equation} \label{eq:V_Inflaton}
	\Delta V\left(\phi_{\text{ISS}},\bar{\phi}_{\text{ISS}}\right) \sim e^{K\left(\phi_{\text{ISS}},\bar{\phi}_{\text{ISS}}\right)}V\left(\eta\right)=cH^{2}\phi_{\text{ISS}}\bar{\phi}_{\text{ISS}}+\cdots~,
\end{equation}
where $H$ is the Hubble parameter, and generically $c = 3$ for $K_{\text{ISS}} = \phi_{\text{ISS}}\bar{\phi}_{\text{ISS}}$
\cite{Randall_inflation_flatdirections}, which is the case for the ISS fields $\phi_{\text{ISS}} = \{S_{ij}, q_{ia}, \tilde{q}_{ai}\}$ --- see equation (\ref{eq:ISS_K}). The effect of $\Delta V\left(\phi_{\text{ISS}},\bar{\phi}_{\text{ISS}}\right)$
is to make the VEVs during inflation --- $\left\langle S_{ij}\right\rangle_{\text{ins}}$ and $\left\langle q_{ia},\tilde{q}_{ai}\right\rangle_{\text{ins}}$ --- to assume smaller values compared with their true minimum given in equation (\ref{eq:VEVs_ISS}), which we call from now on by $\left\langle S_{ij}\right\rangle_{\text{min}}$ and $\left\langle q_{ia},\tilde{q}_{ai}\right\rangle_{\text{min}}$. Namely, for $S$, one has
\begin{align}
	\left\langle S_{ij} \right\rangle_{\textrm{ins}} \simeq{}& \left\langle S_{ij}\right\rangle_{\text{min}}\left(1+\frac{cH^2}{2h^2M^2}\right)^{-1}\simeq \frac{2h^2M^4}{cH^2M_{\rm P}}\ll \left\langle S_{ij}\right\rangle_{\text{min}}~\text{for}~i,j\leq N~,\label{eq:InstVEV_S1}\\
	\left\langle S_{ij} \right\rangle_{\textrm{ins}} \simeq{}& \left\langle S_{ij}\right\rangle_{\text{min}}\left(1+\frac{8\pi^2cH^2}{\left(\textrm{ln}(4)-1\right)h^4M^2}\right)^{-1}\simeq \frac{2h^2M^4}{cH^2M_{\rm P}}\ll \left\langle S_{ij}\right\rangle_{\text{min}}~\text{for}~i,j> N~.\label{eq:InstVEV_S2}
\end{align}
We have assumed high-scale inflation in these equations, i.e. $H\gg M$. And, for $q_{ia}$ and $\tilde{q}_{ai}$, one has
\begin{eqnarray} \label{eq:InstVEV_q}
	\left\langle q_{ia},\tilde{q}_{ai}\right\rangle _{\textrm{ins}} & \simeq & \begin{cases} \frac{1}{h}\sqrt{h^{2}M^{2}-cH^{2}} & ~\text{for}~~ cH^{2} \leq h^{2}M^{2}~,\\
	0 & ~\text{for}~~ cH^{2}>h^{2}M^{2}~.\end{cases}
\end{eqnarray}
The $\left\langle S_{ij}\right\rangle_{\text{ins}}$ and $\left\langle q_{ia},\tilde{q}_{ai}\right\rangle_{\text{ins}}$ now evolve into the direction of the minima $\left\langle S_{ij}\right\rangle_{\text{min}}$ and $\left\langle q_{ia},\tilde{q}_{ai}\right\rangle_{\text{min}}$.

A comment is useful. The inflaton potential introduces the mass contribution $cH^{2}\phi_{\textrm{ISS}}\bar{\phi}_{\textrm{ISS}}$ to the ISS fields, which yields a mass $\sim\sqrt{c}H$ for all the ISS fields during inflation $\left(H\gg M\right)$, including the mass of the Goldstone modes $\textrm{Im}Q_{2}$, $\textrm{Re}Q_{4}$ and $\textrm{Im}Q_{4}$. Although the $S_{ij}$ retain a small VEV, it cannot keep the Goldstone particles from being massive.

Now take some generic field $\varphi$. The fields $\varphi$ that start oscillating after inflation are the ones which possess a non-vanishing difference between the VEV during inflation $\langle \varphi \rangle_{\text{ins}}$ and the VEV well after inflation $\langle \varphi\rangle_{\text{min}}$, i.e.,
\begin{equation}
	|\langle \varphi \rangle|_{\text{amp}} = |\langle \varphi \rangle_{\text{ins}} - \langle \varphi \rangle_{\text{min}}| \neq 0~.
\end{equation}
From this observation, we notice that there are $N_{f}$ oscillating
fields in the $S_{ij}$ sector, $N$ for $i,j\leq N$ and $(N_{f}-N)$ for $i,j>N$. The linear combinations responsible for their oscillations are $\textrm{Re}S_{1}$
and $\textrm{Re}S_2$. Furthermore, assuming that $q$ and $\tilde{q}$ have the same VEV due to the symmetry of the superpotential, there are no oscillations for $Q_2$ due to its definition given by equation (\ref{eq:Lc_2}). Since $Q_3$ and $Q_4$ are defined in the region $i\leq N$ and $N<a\leq N_f$, by the first equation in (\ref{eq:VEVs_ISS}) and their definition in equations (\ref{eq:Lc_3}) and (\ref{eq:Lc_4}), they also do not contribute with any oscillation. Therefore, there are only $N$ oscillating fields from $q$ and $\tilde{q}$, corresponding to the mass eigenstate Re$Q_1$. 

Furthermore, after the end of inflation, the inflaton $\eta$ starts to oscillate about its true minimum \cite{Randall_inflation_flatdirections,Linde_relaxing,Nakayama_Adiabatic} and, since it dominates the energy density of the Universe, this constitutes a matter dominated period. The energy density of the inflaton $\eta$ and the Hubble parameter after inflation are given by
\begin{eqnarray}
	\rho_{\eta} & = & \frac{1}{2}m_{\eta}^{2}\eta^{2}=\frac{1}{2}m_{\eta}^{2}\left\langle \eta\right\rangle _{\textrm{amp}}^{2}\left(\frac{R_{\eta}}{R}\right)^{3}=\frac{4}{3}m_{\eta}^{2}M_{P}^{2}\left(\frac{R_{\eta}}{R}\right)^{3}~, \label{eq:Inflaton_EnergyDensity_AfterInflation}\\
	H & = & \sqrt{\frac{1}{M_{P}^{2}}\frac{\rho}{3}}=\frac{2}{3}m_{\eta}\left(\frac{R_{\eta}}{R}\right)^{3/2}~.\label{eq:Inflaton_HubbleParameter_AfterInflation}
\end{eqnarray}

$R$ denotes the cosmological scale factor in the FLRW metric, and $R_{\eta}$ denotes the cosmological scale factor at the onset of the $\eta$ oscillations. We take the amplitude of the inflaton to be $\left\langle \eta\right\rangle _{\textrm{amp}}=\left\langle \eta\right\rangle _{\textrm{ins}}=\sqrt{8/3}M_{P}$ after inflation because $\left\langle \eta\right\rangle _{\textrm{min}}=0$ now. This amplitude yields $H=\frac{2}{3}m_\eta$ for $R=R_\eta$, condition which determines the beginning of inflaton oscillations. 

About the energy density of Re$S_1$, we may say that it does not dominate the energy content of the universe given its VEV being proportional to $M^2$ with non-relevant prefactors. Even more so because it starts to oscillate at the same epoch or earlier than Re$S_2$ or Re$Q_1$. Regarding the case of $\text{Re}S_{2}$, though it has a VEV proportional to $M^2$, its pre-factors are also important, i.e., $h$ can be such that the energy density of $\text{Re}S_{2}$ is similar to the one of $\text{Re}Q_{1}$. In section \ref{subsec:Evolution}, we study the evolution of Re$Q_1$, Re$S_1$ and $S_2$.

Within the matter dominated period after inflation with dominant oscillations from $\eta$, the Hubble parameter decreases to the point where $\left\langle q_{ia},\tilde{q}_{ai}\right\rangle _{\textrm{ins}}>0$. When this happens, the ISS fields $q$, $\tilde{q}$ and $S$ adiabatically track their instantaneous minima given by equations (\ref{eq:InstVEV_q}), (\ref{eq:InstVEV_S1}) and (\ref{eq:InstVEV_S2}) \cite{Randall_inflation_flatdirections,Linde_relaxing,Nakayama_Adiabatic} until they reach the point $H \sim m_{\phi_\textrm{ISS}}$, where they
start damped oscillations about their true minimum\footnote{The condition   $h^{2}M^{2} \gtrsim \,c\left.H^{2}\right|_{H=2m_{\phi_\textrm{ISS}}/3}$ implies that $\langle \phi_\textrm{ISS}\rangle_\textrm{ins}>0$ and consequently $\phi_\textrm{ISS}$ can indeed start oscillations. This condition is satisfied for any $h$ and $M$, with $c$ being the generic one $c=3$.}  \eqref{eq:VEVs_ISS}. We use $H = \frac{2}{3}m_{\phi_\textrm{ISS}}$.

The cosmological scale factor at the onset of $\phi_\textrm{ISS}$ oscillations is given by
\begin{equation}
	R_{\phi_\textrm{ISS}}=\left(\frac{m_{\eta}}{m_{\phi_\textrm{ISS}}}\right)^{2/3}R_{\eta}~. 
\end{equation}
For the cases of Re$Q_1$, Re$S_1$ and Re$S_2$, it becomes
\begin{eqnarray}
	R_{Q_{1}} & = & \left(\frac{N_{f}-N}{6}\right)^{1/3}\left(\frac{M}{M_{P}}\frac{m_{\eta}}{m_{3/2}}\right)^{2/3}R_{\eta}~,\label{eq:Q1_ScaleFactor_Osc}\\
	R_{S_{1}} & = & \left(\frac{N_{f}-N}{6}\right)^{1/3}\left(\frac{M}{M_{P}}\frac{m_{\eta}}{m_{3/2}}\right)^{2/3}R_{\eta}~,\label{eq:S1_ScaleFactor_Osc}\\
	R_{S_{2}} & = & \left(\frac{N_f-N}{3N}\right)^{1/3}\left(\frac{8\pi^{2}}{\left(\textrm{ln}\left(4\right)-1\right)h^2}\right)^{1/3}\left(\frac{M}{M_{P}}\frac{m_{\eta}}{m_{3/2}}\right)^{2/3}R_{\eta}~.\label{eq:S2_ScaleFactor_Osc}
\end{eqnarray}
From this time on, the ISS fields $\phi_\textrm{ISS}$ start to oscillate about their true minimum. This may happen before or after the inflaton oscillations have decayed.

The energy densities for the $N$ oscillating fields Re$Q_1$, and the $N_f$ oscillating Re$S_1$ and Re$S_2$ are given by
\begin{eqnarray}
	\rho_{Q_{1}} &=& N \times\frac{1}{2}m_{Q_{1}}^{2} \langle Q_{1}\rangle _{\textrm{amp}}^{2}\left(\frac{R_{Q_{1}}}{R}\right)^{3}~,\\
	\rho_{S_{1}} &=& N \times\frac{1}{2}m_{S_{1}}^{2} \langle S_{1}\rangle _{\textrm{amp}}^{2}\left(\frac{R_{S_{1}}}{R}\right)^{3}~,\\
	\rho_{S_{2}} &=& \left(N_f-N\right) \times\frac{1}{2}m_{S_{2}}^{2} \langle S_{2}\rangle _{\textrm{amp}}^{2}\left(\frac{R_{S_{2}}}{R}\right)^{3}~.
\end{eqnarray}
Inserting the relevant quantities, these translate to
\begin{eqnarray}
\rho_{Q_{1}} &=& \frac{12N}{N_f-N} \left(\frac{m_{3/2}}{M_{\rm P}}\right)^2M_{\rm P}^4\left(\frac{R_{Q_{1}}}{R}\right)^{3}~.\\
\rho_{S_{1}} &=& N \left(\frac{m_{3/2}}{M_{\rm P}}\right)^2\left(\frac{M}{M_{\rm P}}\right)^2M_{\rm P}^4\left(\frac{R_{S_{1}}}{R}\right)^{3}~,\\
\rho_{S_{2}} &=& \frac{\left(N_f-N\right)^2}{N}\left(\frac{16\pi^{2}}{3\left(\textrm{ln}\left(4\right)-1\right)h^2}\right) \left(\frac{m_{3/2}}{M_{\rm P}}\right)^2\left(\frac{M}{M_{\rm P}}\right)^2M_{\rm P}^4\left(\frac{R_{S_{2}}}{R}\right)^{3}~.
\end{eqnarray}

We write below the decay rate of the inflaton $\eta$ and recall the relevant decay rates of the ISS fields Re$Q_1$, Re$S_1$, Re$S_2$ as well as Re$Q_2$, which were obtained in section \ref{Sec:Interactions_DecayModes} --- namely by equations (\ref{eq:S_1_decay_total}), (\ref{eq:S_2_decay_total}), (\ref{eq:Q_1_decay_total}) and (\ref{eq:Q_2_decay_total}). Apart from $N=1$, we assume\footnote{Remember that this is the minimal choice for $N_{f}>3N$ (equivalent to $N_f<\frac{3}{2}N_c$), which is required for the ISS model to be infrared free in the magnetic range.} from now on also $N_{f}=4$ whenever numerical values are required. Thus, the expressions of the decay rates can be written as
\begin{eqnarray} 
	\Gamma_{\eta}&\simeq& a_{\eta}^{2}\frac{m_{\eta}^{3}}{M_{P}^{2}}~,\label{eq:Decay_Inflaton}\\
	\Gamma^\textrm{total}_{S_1} & \simeq & 5.63\times 10^{-2}\frac{m_{3/2}^3}{M_{\rm P}^2}\left(\frac{M_{\rm P}}{M}\right)^5~,\\
	\Gamma^\textrm{total}_{S_2} & \simeq & 2.31\times 10^{-9}\frac{m_{3/2}^3 h^5}{M_{\rm P}^2}\left(\frac{M_{\rm P}}{M}\right)^5~,\\
	\Gamma^\textrm{total}_{Q_1} & \simeq & 3.13\times 10^{-3}\frac{m_{3/2}^3}{M_{\rm P}^2}\left(\frac{M_{\rm P}}{M}\right)^3 + 9.80\times 10^{-11}\frac{m_{3/2}^3 h^2}{M_{\rm P}^2}\left(\frac{M_{\rm P}}{M}\right)^5~,\\
	\Gamma^\textrm{total}_{Q_2} & \simeq & 1.44\times 10^{-8}\frac{m_{3/2}^3h^5}{M_{\rm P}^2}\left(\frac{M_{\rm P}}{M}\right)^3 + 2.57\times 10^{-14}\frac{m_{3/2}^3 h^5}{M_{\rm P}^2}\left(\frac{M_{\rm P}}{M}\right)^5~,\label{eq:Q2_total_NUM}
\end{eqnarray}
where we assumed that $\eta$ decays through gravitational interactions. The quantity $a_{\eta}$ in equation (\ref{eq:Decay_Inflaton}) quantifies the couplings of $\eta$ to matter. For example, this quantity can be very small under the conditions assumed in \cite{Inflaton_decay} --- namely that the gauge kinetic function depends linearly on the inflaton, which thus decay to two MSSM gauge
bosons with coupling $d_{\eta}$, yielding the decay rate $\Gamma=\frac{3}{64\pi}d_{\eta}^{2}\frac{m_{\eta}^{3}}{M_{P}^{2}}\sim10^{-2}d_{\eta}^{2}\frac{m_{\eta}^{3}}{M_{P}^{2}} = a_{\eta}^{2}\frac{m_{\eta}^{3}}{M_{P}^{2}}$. In other words, let $d_{\eta}$ be the coupling between the inflaton $\eta$ and two MSSM gauge bosons. The coupling $a_{\eta}$ is given by 
\begin{equation}
	a_{\eta} = 10^{-1}d_{\eta}~.
\end{equation}
Later in this work, the cosmological and phenomenological analysis to obtain constraints on the ISS parameters $M$ and $h$ will be performed for a couple of different couplings $a_{\eta}$.

The decay rates of the heavy fields $\eta$, Re$S_1$, Re$S_2$ and Re$Q_1$ have been given, as well as the one for Re$Q_2$, which is a product of Re$Q_1$. The decay rates of the remaining products, $\psi_{3/2}$ and $\chi_{S1}$, must also be known for the discussions of late entropy production in the current section, and dark matter generation in section \ref{Sec:DarkMatter}. The gravitino decay rate is given by \cite{gravitino_rate} 
 \begin{eqnarray}\label{eq:gravitino_decayrate}
 \bar{\Gamma}_{3/2}\left(\psi_{3/2}\rightarrow\textrm{MSSM}\right) & = & \frac{193}{384\pi}\frac{m_{3/2}^{3}}{M_{P}^{2}}~.
 \end{eqnarray}
The gravitino decays predominantly into an R-parity even MSSM particle and its
 supersymmetric partner\footnote{The decay channels are given by $\psi_{3/2}\rightarrow\lambda+A_{m}$, $\psi_{3/2}\rightarrow\phi_{m}+\bar{\chi}_{m}$ and $\psi_{3/2}\rightarrow\phi_{m}^{*}+\chi_{m}$, where $\lambda$ are gauginos, $A_{m}$ are gauge bosons, $\phi_{m}$ are scalars, and $\chi_{m}$ are left-handed fermions.}, since it possesses R-parity $R=-1$ \cite{R-parity_gravitino}.
 We assumed that $m_{3/2}\gg m_{\textrm{MSSM}}$ with $m_{\textrm{MSSM}}$ being the mass of any MSSM particle. Furthermore, the factor $\frac{193}{384\pi}$ counts the number of the MSSM gauge bosons, fermions ---
 leptons and quarks --- and Higgs scalars. More specifically, we have the following relation $\frac{193}{384\pi}=\frac{1}{384\pi}\left(12\times12_{\textrm{gauge}}+3_{\textrm{families}}\times15_{\textrm{fermions}}+4_{\textrm{Higgs}}\right)$.
 The $\chi_{S1}$ decay rate,
 \begin{equation}\label{eq:chiS1_decayrate}
 \begin{split}
 \overline{\Gamma}_{\chi_{S1}}\left(\chi_{S1}\rightarrow\textrm{Im}Q_{2}+\chi_{\textrm{MSSM}}+\phi_{\textrm{MSSM}}\right) &= 2.38\times10^{-4}\frac{m_{3/2}^{5}}{M_{P}^{4}} \\ &\simeq 1.12\times10^{-8}\frac{m_{3/2}^{3}}{M_{P}^{2}}h^{2}\left(\frac{M}{M_{P}}\right)^{4}~,
 \end{split}
 \end{equation}
 is obtained from the term $e^{G/2}\left(\frac{1}{3}\left(K_{q_{ia}}+K_{\tilde{q}_{ia}}\right)K_{\textrm{MSSM}}\right)\bar{\chi}_{R}^{ia}\chi_{L}^{\textrm{MSSM}}$
 within the scalar potential. Again $m_{3/2}\gg m_{\textrm{MSSM}}$
 was assumed, and the counting over the scalars and fermions has been
 done, i.e., $49=3\times15+4$, and over the contribution from the GM
 term, which yields $2c_{H}\equiv2$. This decay rate is smaller than
 the gravitino one and can pose problems to BBN if the lifetime of $\chi_{S1}$ exceeds $\sim1$ second, or equivalently the temperature $\sim1\textrm{ MeV}$. This point will be discussed in section \ref{Sec:DarkMatter}.

%%%%%%%%%%%%%%%%%%%%%%%%%%%%
\subsection{Evolution of the Universe}
\label{subsec:Evolution}
%%%%%%%%%%%%%%%%%%%%%%%%%%%%

In this section, we discuss how the oscillations and decays referred
to in the previous section account for the evolution of the Universe.

\begin{figure}[h]
	\begin{minipage}[c]{0.45\columnwidth}
		{\includegraphics[height=6.45cm]{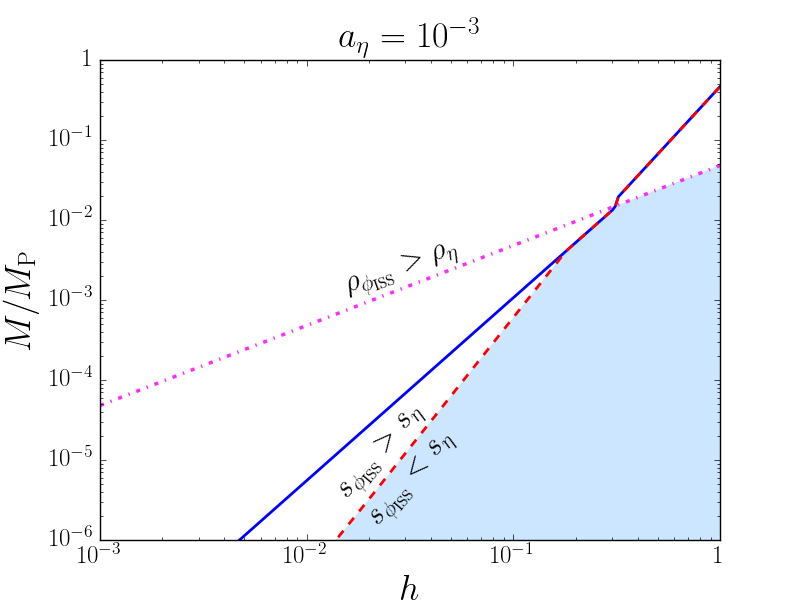}}
	\end{minipage}
	\hspace*{0.5cm}
	\begin{minipage}[c]{0.45\columnwidth}
		{\includegraphics[height=6.45cm]{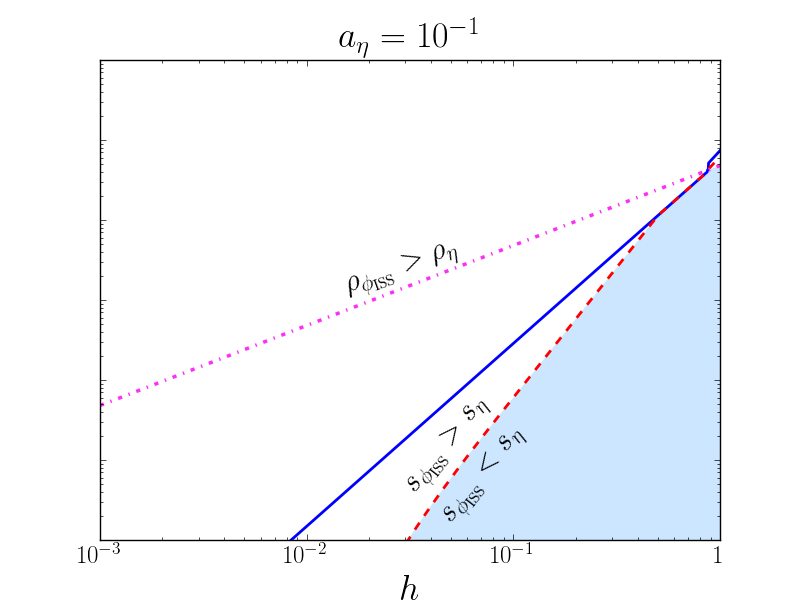}}
	\end{minipage}
	\caption{The curves $s_{\phi_\textrm{ISS}}=s_\eta$ summarizing the constraints obtained in this section --- see table \ref{tab:Constraints_Sec4} --- for small (large) coupling $a_\eta=10^{-3}$ ($a_{\eta}=10^{-1}$). The allowed region is shaded in blue. It also appears in section \ref{Sec:DarkMatter} with the label ``sec. 4", where apart from these constraints on entropy production we also consider constraints on dark matter production. Notice that the blue shaded region should be extended to larger values of $M/M_{\rm P}$ and lower values of $h$ if the entropy of the gravitino is not considered, i.e., the red curve of entropy production is less constrained and we should not consider the red curve anymore. We see in section \ref{Sec:DarkMatter} when this turns out to be the case. \label{fig:final_constraints_sec4}}
\end{figure}

In order to study the evolution of the fields Re$Q_{1}$, Re$S_{1}$,
Re$S_{2}$ and $\eta$, we devise a list of issues we will account
for. 
\begin{itemize}
	\item We assume, as we already mentioned, that the energy of the Universe after the end of inflation is dominated by $\eta$ oscillations --- by $\eta$ decay products after it decays; 
	\item we must know whether the ISS fields decay in the $\eta$ oscillation
	era or in the $\eta$ decay products era, since this has an important
	impact on the energy densities of their products; 
	\item the relativistic decay products may turn non-relativistic as the Universe
	cools down, which means their energy evolves differently than radiation,
	i.e., $\rho_{\textrm{rad}}/\rho_{\textrm{non}}\sim R^{-1}$; 
	\item decay products which are massless ISS particles should be carefully studied since they contribute to the present relativistic degrees of freedom
	$N_{\textrm{eff}}$, which is $N_{\textrm{eff}}=3.15\pm0.23$ \cite{Planck}; 
	\item decay products with small decay rates should be also carefully studied, since they should not decay after the BBN epoch with $T\sim1\textrm{ MeV}$.
\end{itemize}
We start with the first and secoind points. The combined $\phi_{\textrm{ISS}}$ oscillation
energy is given by
\begin{eqnarray}
\rho_{\phi_{\textrm{ISS}}} & = & \rho_{Q_{1}}+\rho_{S_{1}}+\rho_{S_{2}}~\nonumber \\
& = & \left(\frac{m_{3/2}}{M_{P}}\right)^{2}M_{P}^{4}\left\{ \frac{12N}{N_{f}-N}\left(\frac{R_{Q_{1}}}{R}\right)^{3}+N\left(\frac{M}{M_{P}}\right)^{2}\left(\frac{R_{S_{1}}}{R}\right)^{3}\right.\\
&  & \left. +\,\frac{\left(N_{f}-N\right)^{2}}{3N}\left(\frac{16\pi^{2}}{3\left(\textrm{ln}\left(4\right)-1\right)}\right)\left(\frac{M}{M_{P}}\right)^{2}\left(\frac{R_{S_{2}}}{R}\right)^{3}\right\}~.\nonumber 
\end{eqnarray}
In order to compare $\rho_{\phi_{\textrm{ISS}}}$ and $\rho_{\eta}$,
we need to rearrange the energy expression of the former. We rewrite it as
\begin{equation}
\rho_{\phi_{\textrm{ISS}}}=m_{\eta}^{2}M_{P}^{2}\left\{ 2\left(\frac{M}{M_{P}}\right)^{2}+\frac{1}{2}\left(\frac{M}{M_{P}}\right)^{4}+\frac{2.51\times10^{5}}{h^{4}}\left(\frac{M}{M_{P}}\right)^{4}\right\} \left(\frac{R_{\eta}}{R}\right)^{3}~,
\end{equation}
where we have used equations (\ref{eq:Q1_ScaleFactor_Osc}), (\ref{eq:S1_ScaleFactor_Osc}) and (\ref{eq:S2_ScaleFactor_Osc}) with $N_{f}=4$ and $N=1$. When compared to the energy density
of $\eta$, namely
\begin{equation}
\rho_{\eta}=\frac{4}{3}m_{\eta}^{2}M_{P}^{2}\left(\frac{R_{\eta}}{R}\right)^{3}\quad\textrm{for}\quad R_\eta<R<R_{\textrm{d}\eta}~,
\end{equation}
we obtain $\rho_{\eta} > \rho_{\phi_{\textrm{ISS}}}$ at the end of inflation only if
\begin{equation}
\label{eq:Constraint_rhophiOSC_rhoetaOSC}
M<4.80\times10^{-2}hM_{\rm P}~.
\end{equation}
This requirement will be further discussed in figures \ref{fig:entropy_produc-1} and \ref{fig:final_constraints_sec4} when displaying the allowed parameter space $(h,M)$ constrained also by entropy production. 

\begin{figure}[h]
	\begin{minipage}[c]{0.45\columnwidth}
	{\includegraphics[height=6.45cm]{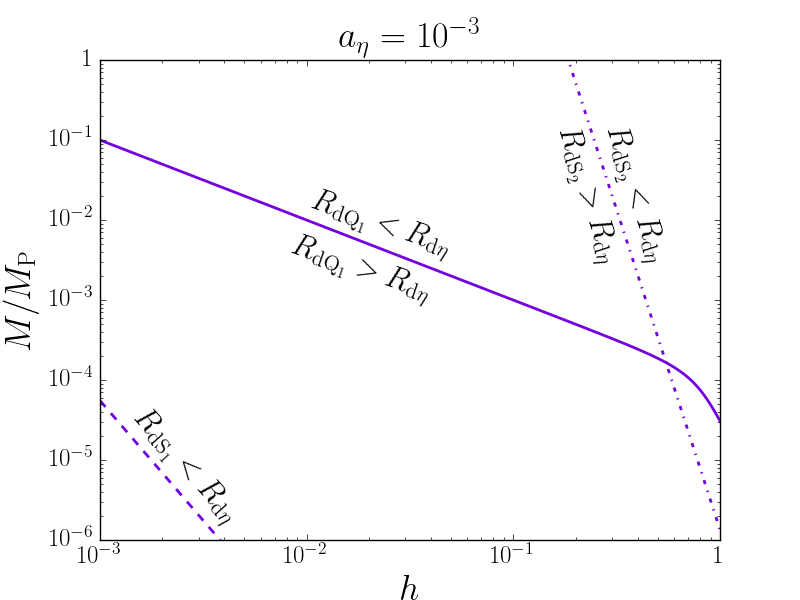}}
	\end{minipage}
	\hspace*{0.5cm}
	\begin{minipage}[c]{0.45\columnwidth}
	{\includegraphics[height=6.45cm]{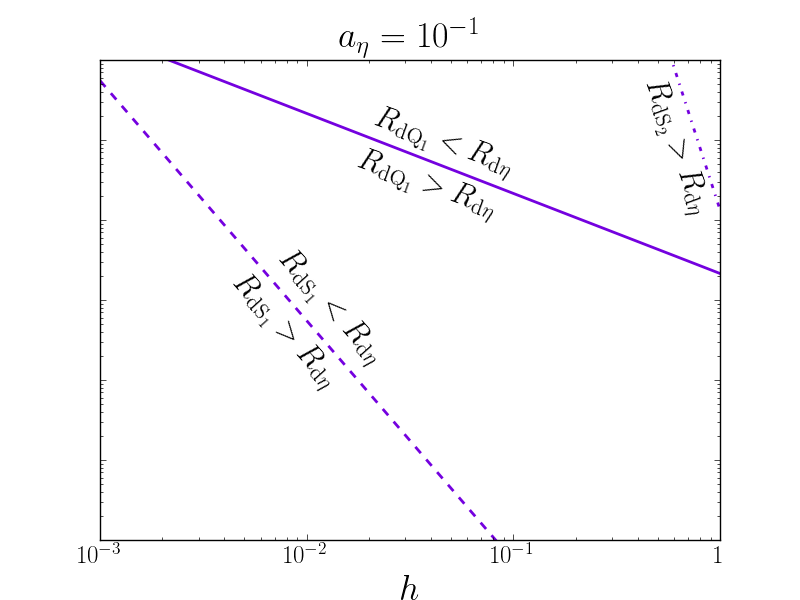}}
	\end{minipage}
	\caption{The curves of $\Gamma^i_{\phi_\textrm{ISS}}=\Gamma_\eta$, where $i=Q_1,S_1,S_2$, for small (large) coupling $a_\eta=10^{-3}$ ($a_{\eta}=10^{-1}$). Above the curves, their decay rates $\Gamma^i_{\phi_\textrm{ISS}}$ becomes larger, implying a smaller scale factor at the time of their decay.\label{fig:decay_epochs}}
\end{figure}

Now we turn to the question
of the decay epoch of $\phi_{\textrm{ISS}}$, which depends on whether
they decay before or after $\eta$ reheating --- i.e., when $\eta$ decays itself. For the
former and the latter, we obtain respectively
\begin{eqnarray}
\frac{R_{\textrm{d}\phi_{\textrm{ISS}}}}{R_{\textrm{d}\eta}} & = & a_{\eta}^{4/3}\left(\frac{M_{P}}{\Gamma_{\phi_{\textrm{ISS}}}}\right)^{2/3}\left(\frac{m_{\eta}}{M_{P}}\right)^{2}~,\\
\frac{R_{\textrm{d}\phi_{\textrm{ISS}}}}{R_{\textrm{d}\eta}} & = & \frac{2}{\sqrt{3}}a_{\eta}\left(\frac{M_{P}}{\Gamma_{\phi_{\textrm{ISS}}}}\right)^{1/2}\left(\frac{m_{\eta}}{M_{P}}\right)^{3/2}~,
\end{eqnarray}
where $R_{\textrm{d}\phi_{\textrm{ISS}}}$ is the scale factor at
$\phi_{\textrm{ISS}}$ decay and similarly for $\eta$. Here we used $H=a\Gamma$
to obtain the decay epochs --- $a=2/3$ for matter domination and $a=1/2$
for radiation domination --- as well as $R_{\eta}/R_{\textrm{d}\eta}=a_{\eta}^{4/3}\left(\frac{m_{\eta}}{M_{P}}\right)^{4/3}$.
If $\frac{R_{\textrm{d}\phi_{\textrm{ISS}}}}{R_{\textrm{d}\eta}}<1\left(>1\right)$,
$\phi_{\textrm{ISS}}$ decays before (after) the inflaton does. We
show in figure \ref{fig:decay_epochs} how the parameters $M$ and
$h$ determine the time of their decays.
Since we want the $\eta$ oscillation energy to be greater than the
one of $\phi_{\textrm{ISS}}$, production of entropy from the latter
fields will only be problematic if one or more of them decays after
$\eta$ has already decayed. The energy density of the decay products
of $\eta$ is given by
\begin{equation} \label{eq:Constraint_rhoeta}
\rho_{\eta}^{r}=\rho_{\textrm{d}\eta}\left(\frac{R_{\textrm{d}\eta}}{R}\right)^{4}=\frac{4}{3}a_{\eta}^{4}M_{P}^{4}\left(\frac{m_{\eta}}{M_{P}}\right)^{6}\left(\frac{R_{\textrm{d}\eta}}{R}\right)^{4}\quad\textrm{for}\quad R>R_{\textrm{d}\eta}~,
\end{equation}
where $\rho_{\textrm{d}\eta}$ is the energy density $\rho_{\eta}$
at the moment of $\eta$ decay. 

\begin{figure}[h!]
	\begin{minipage}[c]{0.45\columnwidth}
		{\includegraphics[height=6.45cm]{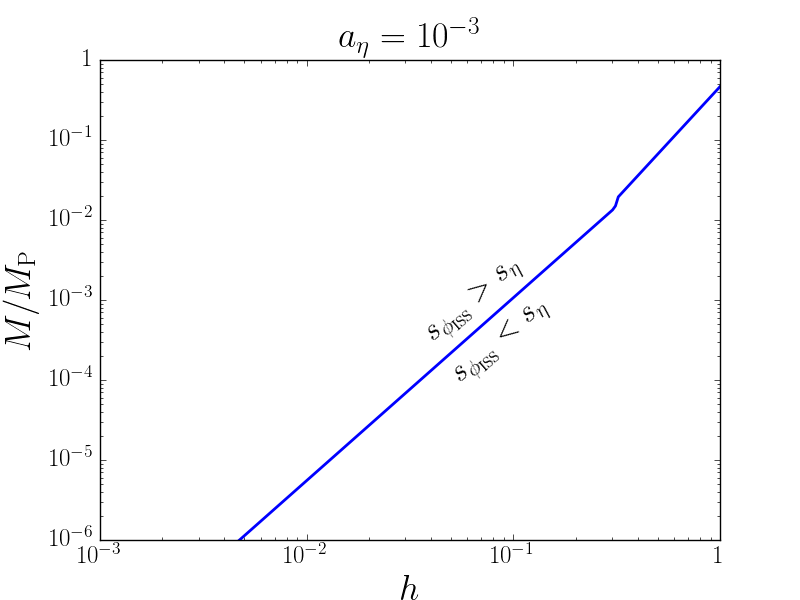}}
	\end{minipage}
	\hspace*{0.5cm}
	\begin{minipage}[c]{0.45\columnwidth}
		{\includegraphics[height=6.45cm]{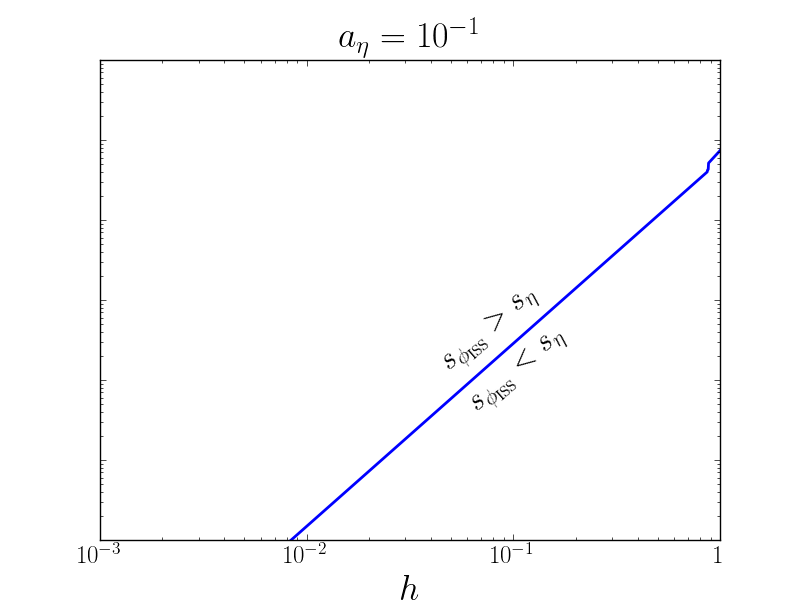}}
	\end{minipage}
	\caption{The curves of $s_{\phi_\textrm{ISS}}=s_\eta$ numerically obtained from equations \eqref{eq:Constraint_rhoeta} and (\ref{eq:Constraint_Entr1}) at the decay epoch of $S_2$, the last decaying particle, for small (large) coupling $a_\eta=10^{-3}$ ($a_{\eta}=10^{-1}$). Above the curves, the energy density $\rho^{r}_{\phi_{\text{ISS}}}$ become larger, implying also larger total entropy density at the decay time of $S_2$. \label{fig:entropy_produc}}
\end{figure}

Now, assuming $\rho_{\phi_{\textrm{ISS}}}^{r}$ is the energy density of radiation from $Q_{1}$, $S_{1}$ and $S_{2}$, we can define a scale factor $R_{1}$ where $\rho_{\eta}^{r}\left(R_{1}\right)=\rho_{\phi_{\textrm{ISS}}}^{r}\left(R_{1}\right)$. Finding $R_{1}$ means there exists a limit on the entropy produced by these ISS fields such that $s_{\eta}>s_{\phi_{\textrm{ISS}}}$ is satisfied, where for any field $i$ the following relation holds, $s_{i}/s_{j}\sim\left(\rho_{i}^{r}/\rho_{j}^{r}\right)^{3/4}$. The energy density $\rho_{\phi_{\textrm{ISS}}}^{r}\left(R\right)$ is given by
\begin{equation}
\rho_{\phi_{\textrm{ISS}}}^{r}\left(R\right)=\rho_{\textrm{d}Q_{1}}\left(\frac{R_{\textrm{d}Q_{1}}}{R}\right)^{4}+\rho_{\textrm{d}S_{1}}\left(\frac{R_{\textrm{d}S_{1}}}{R}\right)^{4}+\rho_{\textrm{d}S_{2}}\left(\frac{R_{\textrm{d}S_{2}}}{R}\right)^{4}~.
\end{equation}
If $\rho_{\eta}^{r}\left(R_{1}\right)>\rho_{\phi_{\textrm{ISS}}}^{r}\left(R_{1}\right)$, then most entropy comes from $\eta$ decays. Otherwise, the ISS decays would provide the most entropy. When $\rho_{\phi_{\textrm{ISS}}}^{r}$ is evaluated at $R_{1}$, the latter is necessary equal to the scale factor of the last decaying field, since otherwise the ISS energy density (if considering matter and radiation together) would overcome $\rho_{\eta}^{r}$ again, since $\rho_{\textrm{rad}}/\rho_{\textrm{matter}}\sim R^{-1}$. 

For example, assuming $R_{\textrm{d}Q_{1}}<R_{\textrm{d}S_{1}}<R_{\textrm{d}S_{2}}$, one would obtain
\begin{equation} \label{eq:Constraint_Entr1}
\rho_{\phi_{\textrm{ISS}}}^{r}\left(R_{1}\right)=\rho_{\textrm{d}Q_{1}}\left(\frac{R_{\textrm{d}Q_{1}}}{R_{1}}\right)^{4}+\rho_{\textrm{d}S_{1}}\left(\frac{R_{\textrm{d}S_{1}}}{R_{1}}\right)^{4}+\rho_{\textrm{d}S_{2}}~,
\end{equation}
where $R_{1}=R_{\textrm{d}S_{2}}$ for $S_{2}$ as the latest decaying particle. Taking these details into account, in figure \ref{fig:decay_epochs} we plot numerically the bound $s_{\eta}>s_{\phi_{\textrm{ISS}}}$.	

A comment is in order. For the region at which the curve is drawn, $S_2$ dominates the energy density compared with the ones from either $Q_1$ or $S_1$ because it has the longest lifetime and a sizable VEV. Both $Q_1$ and $S_1$ become important only for smaller $M$ and larger $h$, i.e., in the lower right corner  of both figures. In addition, there is a noticeable step at the right upper corner for both cases $a_\eta=10^{-1}$ and $a_\eta=10^{-3}$. They form at the point at which $S_2$ turns from decaying after $\eta$ to decaying before $\eta$. This introduces a dip in the energy density function. The step is quite steep only because our analysis assumes instantaneous decays.

To this discussion we must add the behaviour of the decay products of $Q_1$, $S_1$ and $S_2$. These decay products can turn from relativistic to non-relativistic at some point, which could render their energy density bigger than $\rho_{\eta}^{r}$. Below we find constraints such that their energy densities do not surpass $\rho_{\eta}^{r}$. For this, we start by displaying the decay products
of each field,
\begin{eqnarray}
\textrm{Re}S_{1} & : & \left(\chi_{S1}+\bar{\chi}_{S1}\right)~,\\
\textrm{Re}S_{2} & : & \left(\psi_{3/2}+\psi_{3/2}\right)~,\\
\textrm{Re}Q_{1} & : & \left(\psi_{3/2}+\psi_{3/2},\,\chi_{S1}+\bar{\chi}_{S1}+\textrm{Im}Q_{2},\,\chi_{S1}+\bar{\chi}_{S1}+\textrm{Re}Q_{2}\right)~,\\
\textrm{Re}Q_{2} & : & \left(\psi_{3/2}+\psi_{3/2},\,\chi_{S1}+\bar{\chi}_{S1}+\textrm{Im}Q_{2}\right)~.
\end{eqnarray}
Furthermore, we recall the masses of the final particles
\begin{equation}
	\begin{split}
	m_{\chi_{S1}} & = m_{3/2}~,\\
	m_{\textrm{Re}Q_{2}} & =\sqrt{\frac{3\left(\textrm{ln}\left(4\right)-1\right)}{8\pi^{2}}}h\left(\frac{M_{P}}{M}\right)m_{3/2}~,\\ 			m_{\textrm{Im}Q_{2}} & = 0~.
	\end{split}
\end{equation}
Since $m_{\text{Im}Q_2}=0$, $\text{Im}Q_2$ can never become non-relativistic. For massive particles, the scale factor at which they turn non-relativistic
$R_{\textrm{non}}$ is related to the scale factor at the decay of the initial particle. We have the following relations
\begin{equation}
	\begin{split}
& \text{Re}S_1~:~\frac{R_{\text{d}S_{1}}}{R_{\text{non}}^{\chi_{S1}}}=\frac{2m_{\chi_{S1}}}{m_{\text{Re}S_{1}}}~,\\
& \text{Re}S_2~:~
\frac{R_{\text{d}S_{2}}}{R_{\text{non}}^{\psi_{3/2}}}=\frac{2m_{\psi_{3/2}}}{m_{\text{Re}S_{2}}}~, \\
& \text{Re}Q_1~:~\frac{R_{\textrm{d}Q_{1}}}{R_{\textrm{non}}^{\psi_{3/2}}}=\frac{2m_{\psi_{3/2}}}{m_{\textrm{Re}Q_{1}}}~,~~
\frac{R_{\textrm{d}Q_{1}}}{R_{\textrm{non}}^{\chi_{S1}}}=\frac{3m_{\chi_{S1}}}{m_{\textrm{Re}Q_{1}}}~,~~
\frac{R_{\textrm{d}Q_{1}}}{R_{\textrm{non}}^{\text{Re}Q_2}}=\frac{3m_{\text{Re}Q_2}}{m_{\textrm{Re}Q_{1}}}~,~~~\\
& \text{Re}Q_2~:~\frac{R_{\textrm{d}Q_{2}}}{R_{\textrm{non}}^{\psi_{3/2}}}=\frac{2m_{\psi_{3/2}}}{m_{\textrm{Re}Q_{2}}}~,~~
\frac{R_{\textrm{d}Q_{2}}}{R_{\textrm{non}}^{\chi_{S1}}}=\frac{3m_{\chi_{S1}}}{m_{\textrm{Re}Q_{2}}}~.
	\end{split}
\end{equation}

These were computed via $f_{\Phi}^{\varphi_{i}}\cdotp\rho_{\textrm{d}\Phi}\,\cdotp\left(R_{\textrm{d}\Phi}/R^{\varphi_i}_{\textrm{non}}\right)^{4}=\rho_{\varphi_{i}}|_{T=m_{\varphi_{i}}}$, where $\Phi$ is the mother-particle which decays into $\varphi_{i}+\varphi_{j}\left(+\varphi_{k}\right)$, and $f_{\Phi}^{\varphi_{i}}$ is the share of the energy of each $\Phi$ particle given to a product-particle $\varphi_{i}$. For example, $f_{\Phi}^{\varphi_{i}}=1$, for $\Phi=\textrm{Re}Q_{2}$ and $\varphi_{i}=\psi_{3/2}$. We have also considered that the masses of the products are much smaller than the masses of the decaying particles, which leads to the numerical factors in these expressions.	

If the decay rate of a particle is sufficiently large, it decays before it can turn non-relativistic. In that case, there would be no change to the curve we obtained in figure \ref{fig:entropy_produc}. On the other hand, if its decay rate is small, we must change the energy
density equations accordingly
\begin{equation}
\rho_{i}^{\textrm{non}}=\sum_{j}\rho_{\textrm{d}i}\left(\frac{R_{\textrm{d}i}}{R_{\textrm{non}}^{j}}\right)^{4}\left(\frac{R_{\textrm{non}}^{j}}{R}\right)^{3}=\sum_{j}\rho_{\textrm{d}i}\left(\frac{R_{\textrm{d}i}}{R_{\textrm{non}}^{j}}\right)\left(\frac{R_{\textrm{d}i}}{R}\right)^{3}~.
\end{equation}
The decay rates of the products have been given in equations (\ref{eq:Q2_total_NUM}), (\ref{eq:gravitino_decayrate}) and (\ref{eq:chiS1_decayrate}) for Re$Q_{2}$, $\psi_{3/2}$ and $\chi_{S1}$, respectively.

\begin{figure}[h]
	\begin{minipage}[c]{0.45\columnwidth}
		{\includegraphics[height=6.45cm]{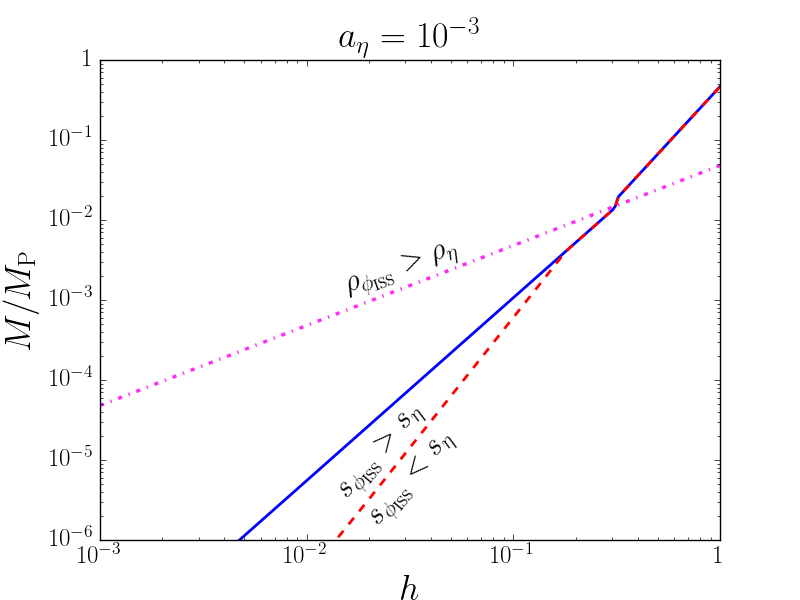}}
	\end{minipage}
	\hspace*{0.5cm}
	\begin{minipage}[c]{0.45\columnwidth}
		{\includegraphics[height=6.45cm]{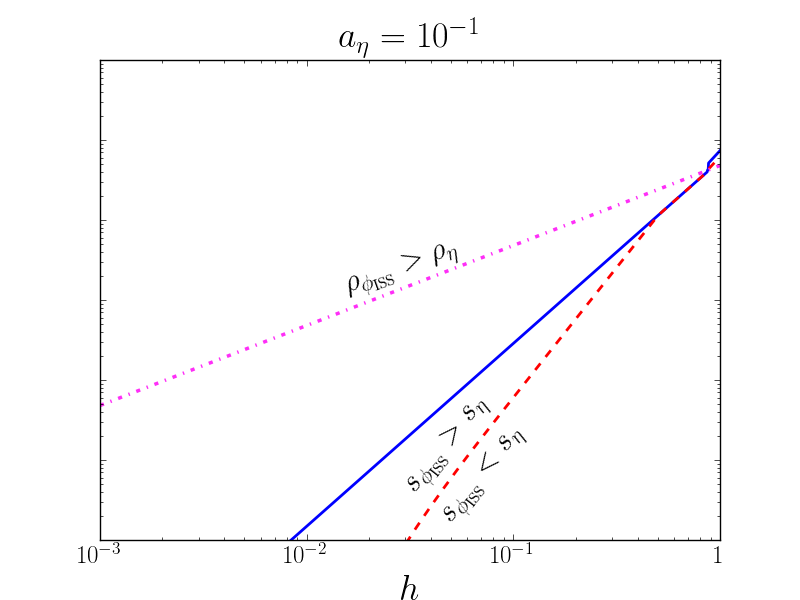}}
	\end{minipage}
	\caption{The curves of $s_{\phi_\textrm{ISS}}=s_\eta$ numerically obtained from equations \eqref{eq:Constraint_rhoeta} and (\ref{eq:Constraint_Entr1}), and the magenta dot-dashed curve $\rho_{\phi_{\text{ISS}}} = \rho_{\eta}$ from equation (\ref{eq:Constraint_rhophiOSC_rhoetaOSC}) for small (large) coupling $a_\eta=10^{-3}$ ($a_{\eta}=10^{-1}$). The blue solid curve is evaluated at the decay epoch of decay of $S_2$, the last decaying ISS particle as already seen in figure \ref{fig:entropy_produc}, and the red dashed curve is evaluated at the decay time of $\chi_{S1}$, which is the last product to decay --- we write legends just for the latter because the one for the blue solid curve has been already given in figure \ref{fig:entropy_produc}. \label{fig:entropy_produc-1}}
\end{figure}

We calculate again the entropy production bound, now taking into account the non-relativistic behaviour of the products $\chi_{S1}$, $\psi_{3/2}$, and $\textrm{Re}Q_{2}$. The obtained bound is shown in the red curve in figure \ref{fig:entropy_produc-1}. It was also obtained numerically from equation (\ref{eq:Constraint_Entr1}). As one can notice, the deviation from the blue dashed curve becomes more pronounced for lower values of $M$ and $h$. This is because the products from ISS decays turn non-relativistic earlier as $M$ and $h$ take such values. At $h\gtrsim 0.1$  and for $a_\eta=10^{-3}$, the ISS products decay when they are still relativistic, thus the agreement on the curves. For $a_\eta=10^{-1}$, the same happens for $h\gtrsim 0.4\,$. From now on, we thus consider the constraint on the entropy of the products from the red curve, taking into account their non-relativistic behaviour.

We address now the issue of $\textrm{Im}Q_{2}$ relativistic degrees of freedom since they are the only massless products in the relevant decays. First we obtain an expression for the energy density from $\textrm{Im}Q_{2}$,
\begin{equation}
	\begin{split}
\rho_{\textrm{Im}Q_{2}}	&=	\frac{\Gamma_{Q_{1}}^{\chi\chi\textrm{Re}Q_{2}}}{\Gamma_{Q_{1}}^{\textrm{total}}}\left[\frac{1}{3}\left(\frac{R_{\textrm{d}Q_{1}}}{R}\right)^{4}+\frac{1}{3}\,\cdotp\frac{1}{3}\frac{\Gamma_{Q_{2}}^{\chi\chi\textrm{Im}Q_{2}}}{\Gamma_{Q_{2}}^{\textrm{total}}}\left(\frac{R_{\textrm{d}Q_{1}}}{R_{\textrm{non}}^{\textrm{Re}Q_{2}}}\right)^{4}\left(\frac{R_{\textrm{non}}^{\textrm{Re}Q_{2}}}{R_{\textrm{d}Q_{2}}}\right)^{3}\left(\frac{R_{\textrm{d}Q_{2}}}{R}\right)^{4}\right]\rho_{\textrm{d}Q_{1}}\\
&=	\frac{\Gamma_{Q_{1}}^{\chi\chi\textrm{Re}Q_{2}}}{\Gamma_{Q_{1}}^{\textrm{total}}}\left[\frac{1}{3}+\frac{1}{9}\frac{\Gamma_{Q_{2}}^{\chi\chi\textrm{Im}Q_{2}}}{\Gamma_{Q_{2}}^{\textrm{total}}}\left(\frac{R_{\textrm{d}Q_{2}}}{R_{\textrm{non}}^{\textrm{Re}Q_{2}}}\right)\right]\rho_{\textrm{d}Q_{1}}\left(\frac{R_{\textrm{d}Q_{1}}}{R}\right)^{4}\\
&\simeq	\frac{\Gamma_{Q_{1}}^{\chi\chi\textrm{Re}Q_{2}}}{\Gamma_{Q_{1}}^{\textrm{total}}}\left[\frac{1}{9}\frac{\Gamma_{Q_{2}}^{\chi\chi\textrm{Im}Q_{2}}}{\Gamma_{Q_{2}}^{\textrm{total}}}\left(\frac{R_{\textrm{d}Q_{2}}}{R_{\textrm{non}}^{\textrm{Re}Q_{2}}}\right)\right]\rho_{\textrm{d}Q_{1}}\left(\frac{R_{\textrm{d}Q_{1}}}{R}\right)^{4}\quad\quad\quad\quad\quad\quad\quad\quad\quad\quad\\
&\leq	\frac{1}{18}\left(\frac{R_{\textrm{d}Q_{2}}}{R_{\textrm{non}}^{\textrm{Re}Q_{2}}}\right)\rho_{\textrm{d}Q_{1}}\left(\frac{R_{\textrm{d}Q_{1}}}{R}\right)^{4}\quad,
	\end{split}
\end{equation}
where the ratios $\frac{1}{3}$ and $\frac{1}{3}\,\cdotp\frac{1}{3}$ correspond to the energy share carried by $\textrm{Im}Q_{2}$ for the two respective sources, $Q_{1}\rightarrow\bar{\chi}_{S1}+\chi_{S1}+\textrm{Im}Q_{2}$ and $Q_{1}\rightarrow\bar{\chi}_{S1}+\chi_{S1}+\textrm{Re}Q_{2}$ followed by $\textrm{Re}Q_{2}\rightarrow\bar{\chi}_{S1}+\chi_{S1}+\textrm{Im}Q_{2}$, assuming massless products when compared to $\textrm{Re}Q_{1}$. From the second to the third line we used $\frac{R_{\textrm{d}Q_{2}}}{R_{\textrm{non}}^{\textrm{Re}Q_{2}}}>100$ --- which can be proved for the parameter ranges $M\in\left[10^{-6},\,1\right]$ and $h\in\left[10^{-3},\,1\right]$. From the third to the fourth line, we used the maximum values for the branching ratios, i.e. $\frac{\Gamma_{Q_{1}}^{\chi\chi\textrm{Re}Q_{2}}}{\Gamma_{Q_{1}}^{\textrm{total}}}=\frac{1}{2}$ and $\frac{\Gamma_{Q_{2}}^{\chi\chi\textrm{Im}Q_{2}}}{\Gamma_{Q_{2}}^{\textrm{total}}}=1$, for simplicity. Next, we know dark radiation --- neutrinos plus other unknown degrees of freedom --- to be parametrized by the effective degrees of freedom $N_{\textrm{eff}}$, which yields the energy density for dark radiation (for $T\ll1\,\textrm{MeV}$),
\begin{equation}
\rho_{\textrm{dark}}=N_{\textrm{eff}}\,\frac{7}{8}\left(\frac{4}{11}\right)^{4/3}\rho_{\gamma}\quad,
\end{equation}
where $\rho_{\gamma}$ is the photon energy density. The observational parameter $N_{\text{eff}}=3.15\pm0.23$ \cite{Planck} allows for a significant additional radiation density, if one takes the contribution from the SM three neutrinos to be $N_{\nu}\simeq3.046$ \cite{mangano_miele_neutrinoEff}. Comparing the two last equations, one can obtain that for the allowed parameter space shown in figure \ref{fig:final_constraints_sec4}, $\rho_{\textrm{Im}Q_{2}}<\rho_{\textrm{dark}}-\rho_{\textrm{SM}\,\nu}$.

To summarize the constraints regarding entropy production obtained in this section that will be relevant for section \ref{Sec:DarkMatter}, we collect them in table \ref{tab:Constraints_Sec4} with their depiction in figure \ref{fig:final_constraints_sec4}.
\begin{table}[H]
\begin{center}
	\begin{tabular}{|c|c|c|c|}
		\hline 
		Location & Constraint & Meaning & Legend\tabularnewline
		\hline 
		\hline 
		Eq. (\ref{eq:Constraint_rhophiOSC_rhoetaOSC}) & $M < 4.80 \times 10^{-2}hM_{\rm P}$ & $\rho_{\eta} > \rho_{\phi_{\text{ISS}}}$ & magenta curve \\
		& & at the end of inflation & \\  
		\hline 
		Eqs. (\ref{eq:Constraint_rhoeta}) & Numerical & $s_{\eta} > s_{\phi_{\text{ISS}}}$   at decay epoch & blue curve \\
		and (\ref{eq:Constraint_Entr1}) & & of last decaying particle $S_2$ & \\
		\hline 
		Eqs. (\ref{eq:Constraint_rhoeta})  & Numerical & $s_{\eta} > s_{\phi_{\text{ISS}}}$ at decay epoch & red curve \\
		and (\ref{eq:Constraint_Entr1}) & & of last decaying product $\chi_{S1}$ &  \\
		\hline 
	\end{tabular}
	
	\caption{The constraints on the ISS parameters $M$ and $h$ obtained in this section to have acceptable entropy production, their location in the text, their meaning, and their depiction in the figures of this section.}
	\label{tab:Constraints_Sec4}
	\end{center}
\end{table}

\vspace*{-0.5cm}
%%%%%%%%%%%%%%%%%%%%%%%%%%%%
%%%%%%%%%%%%%%%%%%%%%%%%%%%%
\section{Dark matter production} \label{Sec:DarkMatter}
%%%%%%%%%%%%%%%%%%%%%%%%%%%%
%%%%%%%%%%%%%%%%%%%%%%%%%%%%

After finding a protocol for solving the potential problem of entropy production, in this section we turn to the production of dark matter. We consider both the production from gravitinos produced thermally within the reheating phase of the inflaton $\eta$, and the purely  non-thermal production\footnote{Notice that for this case the late thermalization of neutralinos $\chi$ will also be studied.} from the chain of decays from the ISS fields  $\phi_{\text{ISS}}=\{\textrm{Re}S_1,\textrm{Re}S_2,\textrm{Re}Q_1\}$ through gravitinos $\psi_{3/2}$, small mass fermions $\chi_{S1}$, the real part $\textrm{Re}Q_2$ of $Q_2$ and then to the neutralino dark matter candidate. The overproduction of non-thermal dark matter is known to be a delicate issue for moduli, of which the Polonyi model \cite{Polonyi_original, Polonyi_problem, Note_polonyi} is a prime example. We discuss how to take care of this issue within our context.

From decay rate results in section \ref{Sec:Interactions_DecayModes} and subsection \ref{subsec:Evolution}, we know that $S_2$, $Q_1$ and $Q_2$ can decay into $\bar{\psi}_{3/2}+\psi_{3/2}$. When this is the case, we have the following relation between the number density of ISS particles $n^i_{\phi_\textrm{ISS}}$ and the number density of the final gravitinos: $n_{3/2}=2\sum_{i}n_{\phi_{\textrm{ISS}}}^{i}\,\cdotp\textrm{Br}\left(\phi_{\textrm{ISS}}^{i}\rightarrow\bar{\psi}_{3/2}+\psi_{3/2}\right)$, where $\textrm{Br}\left(\Phi_{i}\rightarrow\sum_{j}\varphi_{j}\right)$ is the branching ratio of $\Phi_{i}$ into a particular channel $\sum_{j}\varphi_{j}$. As another source of gravitinos, one should consider the inflaton, either through direct decay or through thermal freezout after $\eta$-reheating.\footnote{For an exact treatment, one should consider ISS-reheating besides $\eta$-reheating, since when the ISS fields decay, they could generate a thermal bath with temperature $T_R^{\textrm{ISS}}$, which would produce gravitinos. We assumed in section \ref{Sec:Post-inflation_Entropy} that the inflaton $\eta$ is the field responsible for the current particle content of the Universe and the entropy provenient from ISS decays could never be greater than the one from $\eta$ decay products. Therefore, constraining entropy production means one does not have to be concerned theoretically about the ISS-reheating gravitino production. Its thermal production comes predominantly from the inflaton decay.} After the gravitinos are produced, they decay when $H\sim \Gamma_{3/2}$ into an odd number of lightest supersymmetric particles (LSP), as already explained. 

On the other hand, the ISS fields can also decay into $\chi_{S1}+\bar{\chi}_{S1}$. In this case, the relation between number densities is a little more diverse. In the case of $S_{1}$, $\textrm{Re}S_{1}$ decays directly to $\chi_{S1}+\bar{\chi}_{S1}$. For $Q_{1}$, $\textrm{Re}Q_{1}$ decays into $\chi_{S1}+\bar{\chi}_{S1}+\left\{ \textrm{Re}Q_{2},\,\textrm{Im}Q_{2}\right\}$ , where $\textrm{Re}Q_{2}$ can subsequently decay to $\chi_{S1}+\bar{\chi}_{S1}+\textrm{Im}Q_{2}$. In the case of $Q_{2}$, $\textrm{Re}Q_{2}$ decays only into $\chi_{S1}+\bar{\chi}_{S1}+\textrm{Im}Q_{2}$. $n_{\chi_{S1}}/2$ or $n_{\chi_{S1}}/4$ depending on the path one looks in. For example, the path $\textrm{Re}Q_{1}\rightarrow\chi_{S1}+\bar{\chi}_{S1}+\textrm{Re}Q_{2}\rightarrow\chi_{S1}+\bar{\chi}_{S1}+\textrm{Im}Q_{2}$ leads to $n_{\chi_{S1}}/4$, while $\textrm{Re}S_{1}\rightarrow\chi_{S1}+\bar{\chi}_{S1}$ yields $n_{\chi_{S1}}/2$. Every path will be weighed by the respective branching ratio. Each $\chi_{S1}$ will in turn contribute to the production of one neutralino at the epoch $H\sim\Gamma_{\chi_{S1}}$.

If the ISS decays into gravitinos or $\chi_{S1}$ are too efficient or if the reheating temperature after inflation is sufficiently high, the high gravitino or $\chi_{S1}$ density may generate an LSP abundance which may overclose the universe. Since we assume that the LSP is a neutralino, LSP production via direct decays of $\phi_{\text{ISS}}$ is negligible compared to its production after the ISS fields have first decayed to gravitinos or $\chi_{S1}$.

%%%%%%%%%%%%%%%%%%%%%%%%%%%%
\subsection{Thermal gravitino production} \label{Subsec:Thermal}
%%%%%%%%%%%%%%%%%%%%%%%%%%%%

We start the analysis of dark matter production via the thermal gravitino source \cite{thermal_gravitino_1}, from the thermal plasma created by the decay of the inflaton field $\eta$. This production depends on the reheating temperature of the universe dominated by the inflaton after its decay, which is given by
\begin{equation}
T_{R}=\left(\frac{40}{\pi^{2}g_{\eta}}\right)^{1/4}a_{\eta}\left(\frac{m_{\eta}}{M_{P}}\right)^{3/2}M_{P}~,
\end{equation}
where $g_{\eta}$ is the number of thermalized relativistic degrees of freedom at temperature $T_R$. 

We assume the inflaton to decay mainly into MSSM particles. If the MSSM and the ISS sector ever had sizeable interactions with each other, that would mean the ISS fields could thermalize with a temperature $T_{R}$, and that would pose a problem if $T_{R}>M_{\textrm{P}}$, since the reheating temperature would then melt the condensed ISS sector (recall that the ISS sector considered here is a description of SQCD at low temperatures) and then one would have to deal with the original quarks, squarks, gauge bosons and gauginos. Neither of the two hypothesis are true: the ISS sector and the MSSM do not have as a sizeable interaction as to thermalize the latter with a temperature $T_{\textrm{R}}$; second, we note that $T_{R}\lesssim10^{-9}M_{\textrm{P}}$ for $a_{\eta}\leq10^{-1}$, $g_{\eta}=100$ and $m_{\eta}=10^{-5}M_{\textrm{P}}$, therefore $T_{R}$ is below the energy scale of $M/M_{\textrm{P}}$ considered in the figures of section \ref{Sec:Post-inflation_Entropy} and further in the plots of the current section.

The ratio of the gravitino number density to the entropy density in the reheating phase is given by, for $m_{1/2}\ll m_{3/2}\ll T_R$ \cite{thermal_gravitino_3} ($m_{1/2}$ stands for the gaugino masses),
\begin{equation}
\left(\frac{n_{3/2}}{s}\right)_{\textrm{rh}}=2.3\times10^{-12}\left(\frac{T_{R}}{10^{10}\textrm{ GeV}}\right)=2.5\times10^{-11}a_{\eta}g_{\eta}^{-1/4}\left(\frac{m_{\eta}}{10^{-5}M_{P}}\right)^{3/2}~.\label{eq:2}
\end{equation}
For high values of $T_{R}$, gravitino to LSP decay may potentially overclose the universe. Thus a further investigation of this type of process is necessary. 

If one assumes that the number density of neutralinos is given by\footnote{This assumption is feasible. The R-parity of the gravitino is odd,
thus its largest decay rates are for channels $\phi_{i}^{\textrm{odd}}+\phi_{j}^{\textrm{even}}$,
$\phi_{i}^{\textrm{odd}}+\phi_{j}^{\textrm{even}}+\phi_{k}^{\textrm{even}}$,
$\phi_{i}^{\textrm{odd}}+\phi_{j}^{\textrm{odd}}+\phi_{k}^{\textrm{odd}}$.
However, given a number $n$ of final particles, $\left(2n-3\right)$
final particles can be fermions since $\psi_{3/2}$ has mass dimension $3/2$. The lowest order process is $\psi_{3/2}\rightarrow\phi_{i}^{\textrm{odd}}+\phi_{j}^{\textrm{even}}$, which leads to $n_{\textrm{odd}}=n_{3/2}$.} $n_{\chi}\simeq n_{3/2}$, their mass density in units of the critical density $\rho_{c}$ will be given by
\begin{equation}
\Omega_{\chi}^{\textrm{rh}}h_{\textrm{d}}^{2}\simeq\frac{7n_{\gamma}}{s}\frac{m_{\chi}n_{3/2}}{\rho_{c}}h_{\textrm{d}}^{2}\simeq 2.78\times10^{10}\left(\frac{m_{\chi}}{100\textrm{ GeV}}\right)\left(\frac{n_{3/2}}{s}\right)_{\textrm{rh}}~,
\end{equation}
where today's density of photons is related to today's entropy density by $7n_{\gamma}\simeq s_{0}$, and $h_{\textrm{d}}=H_0/\left(100\textrm{ km}\textrm{s}^{-1}\textrm{Mpc}^{-1}\right)$
is the dimensionless Hubble parameter with $H_0$ being the today's dimensionful Hubble parameter. Therefore, for $\Omega_{\chi}^{\textrm{rh}}h_{\textrm{d}}^{2}\apprle0.12$,
the allowed gravitino to entropy ratio is 
\begin{equation}
\left(\frac{n_{3/2}}{s}\right)_{\textrm{rh}}\apprle 4.32\times10^{-12}\left(\frac{100\textrm{ GeV}}{m_{\chi}}\right)~.\label{eq:4-1}
\end{equation}
By combining equations (\ref{eq:2}) and (\ref{eq:4-1}), we have the upper bound
\begin{equation} \label{eq:LowerBound}
a_{\eta}g_{\eta}^{-1/4}\left(\frac{m_{\eta}}{10^{-5}M_{p}}\right)^{3/2}\left(\frac{m_{\chi}}{100\textrm{ GeV}}\right)\apprle 0.17~.
\end{equation}
This bound can be evaded in the following distinct situations, namely:

$\bullet$ when the inflaton decays into a gravitino plus an inflatino. However, this channel may be kinematically forbidden if $\left|m_{\eta}-m_{\tilde{\eta}}\right|<m_{3/2}$, where $\tilde{\eta}$ is the inflatino, or kinematically suppressed if the inflaton(ino) scale is much higher than the gravitino scale, $\mathcal{O}\left(m_{3/2}\right)\ll\mathcal{O}\left(m_{\eta},m_{\tilde{\eta}}\right)$\cite{Nilles_inflatino}.

$\bullet$ when the inflaton decays into a pair of gravitinos through the interaction term 
\begin{equation}
	\begin{split}
	\mathcal{\ensuremath{L}}^{3/2}_{\eta}=-\frac{i}{8}\epsilon^{\mu\nu\rho\sigma}\bar{\psi}_{\mu}\gamma_{\nu}\psi_{\rho} & \left(G_{\eta}\partial_{\sigma}\eta-G_{\eta^{*}}\partial_{\sigma}\eta^{*}\right) \\ & +\frac{i}{4}\left(1+K\left(\eta,\bar{\eta}\right)\right)m_{3/2}\left(2+\frac{W\left(\eta\right)}{W}\right) M_{P}\bar{\psi}_{\mu}\sigma^{\mu\nu}\psi_{\nu}~,\label{eq:inflaton_avoid}
	\end{split}
\end{equation} 

where $W = W_{\text{KL-ISS}} + W_{\text{MSSM}} + W(\eta)$ is the total superpotential, $G_{\eta}$ is the derivative of $G=K+\text{ln}(W\overline{W})$ with respect to $\eta$ and $K = K\left({\eta},\bar{\eta}\right) + K_{\text{KL-ISS}} + K_{\text{MSSM}}$ is the total K\"ahler potential.

As said earlier in subsection \ref{Subsec:F-term_Consequences}, the K\"ahler potential and superpotential in \cite{Inflaton_decay} imply equation (\ref{eq:inflaton_avoid}) yields a null decay rate $\eta\rightarrow2\psi_{3/2}$. There, the K\"ahler potential is given as  $K\left(\eta,\bar{\eta}\right)=K\left(\left(\eta-\bar{\eta}\right)^2,S\bar{S}\right)$, where $S$ is a stabilizer field with null expectation value; also $W\left(\eta\right)=Sf(\eta)$. The form of these functions imply no interaction terms in (\ref{eq:inflaton_avoid}) between the inflaton and the gravitinos.

$\bullet$ When gravitinos decay at a temperature which is higher than the freezout temperature of the neutralinos $T_{\chi}^{f}\sim m_{\chi}/20$.

Based on these arguments, the condition $(\ref{eq:LowerBound})$ --- which has to be satisfied in order not to close the universe --- applies if none of the conditions discussed above applies. In this work, we assume the production of gravitinos via direct inflaton/inflatino decays is negligible, while the condition for gravitinos to decay after neutralino freezout is considered in figures \ref{fig:DM_production} and \ref{fig:ann_crosssections}. Said that, we consider thermal gravitinos --- from $\eta$ reheating --- as one of the sources to produce dark matter. 

%%%%%%%%%%%%%%%%%%%%%%%%%%%%
\subsection{Mixture of thermal and non-thermal production} \label{Subsec:Thermal+NonThermal}
%%%%%%%%%%%%%%%%%%%%%%%%%%%%

We now analyze dark matter production via a mixture of thermal and non-thermal processes, from gravitinos produced during the reheating phase of $\eta$ and from $\phi_\textrm{ISS}$ decays, respectively.

First we study the production of neutralinos from $\phi_\textrm{ISS}$ decays. The number density of neutralinos $\chi$ differs whether $R_{\textrm{d}\phi_\textrm{ISS}}>R_{\text{d}\eta}$
or $R_{\textrm{d}\phi_\textrm{ISS}}<R_{\text{d}\eta}$, since in the former scenario each $\phi_\textrm{ISS}$ decays are within the $\eta$-reheated universe, and in the latter scenario the universe is dominated by matter from $\eta$ oscillations.  However, the same ratios of $\psi_{3/2}$ and $\chi_{S1}$ number densities, $n_{3/2}$ and $n_{\chi_{S1}}$, to entropy density $s$ are produced in the end. They are given by
\begin{align}
	\begin{split}\label{eq:n_grav}
\frac{n_{3/2}}{s} ={}& a_{\eta}g_{\eta}^{-1/4}\left(\frac{m_{\eta}}{M_{P}}\right)^{3/2}\left(\frac{M_{P}}{m_{3/2}}\right)\left(\frac{M}{M_{P}}\right)^{3} \\
 & \times \left\{ 2.26\left(\frac{\Gamma_{Q_{1}}^{2\psi_{3/2}}}{\Gamma_{Q_{1}}^{\textrm{total}}}+\frac{\Gamma_{Q_{1}}^{\chi\chi\textrm{Re}Q_{2}}}{\Gamma_{Q_{1}}^{\textrm{total}}}\frac{\Gamma_{Q_{2}}^{2\psi_{3/2}}}{\Gamma_{S_{2}}^{\textrm{total}}}\right)+\,\frac{5.72\times10^{6}}{h^{5}}\left(\frac{M}{M_{P}}\right)^{2}\frac{\Gamma_{S_{2}}^{2\psi_{3/2}}}{\Gamma_{S_{2}}^{\textrm{total}}}\right\}~,
	\end{split}\\
	\begin{split}\label{eq:n_chiS1}
\frac{n_{\chi_{S1}}}{s} ={}& a_{\eta}g_{\eta}^{-1/4}\left(\frac{m_{\eta}}{M_{P}}\right)^{3/2}\left(\frac{M_{P}}{m_{3/2}}\right)\left(\frac{M}{M_{P}}\right)^{3} \\
& \times \left\{2.26\frac{\Gamma_{Q_{1}}^{\chi\chi\textrm{Re}Q_{2}}}{\Gamma_{Q_{1}}^{\textrm{total}}}\left(\frac{\Gamma_{Q_{2}}^{2\psi_{3/2}}}{\Gamma_{Q_{2}}^{\textrm{total}}}+2\frac{\Gamma_{Q_{2}}^{\chi\chi\textrm{Im}Q_{2}}}{\Gamma_{Q_{2}}^{\textrm{total}}}+1\right)+\,0.56\left(\frac{M}{M_{P}}\right)^{2}\frac{\Gamma_{S_{1}}^{\chi\chi}}{\Gamma_{S_{1}}^{\textrm{total}}}\right\}~,
	\end{split}
\end{align}
where we used $n_{3/2}\simeq2\left(\Gamma_{\phi_{\textrm{ISS}}}^{2\psi_{3/2}}/\Gamma_{\phi_{\textrm{ISS}}}^{\textrm{total}}\right)n_{\phi_{\textrm{ISS}}}$ with $n_{\phi_{\textrm{ISS}}}=\rho_{\phi_{\textrm{ISS}}}/m_{\phi_{\textrm{ISS}}}$ and similarly for $\chi_{S1}$. Furthermore, we use $\Gamma_{Q_{1}}^{\textrm{Im}Q_{2}}\simeq\Gamma_{Q_{1}}^{\textrm{Re}Q_{2}}$. Notice that, for $Q_{1}$, there is a possibility of a decay chain generating $n_{\chi_{S1}}$, which we also take into account, translated in the term $\propto\frac{\Gamma_{Q_{1}}^{\chi\chi\textrm{Re}Q_{2}}}{\Gamma_{Q_{1}}^{\textrm{total}}}\frac{\Gamma_{Q_{2}}^{2\psi_{3/2}}}{\Gamma_{Q_{2}}^{\textrm{total}}}$. With $n_{\chi}\simeq n_{3/2}$ and $n_{\chi}\simeq n_{\chi_{S1}}$ applied in (\ref{eq:n_grav}) and (\ref{eq:n_chiS1}) separately, the neutralino relic densities from both sources are then given by

\begin{eqnarray}
\Omega_{\chi}^{3/2}h_{\textrm{d}}^{2}	&\simeq&	\frac{7n_{\gamma}}{s}\frac{m_{\chi}n_{3/2}}{\rho_{c}}h_{\textrm{d}}^{2}\nonumber\\
&\simeq&	0.12\,\cdotp\left(\frac{a_{\eta}}{10^{-2}}\right)\left(\frac{100}{g_{\eta}}\right)^{1/4}\left(\frac{m_{\eta}}{10^{-5}M_{P}}\right)^{3/2}\left(\frac{m_{\chi}}{100\textrm{ GeV}}\right)f_{\psi_{3/2}}\left(h,M\right)\,\,,\quad\quad\quad\quad \label{eq:omegadecay_grav}\\
\Omega_{\chi}^{\chi_{S1}}h_{\textrm{d}}^{2}	&\simeq&	\frac{7n_{\gamma}}{s}\frac{m_{\chi}n_{\chi_{S1}}}{\rho_{c}}h_{\textrm{d}}^{2}\nonumber\\
&\simeq&	0.12\,\cdotp\left(\frac{a_{\eta}}{10^{-2}}\right)\left(\frac{100}{g_{\eta}}\right)^{1/4}\left(\frac{m_{\eta}}{10^{-5}M_{P}}\right)^{3/2}\left(\frac{m_{\chi}}{100\textrm{ GeV}}\right)f_{\chi_{S1}}\left(h,M\right)\,\,,\quad\quad\quad\quad\label{eq:omegadecay_chiS1}
\end{eqnarray}
where we have defined the functions $f_{i}$ as
\begin{eqnarray}
f_{\psi_{3/2}}\left(h,M\right)	&=&	406.56\,\left(\frac{1}{h}\right)\left(\frac{M}{M_{P}}\right)\times\nonumber\\
&\,&\quad\left\{ 2.26\left(\frac{\Gamma_{Q_{1}}^{2\psi_{3/2}}}{\Gamma_{Q_{1}}^{\textrm{total}}}+\frac{\Gamma_{Q_{1}}^{\chi\chi\textrm{Re}Q_{2}}}{\Gamma_{Q_{1}}^{\textrm{total}}}\frac{\Gamma_{Q_{2}}^{2\psi_{3/2}}}{\Gamma_{Q_{2}}^{\textrm{total}}}\right)+\frac{5.72\,\cdotp10^{6}}{h^{5}}\left(\frac{M}{M_{P}}\right)^{2}\right\} \,\,,\quad\quad\quad\\
f_{\chi_{S1}}\left(h,M\right)	&=&	406.56\,\left(\frac{1}{h}\right)\left(\frac{M}{M_{P}}\right)\times\nonumber\\
&\,&\quad\left\{ 2.26\frac{\Gamma_{Q_{1}}^{\chi\chi\textrm{Re}Q_{2}}}{\Gamma_{Q_{1}}^{\textrm{total}}}\left(\frac{\Gamma_{Q_{2}}^{2\psi_{3/2}}}{\Gamma_{Q_{2}}^{\textrm{total}}}+2\frac{\Gamma_{Q_{2}}^{\chi\chi\textrm{Im}Q_{2}}}{\Gamma_{Q_{2}}^{\textrm{total}}}+1\right)+0.56\,\left(\frac{M}{M_{P}}\right)^{2}\right\} \,\,.\quad\quad\quad
\end{eqnarray}
They yield $f_{\psi_{3/2}}\left(\alpha_{h},\alpha_{M}M_{\textrm{P}}\right)=1$ and $f_{\chi_{S1}}\left(\alpha_{h},\lambda_{M}M_{\textrm{P}}\right)=1$ for $\alpha_{h}$, $\lambda_{M}$ and $\alpha_{M}$. The latter parameters respect the following relations:
\begin{eqnarray}
406.56\,\alpha_{h}^{-1}\alpha_{M}\left\{ 2.26\left[\frac{\Gamma_{Q_{1}}^{2\psi_{3/2}}}{\Gamma_{Q_{1}}^{\textrm{total}}}+\frac{\Gamma_{Q_{1}}^{\chi\chi\textrm{Re}Q_{2}}}{\Gamma_{Q_{1}}^{\textrm{total}}}\frac{\Gamma_{Q_{2}}^{2\psi_{3/2}}}{\Gamma_{Q_{2}}^{\textrm{total}}}\right]_{h=\alpha_{h}}^{M=\alpha_{M}M_{\textrm{P}}}+5.72\,\cdotp10^{6}\frac{\alpha_{M}^{2}}{\alpha_{h}^{5}}\right\} =1\quad\quad\,\,\,\,	&,&\nonumber\\
406.56\,\alpha_{h}^{-1}\lambda_{M}\left\{ 2.26\left[\frac{\Gamma_{Q_{1}}^{\chi\chi\textrm{Re}Q_{2}}}{\Gamma_{Q_{1}}^{\textrm{total}}}\left(\frac{\Gamma_{Q_{2}}^{2\psi_{3/2}}}{\Gamma_{Q_{2}}^{\textrm{total}}}+2\frac{\Gamma_{Q_{2}}^{\chi\chi\textrm{Im}Q_{2}}}{\Gamma_{Q_{2}}^{\textrm{total}}}+1\right)\right]_{h=\alpha_{h}}^{M=\lambda_{M}M_{\textrm{P}}}+0.56\,\lambda_{M}^{2}\right\} =1	&.&\nonumber\\
\end{eqnarray}	
For instance, we can obtain $\alpha_{M}=3.72\,\cdotp10^{-8}$ and $\lambda_{M}=\left(1.74\,\cdotp10^{-2},\,1.48\,\cdotp10^{-6},\,2.12\,\cdotp10^{-6}\right)$ for $\alpha_{h}=10^{-2}$; and $\alpha_{M}=1.99\,\cdotp10^{-4}$ and $\lambda_{M}=\left(8.05\,\cdotp10^{-2},\,1.48\,\cdotp10^{-4},\,2.12\,\cdotp10^{-4}\right)$ for $\alpha_{h}=1$. Thus, for $a_\eta=10^{-2}$, $g_\eta=100$, $m_\eta=10^{-5}M_P$ and $m_\chi=100\textrm{ GeV}$, one should use $h=\alpha_{h}$ and $M=\left(\alpha_{M};\,\lambda_{M}\right)M_{\textrm{P}}$ to obtain $\Omega_{\chi}^{i}h_{\textrm{d}}^{2}\simeq0.12$, where $M$ admits three solutions for $\chi_{S1}$. 

\begin{figure}[h!]
	\begin{minipage}[c]{0.45\columnwidth}
		{\includegraphics[height=6.47cm]{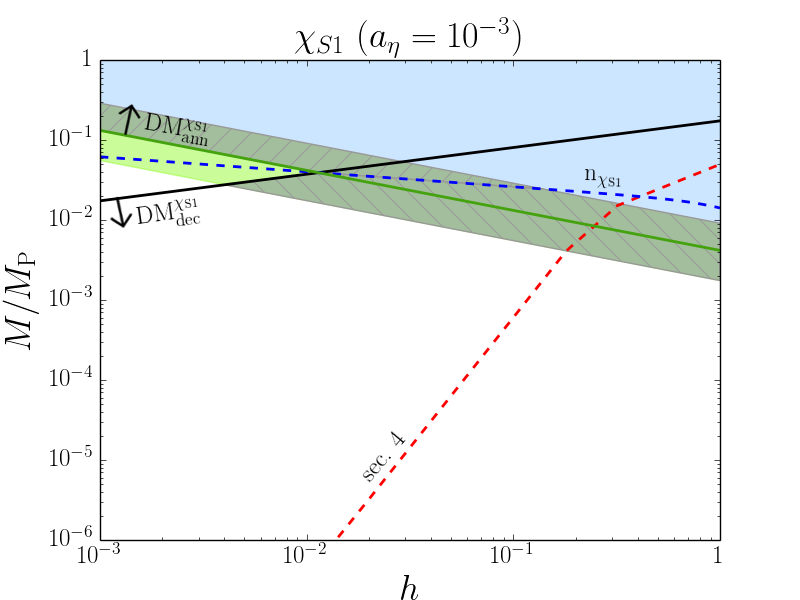}}
	\end{minipage}
	\hspace*{0.5cm}
	\begin{minipage}[c]{0.45\columnwidth}
		{\includegraphics[height=6.47cm]{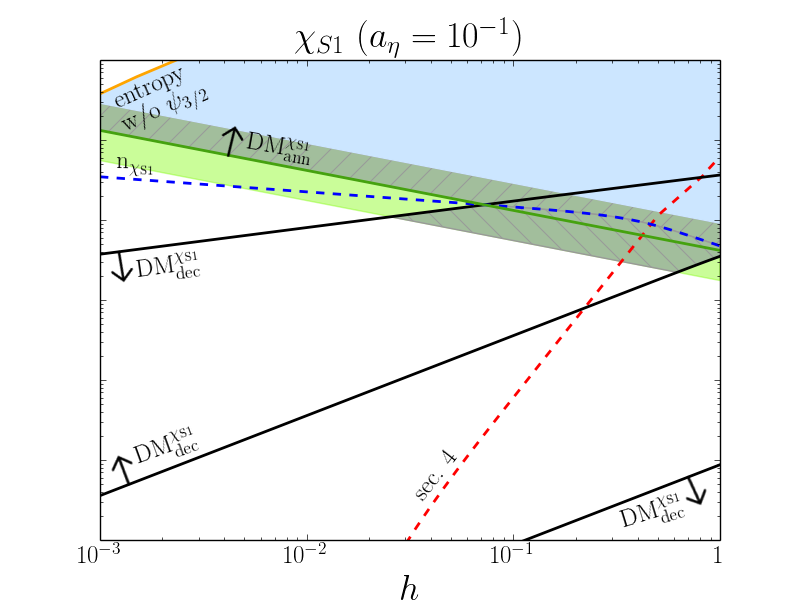}}
	\end{minipage}
	\\
	\bigskip
	\\
	\begin{minipage}[c]{0.45\columnwidth}
		{\includegraphics[height=6.47cm]{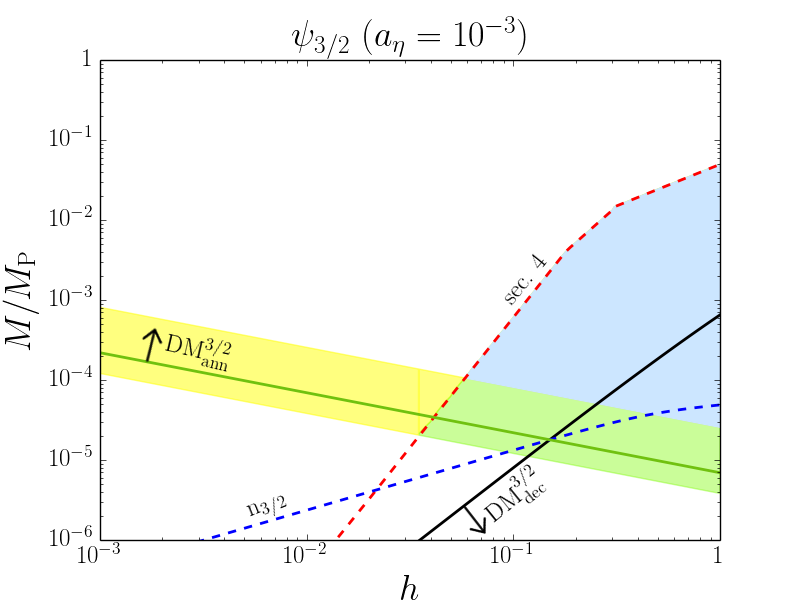}}
	\end{minipage}
	\hspace*{0.5cm}
	\begin{minipage}[c]{0.45\columnwidth}
		{\includegraphics[height=6.47cm]{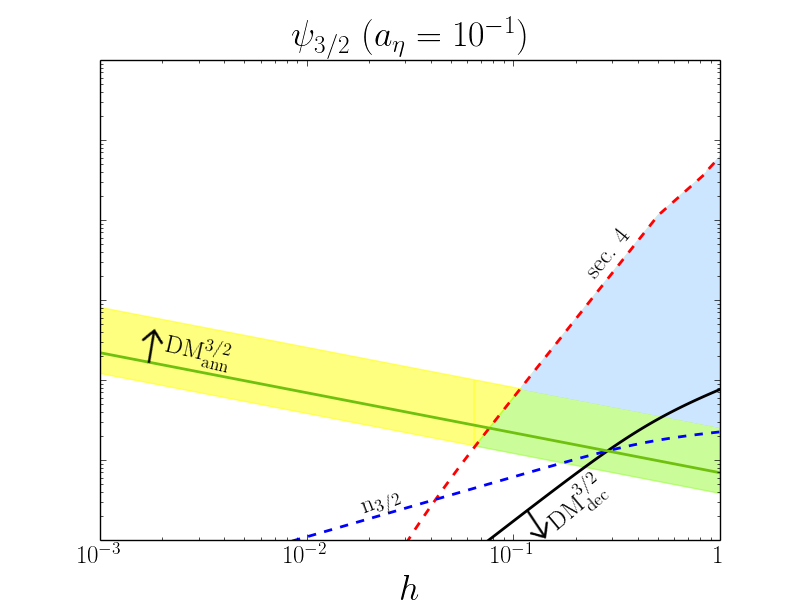}}
	\end{minipage}
	\caption{All the constraints on the ISS parameters $M$ and $h$ for dark matter production from either direct decays $\rm{DM}_{\rm{dec}}$ or decays followed by annihilations  $\rm{DM}_{\rm{ann}}$ for $\chi_{S1}$ and $\psi_{3/2}$ --- for a thermal cross section $\left\langle \sigma_{\text{ann}}v_{\text{M\o{}l}}\right\rangle=10^{-7}\textrm{GeV}^{-2}$ and for small (large) coupling $a_\eta=10^{-3}$ ($a_{\eta}=10^{-1}$). The arrows for the green and black lines point in the direction where $\Omega_\chi h_\textrm{d}^2< 0.12$.
	We hatched the areas where $\Omega_\chi h_\textrm{d}^2\leq 0.12$ is obtained for before BBN decays, essentially the upper green bands.
	Also, notice that the blue shaded regions combine with the yellow bands for $\chi_{S1}$ or $\psi_{3/2}$ to form the green regions. That is, the yellow band for $\chi_{S1}$ does not appear separately from the green region, but is essentially ``hidden" behind the green region. For $\psi_{3/2}$ one can still see part of its yellow band because the red curve which applies for $\psi_{3/2}$ is more constrained than the orange curve for $\chi_{S1}$. A more careful explanation is given in the main text. \label{fig:DM_production}}
\end{figure}

We replaced $m_{3/2}$ by its function depending on both $h$ and $M$ via equation (\ref{eq:Mass_Gravitino}). The equations (\ref{eq:omegadecay_grav}) and (\ref{eq:omegadecay_chiS1}) will be later drawn (for $a_\eta=10^{-3}\textrm{ and }10^{-1}$) in figure \ref{fig:DM_production} for comparison with the cosmological constraints.

We digress a little about the possibility of dark matter being constituted by $\chi_{S1}$. For $\chi_{S1}$ to decay before BBN happens ($T\sim1\textrm{ MeV}$),
$M$ has to satisfy 
\begin{equation}
\label{eq:chi_S1_BBN}
M\gtrsim1.75\times10^{-3}h^{-1/2}M_{P}~,
\end{equation}
which implies a quite heavy gravitino mass. If $\chi_{S1}$ is allowed to decay after the
present time $\sim 10^{18}$ s, we have to assume $M\lesssim2.3\times10^{-5}h^{-1/2}M_{P}$. 
The relic density in the case where $\chi_{S1}$ decays after the present time and can then constitute dark matter is written as --- simplifying equation (\ref{eq:n_chiS1}) and using $m_{\chi_{S1}}=m_{3/2}$ ---

\begin{eqnarray}
\Omega_{\chi_{S1}}h_{\textrm{d}}^{2}	&\simeq&	\frac{7n_{\gamma}}{s}\frac{m_{3/2}n_{\chi_{S1}}}{\rho_{c}}h_{\textrm{d}}^{2}\nonumber\\
&\simeq&	0.12\,\cdotp\left(\frac{a_{\eta}}{10^{-2}}\right)\left(\frac{100}{g_{\eta}}\right)^{1/4}\left(\frac{m_{\eta}}{10^{-5}M_{\textrm{P}}}\right)^{3/2}v_{\chi_{S1}}\left(h,M\right)\quad,
\end{eqnarray}
where we have defined the function $v_{\chi_{S1}}$ as $v_{\chi_{S1}}\left(h,M\right)=1.67\,\cdotp10^{14}h\left(\frac{M}{M_{\textrm{P}}}\right)^{2}f_{\chi_{S1}}\left(h,M\right)$. It yields $v_{\chi_{S1}}\left(\alpha_{h},\beta_{M}M_{\textrm{P}}\right)=1$ for $\alpha_{h}$ and $\beta_{M}$, which respect the equation
\begin{equation}
6.80\,\cdotp10^{16}\,\beta_{M}^{3}\left\{ 2.26\left[\frac{\Gamma_{Q_{1}}^{\chi\chi\textrm{Re}Q_{2}}}{\Gamma_{Q_{1}}^{\textrm{total}}}\left(3-\frac{\Gamma_{Q_{2}}^{2\psi_{3/2}}}{\Gamma_{Q_{2}}^{\textrm{total}}}\right)\right]_{h=\alpha_{h}}^{M=\beta_{M}M_{\textrm{P}}}+0.56\,\beta_{M}^{2}\right\} =1~.
\end{equation}
For example, we obtain $\beta_{M}=8.64\,\cdotp10^{-7}$ for $\alpha_{h}=10^{-2}$ and $\beta_{M}=8.05\,\cdotp10^{-7}$ for $\alpha_{h}=1$. When $a_{\eta}=10^{-2}$, $g_{\eta}=10^{-2}$, ${\eta}=10^{-5}M_{\textrm{P}}$ and $m_\chi=100\textrm{ GeV}$, we have $\Omega_{\chi_{S1}}h_{\textrm{d}}^{2}\simeq0.12$. Here we have replaced $\Gamma_{Q_{2}}^{\textrm{total}}-\Gamma_{Q_{2}}^{2\psi_{3/2}}=\Gamma_{Q_{2}}^{\chi\chi\textrm{Im}Q_{2}}$. From the last equation, it is possible to obtain that $\chi_{S1}$ does not close the universe only for $M$ values (within $a_{\eta}\in\left[10^{-3},\,10^{-1}\right]$ and $h\in\left[10^{-2},\,1\right]$) which are below the bound on the gravitino decay rate in equation (\ref{eq:Constraint_GravM}). Therefore this scenario is impossible and we then assume $\Gamma_{\chi_{S1}}>t_{\textrm{BBN}}^{-1}$, i.e., $\chi_{S1}$ decays before BBN).

Equations (\ref{eq:omegadecay_grav}) and (\ref{eq:omegadecay_chiS1}) depend on whether the entropy production from $\psi_{3/2}$ and $\chi_{S1}$ is negligible. As discussed in section \ref{Sec:Post-inflation_Entropy}, we are working within the shaded parameter region of figure \ref{fig:entropy_produc-1}, which implies no significant entropy production from all $\phi_\textrm{ISS}$.

We now compare the neutralino comoving number produced by $\phi_\textrm{ISS}$ decays and by reheating of $\eta$. We obtain
\begin{eqnarray}
\frac{\left(n_{\chi}/s\right)_{\phi_{\textrm{ISS}}\textrm{ decays}}}{\left(n_{\chi}/s\right)_{\eta\textrm{ reheating}}}	\simeq	1.83\,\cdotp10^{5}\left(\frac{1}{h}\right)\left(\frac{M}{M_{P}}\right)\left\{ 2.26\left(1+\frac{\Gamma_{Q_{1}}^{\chi\chi\textrm{Re}Q_{2}}}{\Gamma_{Q_{1}}^{\textrm{total}}}\right)+\,\frac{5.72\,\cdotp10^{6}}{h^{5}}\left(\frac{M}{M_{P}}\right)^{2}\right\},
\quad\quad\quad\quad\quad\quad\quad\quad\quad
\end{eqnarray}
where the right-hand side yields $1$ at $h=\alpha_{h}$ and $M=\gamma_{M}M_{\textrm{P}}$, since $\alpha_{h}$ and $\gamma_{M}$ are constrained by
\begin{equation}
1.83\,\cdotp10^{5}\alpha_{h}^{-1}\gamma_{M}\left\{ 2.26\left[1+\frac{\Gamma_{Q_{1}}^{\chi\chi\textrm{Re}Q_{2}}}{\Gamma_{Q_{1}}^{\textrm{total}}}\right]_{h=\alpha_{h}}^{M=\gamma_{M}M_{\textrm{P}}}+\,5.72\,\cdotp10^{6}\frac{\gamma_{M}^{2}}{\alpha_{h}^{5}}\right\} =1~.
\end{equation}
As an specific case, $\gamma_{M}=7.87\,\cdotp10^{-9}$ for $\alpha_{h}=10^{-2}$ and $\gamma_{M}=1.61\,\cdotp10^{-6}$ for $\alpha_{h}=1$. Here we have replaced $\Gamma_{Q_{1}}^{2\psi_{3/2}}=\Gamma_{Q_{1}}^{\textrm{total}}-2\Gamma_{Q_{1}}^{\chi\chi\textrm{Re}Q_{2}}$. Furthermore we neglected the last term in equation (\ref{eq:n_chiS1}) when compared to equation (\ref{eq:n_grav}). Thus unless $M$ assumes very small values, in violation of the bound from equation (\ref{eq:Constraint_GravM}), $\left(n_{\chi}/s\right)_{\phi_{\textrm{ISS}}\textrm{ decays}}$ is dominant over the one from thermal gravitinos. Therefore, we assume the neutralino number density to be given by $\phi_{\textrm{ISS}}$ decay from now on.

An important issue we have to treat now is the annihilation of neutralinos after their production \cite{Olive_non-problems}. If the number density of neutralinos produced from $\psi_{3/2}$ or $\chi_{S1}$ decays is high enough, they can annihilate each other and, in turn, decrease their number density. Technically stated, the neutralino number density is governed by the Boltzmann equation
\begin{equation}
\frac{dn_{\chi}}{dt}+3Hn_{\chi}=-\left\langle \sigma_{\textrm{ann}}v_{\textrm{M\o{}l}}\right\rangle n_{\chi}^{2}~,
\end{equation}
where $\left\langle \sigma_{\text{ann}}v_{\text{M\o{}l}}\right\rangle $
is the thermally averaged annihilation cross section of the neutralinos with $v_{\text{M\o{}l}}$ being the M\o{}ller velocity involving the initial particles. The equilibrium number density $n_{\chi,\text{eq}}$ of the neutralinos was neglected since we look at the epoch soon after they decouple, for which $n_{\chi}>n_{\chi,\text{eq}}$ is satisfied. 

If $n_{\chi}\sim H/\left\langle \sigma_{\text{ann}}v_{\text{M\o{}l}}\right\rangle $, we see that the Hubble term $3Hn_{\chi}$ and the annihilation term $\left\langle \sigma_{\text{ann}}v_{\text{M\o{}l}}\right\rangle n_{\chi}^{2}$
are of the same order of magnitude. In this case, the neutralino freezes out. If $n_{\chi}>H/\left\langle \sigma_{\text{ann}}v_{\text{M\o{}l}}\right\rangle $,
the neutralino annihilates after gravitino or $\chi_{S1}$ decays, and can eventually freeze out when $n_{\chi}\sim H/\left\langle \sigma_{\text{ann}}v_{\text{M\o{}l}}\right\rangle $.
On the other hand, if $n_{\chi}<H/\left\langle \sigma_{\text{ann}}v_{\text{M\o{}l}}\right\rangle $,
the neutralino density is given at the time of decay of the gravitino or the $\chi_{S1}$. An approximate expression for the relic abundance of the neutralino can be written as
\cite{annihilation_after_1,annihilation_after_2,annihilation_after_3}
\begin{equation} \label{eq:relicabundance}
\left(\frac{n_{\chi}}{s}\right)^{-1}\simeq\left(\frac{n_{\chi}}{s}\right)_{\textrm{decay}}^{-1}+\left(\frac{H}{s\left\langle \sigma_{\text{ann}}v_{\text{M\o{}l}}\right\rangle }\right)_{\textrm{decay}}^{-1}~,
\end{equation}
where the lower index $\textrm{decay}$ means evaluation at the time of $\psi_{3/2}$ or $\chi_{S1}$ decay. Therefore, we have an upper limit on $n_{\chi}/s$, i.e., $n_{\chi}/s\lesssim H\left\langle \sigma_{\text{ann}}v_{\text{M\o{}l}}\right\rangle ^{-1}/s$. The following ratio compares both quantities on the right-hand side of equation (\ref{eq:relicabundance}), both for the gravitino $\psi_{3/2}$ and $\chi_{S1}$,

\begin{equation}
\hspace{-0.5cm}\left(\frac{H\left\langle \sigma_{\textrm{ann}}v_{\textrm{M\o{}l}}\right\rangle ^{-1}/s}{n_{i}/s}\right)_{i}	\simeq	\left(\frac{10^{-2}}{a_{\eta}}\right)\left(\frac{10^{-5}M_{P}}{m_{\eta}}\right)^{3/2}\left(\frac{10^{-7}\textrm{GeV}^{-2}}{\left\langle \sigma_{\textrm{ann}}v_{\textrm{M\o{}l}}\right\rangle }\right)w_{\psi_{i}}^{-1}\left(h,M\right)~,\label{eq:Hsigmav_n_gravchiS1}
\end{equation}
with $i=\psi_{3/2},\,\chi_{S1}$. We have defined $w_{\psi_{3/2}}\left(h,M\right)=6.64\,\cdotp10^{12}h^{3/2}\left(\frac{M}{M_{\textrm{P}}}\right)^{3}f_{\psi_{3/2}}\left(h,M\right)$ and $w_{\chi_{S1}}\left(h,M\right)=2.49\,\cdotp10^{16}h^{5/2}\left(\frac{M}{M_{\textrm{P}}}\right)^{5}f_{\chi_{S1}}\left(h,M\right)$.  $w_{\psi_{3/2}}=1$ and $w_{\chi_{S1}}=1$ if evaluated at $h=\alpha_{h}$ and $M=\left(\kappa_{M};\delta_{M}\right)M_{\textrm{P}}$, which in turn are constrained by
\begin{eqnarray}
2.70\,\cdotp10^{15}\alpha_{h}^{1/2}\kappa_{M}^{4}\left\{ 2.26\left[\frac{\Gamma_{Q_{1}}^{2\psi_{3/2}}}{\Gamma_{Q_{1}}^{\textrm{total}}}+\frac{\Gamma_{Q_{1}}^{\chi\chi\textrm{Re}Q_{2}}}{\Gamma_{Q_{1}}^{\textrm{total}}}\frac{\Gamma_{Q_{2}}^{2\psi_{3/2}}}{\Gamma_{Q_{2}}^{\textrm{total}}}\right]_{h=\alpha_{h}}^{M=\kappa_{M}M_{\textrm{P}}}+5.72\,\cdotp10^{6}\frac{\kappa_{M}^{2}}{\alpha_{h}^{5}}\right\} =1\quad,&&	\nonumber\\	
1.01\,\cdotp10^{19}\alpha_{h}^{3/2}\delta_{M}^{6}\left\{ 2.26\left[\frac{\Gamma_{Q_{1}}^{\chi\chi\textrm{Re}Q_{2}}}{\Gamma_{Q_{1}}^{\textrm{total}}}\left(3-\frac{\Gamma_{Q_{2}}^{2\psi_{3/2}}}{\Gamma_{Q_{2}}^{\textrm{total}}}\right)\right]_{h=\alpha_{h}}^{M=\delta_{M}M_{\textrm{P}}}+0.56\,\delta_{M}^{2}\right\} =1\quad.\,\,\,\,\,\quad\quad&&\nonumber\\
&&
\end{eqnarray}	
For instance, we obtain $\delta_{M}=3.00\,\cdotp10^{-2}$ and ${M}=1.61\,\cdotp10^{-6}$ for $\alpha_{h}=10^{-2}$; and $\delta_{M}=8.42\,\cdotp10^{-3}$ and $\kappa_{M}=3.31\,\cdotp10^{-5}$ for $\alpha_{h}=1$. These values yield the ratios (\ref{eq:Hsigmav_n_gravchiS1}) equal to $1$, when considered together with $a_\eta=10^{-2}$, $m_\eta=10^{-5}M_P$, $\left\langle \sigma_{\textrm{ann}}v_{\textrm{M\o{}l}}\right\rangle=10^{-7}\textrm{GeV}^{-2}$.

Therefore, if $M\lesssim \left(\delta_M,\,\kappa_M\right)M_{\rm P}$, we obtain $n_{\chi}/s<H\left\langle \sigma_{\text{ann}}v_{\text{M\o{}l}}\right\rangle ^{-1}/s$. In this case, the neutralinos do not annihilate themselves and $\left(n_{\chi}/s\right)_\textrm{decay}$ stays constant. Hence, equations (\ref{eq:omegadecay_grav}) and (\ref{eq:omegadecay_chiS1}) are valid for obtaining the neutralino relic density. However, if $M\gtrsim \left(\delta_M,\,\kappa_M\right)M_{\rm P}$, we obtain $n_{\chi}/s>H\left\langle \sigma_{\text{ann}}v_{\text{M\o{}l}}\right\rangle ^{-1}/s$. In this case, the neutralinos annihilate themselves until they reach $n_{\chi}/s\sim H\left\langle \sigma_{\text{ann}}v_{\text{M\o{}l}}\right\rangle ^{-1}/s$. The neutralino relic density is then $\left(\textrm{for }i=\psi_{3/2},\,\chi_{S1}\right)$

\begin{eqnarray}\label{eq:grav_chiS1_ann}
\hspace{-2cm}\Omega_{\chi}^{i}h_{\textrm{d}}^{2}	&=&	\frac{7n_{\gamma}}{s}\frac{m_{\chi}n_{\chi}^{i}}{\rho_{c}}h_{\textrm{d}}^{2}\nonumber\\
&\simeq&	0.12\left(\frac{100}{g_{\eta}}\right)^{1/4}\left(\frac{\alpha_{h}}{h}\right)^{c_{i}/2}\left(\frac{\varepsilon_{M}^{i}M_{P}}{M}\right)^{c_{i}}\left(\frac{m_{\chi}}{100\textrm{ GeV}}\right)\left(\frac{10^{-7}\textrm{GeV}^{-2}}{\left\langle \sigma_{\textrm{ann}}v_{\textrm{M\o{}l}}\right\rangle }\right)~.
\end{eqnarray}
$c_{i}$ stands for the exponents associated with $h$ and $M$ for $\psi_{3/2}$ or $\chi_{S1}$ and have the values $c_{\psi_{3/2}}=3$ and $c_{\chi_{S1}}=5$. The parameters $\varepsilon^{i}$ respect the equations
\begin{eqnarray}
3.33\,\cdotp10^{-16}\left(\alpha_{h}^{1/2}\varepsilon_{M}^{\psi_{3/2}}\right)^{-3}	&=&	1\quad,\\
7.25\,\cdotp10^{-8}\left(\alpha_{h}^{1/2}\varepsilon_{M}^{\chi_{S1}}\right)^{-5}	&=&	1\quad.
\end{eqnarray}
Putting in some numerical values , we have $\varepsilon_{M}^{\chi_{S1}}=4.17\,\cdotp10^{-2}$ and $\varepsilon_{M}^{\psi_{3/2}}=6.93\,\cdotp10^{-5}$ for $\alpha_{h}=10^{-2}$; and $\varepsilon_{M}^{\chi_{S1}}=4.17\,\cdotp10^{-3}$ and $\varepsilon_{M}^{\psi_{3/2}}=6.93\,\cdotp10^{-6}$ for $\alpha_{h}=1$. The dependence on $h$ and $M$ stems from $m_{3/2}$ which comes from $R_{\textrm{d}\chi_{S1}}$ and $R_{\textrm{d}3/2}$. In other words, replacing $n_{\chi}^{i}/s\sim H\left\langle \sigma_{\textrm{ann}}v_{\textrm{M\o{}l}}\right\rangle ^{-1}/s$ by $\left.\left(H_{\eta}/s_{\eta}\right)\right|_{R_{\textrm{d}i}}\propto\left(m_{3/2}\right)^{-c_{i}/2}\propto h^{-c_{i}/2}M^{-c_{i}}$. If we substitute $g_\eta=100$, $m_\chi=100\textrm{ GeV}$, $\left\langle \sigma_{\textrm{ann}}v_{\textrm{M\o{}l}}\right\rangle=10^{-7}\textrm{GeV}^{-2}$, as well as $M=\varepsilon_{M}^{i}$ with the correspondent $h=\alpha_h$, into equation (\ref{eq:grav_chiS1_ann}), we obtain $\Omega_\chi^i h^2_\textrm{d}\simeq 0.12$.

There are four types of neutralinos, namely Winos, Binos and two neutral Higgsinos, which possess the following thermally averaged annihilation cross sections\footnote{The Wino thermal cross
section can be found from anomaly mediated SUSY breaking \cite{annihilation_after_3}, while the Bino and the Higgsinos cross sections have been given in \cite{Bino_Higgsinos_crosssections}.}
\begin{eqnarray}
\left\langle \sigma_{\textrm{ann}}v_{\textrm{M\o{}l}}\right\rangle _{\textrm{Wino}} & \simeq & \frac{g_{2}^{4}}{2\pi}\frac{1}{m_{\chi}^{2}}\frac{\left(1-x_{W}^{2}\right)^{3/2}}{\left(2-x_{W}^{2}\right)^{2}} \hspace*{0.1cm} \overset{m_\chi=100\textrm{ GeV}}{\longrightarrow} 3.33\times10^{-7}\textrm{GeV}^{-2}~,\hspace{2cm}\\
\left\langle \sigma_{\textrm{ann}}v_{\textrm{M\o{}l}}\right\rangle _{\textrm{Bino}} & \simeq & \frac{g_{1}^{4}}{16\pi}\frac{1}{m_{\chi}^{2}}\left(\frac{6T_\chi}{m_{\chi}}\right) \hspace*{0.6cm} \overset{m_\chi=100\textrm{ GeV}}{\longrightarrow} 1.79\times10^{-9}\,T_\chi\textrm{ GeV}^{-3}~,\\
\left\langle \sigma_{\textrm{ann}}v_{\textrm{M\o{}l}}\right\rangle _{\textrm{Higgsino}} & \simeq & \frac{g_{2}^{4}}{32\pi}\frac{1}{m_{\chi}^{2}}\frac{\left(1-x_{W}^{2}\right)^{3/2}}{\left(2-x_{W}^{2}\right)^{2}} \hspace*{-0.1cm}  \overset{m_\chi=100\textrm{ GeV}}{\longrightarrow} \hspace*{-0.1cm} 2.08\times10^{-8}\textrm{ GeV}^{-2}~,
\end{eqnarray}
where $x_{W}\equiv m_{W}/m_{\chi}$, and $g_{1}$ and $g_{2}$ are the
couplings of the $U\left(1\right)_{Y}$ and $SU\left(2\right)_{L}$
gauge groups, respectively. The Wino and Higgsino interact mainly mainly annihilate from an S-wave inital state,
while the Bino does from the P-wave, thus the thermally averaged square velocity $\left\langle v^{2}\right\rangle =\frac{6T}{m}$ is important in this last case. Wino pairs annihilate into $W^\pm$ pairs through the mediation of charged Winos\footnote{We disregard coannihilations. If one does consider them, they end up increasing $\left\langle \sigma_{\textrm{ann}}v_{\textrm{M\o{}l}}\right\rangle$ (though not necessarily for Winos), which in turn decreases their relic density.}. Bino pairs annihilate into lepton pairs via right-handed slepton mediation\footnote{We defined the right-handed slepton mass $m_{\tilde{l}_R}$ to be equal to $m_\chi$. Considering a greater $m_{\tilde{l}_R}$ decreases its $\left\langle \sigma_{\textrm{ann}}v_{\textrm{M\o{}l}}\right\rangle$, increasing then its relic density.}. Finally, Higgsinos pairs annihilate mainly into $W^\pm$ and $Z$ pairs.

Before we summarize the constraints obtained so far, we derive weak constraints, which are upper bounds on $M$ such that both $\psi_{3/2}$ and $\chi_{S1}$ do not decay before the neutralino freezes out of the $\eta$ plasma. This, together with the bounds $\Gamma_{3/2} > t^{-1}_{\text{BBN}}$ and $\Gamma_{\chi_{S1}}>t^{-1}_{\textrm{BBN}}$ from equations (\ref{eq:Constraint_GravM}) and (\ref{eq:chi_S1_BBN}), form a range in which the particle can decay so that its decay is safe and in principle non-negligible. For the thermal cross sections of the Wino, Bino and Higgsinos, we know that they freeze out at the values $T^f_\chi\simeq\left(3.69,4.27,4.10\right)\textrm{GeV}$, respectively \cite{Omega_calculation}. However, without damaging our conclusions, we take the reference value $\left\langle \sigma_{\textrm{ann}}v_{\textrm{M\o{}l}}\right\rangle=10^{-7}\textrm{ GeV}^{-2}$. This yields \\ $T_\chi^f=3.86\textrm{ GeV}$. Therefore, for both $\chi_{S1}$ and $\psi_{3/2}$ to decay after the neutralino freezout, $M$ must assume the values, respectively\footnote{These results come from $\rho^r_{\eta}(R_{\text{d}\psi_{3/2}}~\text{or}~R_{\text{d}\chi_{S1}}) = \frac{\pi}{30} g_{\eta} (T^f_{\chi})^4$. One should replace the left-hand side by either $\rho_{\eta}\left(R_{\text{d}\eta}/R_{\text{d}\psi_{3/2}}\right)^4$ or $\rho_{\eta}\left(R_{\text{d}\eta}/R_{\text{d}\psi_{\chi_{S1}}}\right)^4$, and the $M$ dependence will show up once we replace $R_{\text{d}\eta}/R_{\text{d}\psi_{3/2}} \propto (\Gamma^{3/2}/\Gamma_{\eta})^{1/2}$ or, similarly, $R_{\text{d}\eta}/R_{\text{d}\chi_{S1}}\propto (\Gamma^{\chi_{S1}}/\Gamma_{\eta})^{1/2}$.},
\begin{eqnarray}
M & \lesssim & 9.12\times10^{-3}h^{-1/2}M_\textrm{P}~,\label{eq:yellowband_1}\\
M & \lesssim & 2.56\times10^{-5}h^{-1/2}M_\textrm{P}~.\label{eq:yellowband_2}
\end{eqnarray}

For a better understanding of the constraints on the parameter space in figure \ref{fig:DM_production}, we bring them into table \ref{tab:Constraints_Sec5} and take them into account for the following figures.

\begin{figure}[h!]
	\begin{minipage}[c]{0.45\columnwidth}
		{\includegraphics[height=6.47cm]{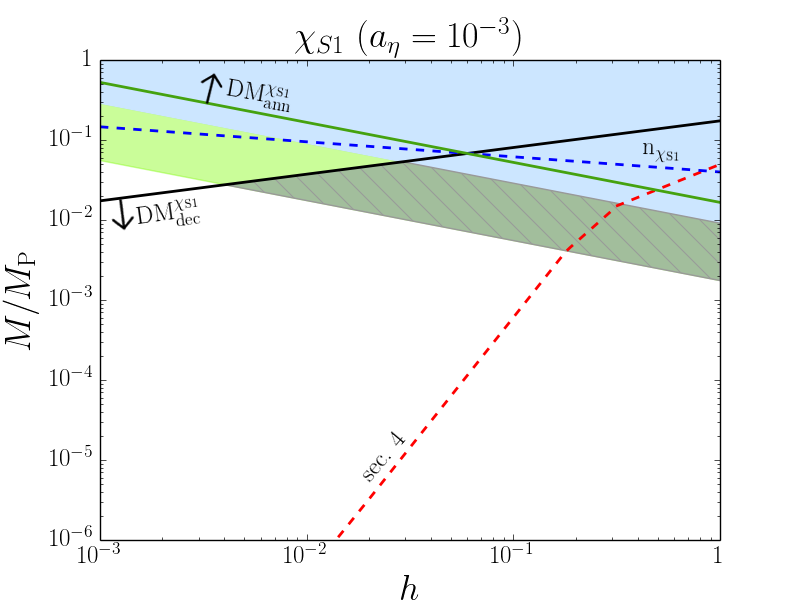}}
	\end{minipage}
	\hspace*{0.5cm}
	\begin{minipage}[c]{0.45\columnwidth}
		{\includegraphics[height=6.47cm]{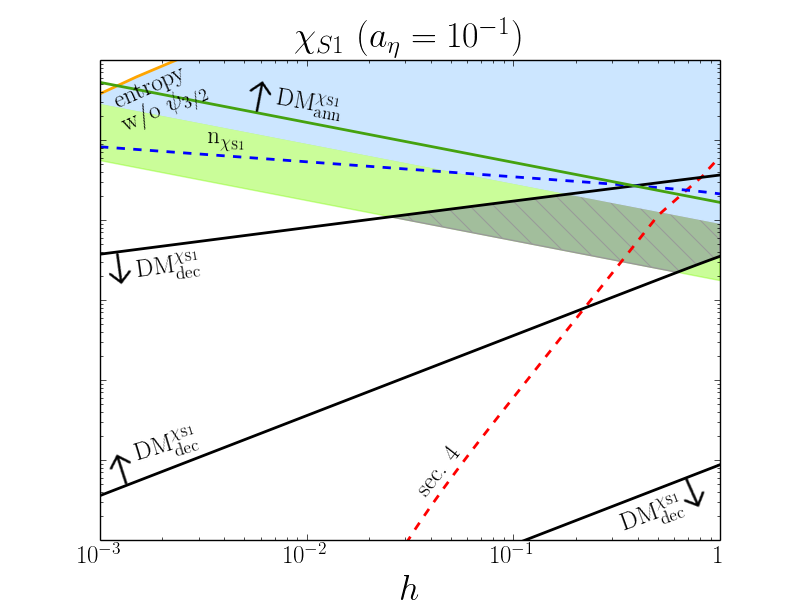}}
	\end{minipage}
	\\
	\bigskip
	\\
	\begin{minipage}[c]{0.45\columnwidth}
		{\includegraphics[height=6.47cm]{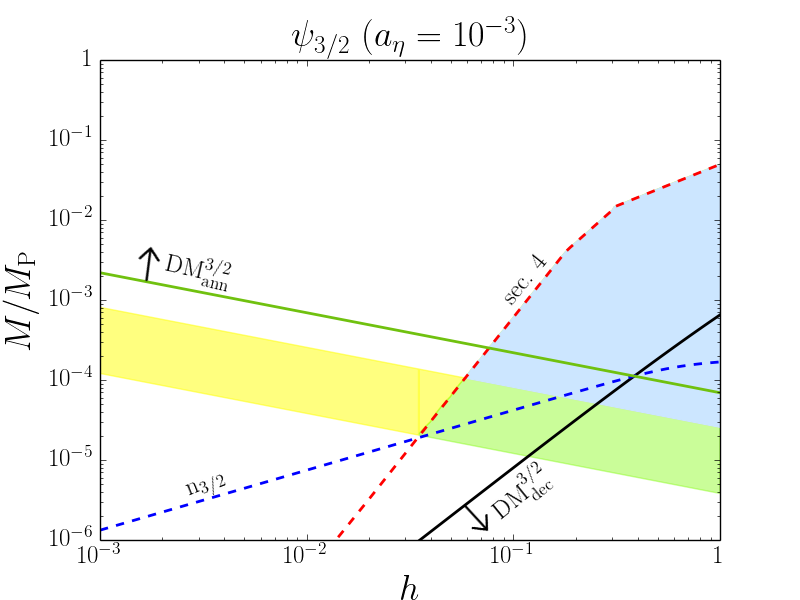}}
	\end{minipage}
	\hspace*{0.5cm}
	\begin{minipage}[c]{0.45\columnwidth}
		{\includegraphics[height=6.47cm]{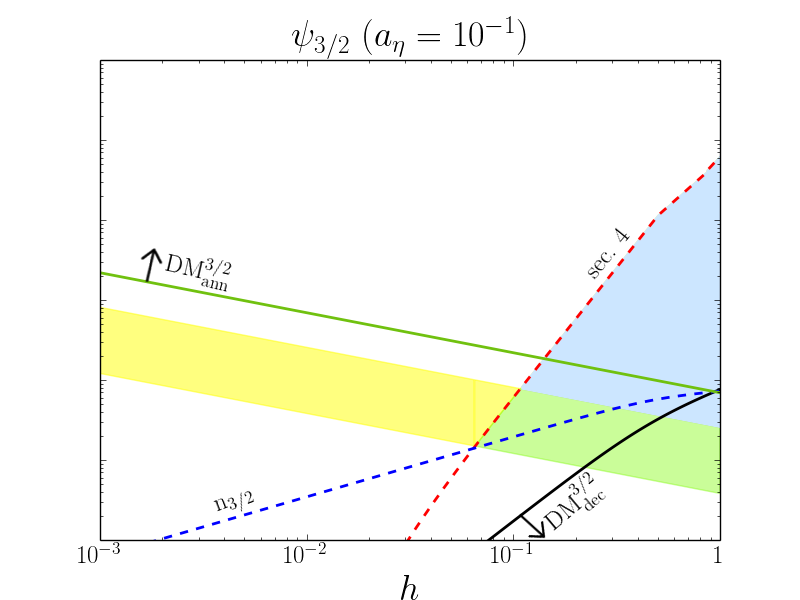}}
	\end{minipage}
	\caption{The same as in figure \ref{fig:DM_production} for a thermal cross section $\left\langle \sigma_{\text{ann}}v_{\text{M\o{}l}}\right\rangle~=~10^{-10}\textrm{GeV}^{-2}$, and for small (large) coupling $a_\eta=10^{-3}$ ($a_{\eta}=10^{-1}$). 
\label{fig:ann_crosssections}}
\end{figure}

\begin{table}[h]\footnotesize
	\center
	\begin{tabular}{|c|c|c|c|}
		\hline 
		Location & Constraint & Meaning & Legend\tabularnewline
		\hline 
		\hline 
		Eq. (\ref{eq:Constraint_GravM}) & $M\gtrsim 3.82\times 10^{-6}h^{-1/2}M_{\rm P}$ & $\psi_{3/2}$ decays before BBN &  \\
		and & and & and & lower yellow band \\
		eq. (\ref{eq:yellowband_2}) & $M \lesssim 2.56\times10^{-5}h^{-1/2}M_{\rm P}$ & $\psi_{3/2}$ decays after neutralino freezeout & \\
		\hline 
		 & Numerical & $s_{\eta}> s_{\phi_{\text{ISS}}}$&  \\
		Fig. \ref{fig:final_constraints_sec4} & and & and & blue shaded region \\
		& $M<4.80\times 10^{-2}h M_{\rm P}$ & $\rho_{\eta}> \rho_{\phi_{\text{ISS}}}$ & \\
		\hline
		Eq. (\ref{eq:chi_S1_BBN}) & $M\gtrsim 1.75\times 10^{-3}h^{-1/2}M_{\rm P}$ & $\chi_{S1}$ decays before BBN & \\
		and & and & and & upper yellow band \\
		eq. (\ref{eq:yellowband_1}) & $M \lesssim 9.12\times10^{-3}h^{-1/2}$ & $\chi_{S1}$ decays after neutralino freezeout & \\
		\hline 
	\end{tabular}	
	\caption{All constraints on the ISS parameters $M$ and $h$ we take into account in this section. Their location in the text, their meaning, and their depiction in the figures of this section are also given. Here we allow for the extension of the blue shaded region for the plots of $\chi_{S1}$ when the entropy production from gravitinos should not be considered, which has already been mentioned earlier in figure \ref{fig:final_constraints_sec4} of section 4.}
	\label{tab:Constraints_Sec5}
\end{table}
In figures \ref{fig:DM_production} and \ref{fig:ann_crosssections}, we draw the following curves
\begin{itemize}
	\item DM$_{\text{dec}}^{3/2}$ and DM$_{\text{dec}}^{\chi_{S1}}$ when $\Omega^i_{\chi}h_d^2 = 0.12$ without further annihilations of neutralinos, obtained from equations \eqref{eq:omegadecay_grav} and \eqref{eq:omegadecay_chiS1} --- these are represented by black solid lines;
	
	\item DM$_{\text{ann}}^{3/2}$ and DM$_{\text{ann}}^{\chi_{S1}}$ when $\Omega^i_{\chi}h_d^2 = 0.12$ with further annihilations of neutralinos, obtained from equation \eqref{eq:grav_chiS1_ann} --- these are represented by green solid lines;
	
	\item $n_{3/2}$ and $n_{\chi_{S1}}$ separating the parameter space into a region where the neutralinos do and do not annihilate. Above $n_{3/2}$ and $n_{\chi_{S1}}$ we should consider DM$_{\text{ann}}^{3/2}$ and DM$_{\text{ann}}^{\chi_{S1}}$, respectively, whereas below $n_{3/2}$ and $n_{\chi_{S1}}$ we should consider DM$_{\text{dec}}^{3/2}$ and DM$_{\text{dec}}^{\chi_{S1}}$, respectively. These are represented by blue dashed lines.
\end{itemize}
Apart from these curves, we also depict the constraints summarized in table \ref{tab:Constraints_Sec5}.

First, we discuss the meaning of the yellow bands. They stand for the region where $\chi_{S1}$ or $\psi_{3/2}$ decay before BBN and after neutralino freezout. While the former condition must be respected, the latter is somewhat a weaker condition, since its violation does not pose problems to cosmological evolution. In fact, if $\chi_{S1}$ or $\psi_{3/2}$ do decay before neutralino freezout, that just means that the $\chi$ relic density is given by the standard thermal decoupling. The other important constraint is given by the entropy production, represented by the red dashed (orange) curve, where the allowed region is in blue. Note that if the blue and yellow regions are superposed, they yield a green area. Thus, the allowed parameter region we should look into if we want to check for dark matter production --- due to the constraints in table \ref{tab:Constraints_Sec5} --- is given by the green regions within the allowed yellow bands constrained further by either the orange curve in the case of $\chi_{S1}$ or the red curve in the case of $\psi_{3/2}+\chi_{S1}$.

Furthermore, looking at the case of $\psi_{3/2}$ for the two lower plots, if the gravitino decays within its yellow band, $\chi_{S1}$ consequently decays after BBN, which cannot happen. The only allowed case is to permit the gravitino to decay within the yellow band of $\chi_{S1}$, which consequently implies that the gravitino decays before neutralino freezeout. The associated entropy from $\psi_{3/2}$ in or above the upper yellow band should therefore not be calculated with the entropy from $\psi_{3/2}$ decays. Thus, the red dashed curve --- which considered entropy production from both $\psi_{3/2}$ and $\chi_{S1}$ --- is replaced by the orange curve, which can be seen on the upper left corner of the figure for $\chi_{S1}$ at $a_{\eta}=10^{-1}$. A similar curve should also appear in the figure for $\chi_{S1}$ at $a_{\eta}=10^{-3}$, however it shows up at even larger values of $M/M_{\rm P}$ and lower values of $h$ and therefore cannot be drawn within the ranges we considered.

For $\left\langle \sigma_{\text{ann}}v_{\text{M\o{}l}}\right\rangle=10^{-7}\textrm{GeV}^{-2}$, we start by discussing the upper plots for $\chi_{S1}$. Again, notice that the green region is formed by the blue shaded region within the yellow band. As mentioned before, above (below) the blue dashed curve we should take into account the green (black) curve for a correct dark matter relic density $\text{DM}_{\text{ann\,(dec)}}^{\chi_{S1}}$. One can see that we can obtain $\Omega_\chi h_\textrm{d}^2\leq0.12$ for both processes, and for both $a_\eta=10^{-3}$ and $a_\eta=10^{-1}$. In fact, above the blue dashed line, we have  $\Omega_\chi h_\textrm{d}^2=0.12$ on the green line and $\Omega_\chi h_\textrm{d}^2<0.12$ above it. On the other hand, below the blue dashed line, we have $\Omega_\chi h_\textrm{d}^2=0.12$ on the black line and $\Omega_\chi h_\textrm{d}^2<0.12$ below it (in the case where there are three possible black lines, $\textrm{d}^2<0.12$ above it). Finally, the orange line is above $M=M_P$ for $a_\eta=10^{-3}$, but visible for $a_\eta=10^{-1}$.
	
We now discuss the lower plots for $\psi_{3/2}$. Here we again have that above (below) the blue dashed curve we should take into account the green (black) curve for a correct dark matter relic density $\text{DM}_{\text{ann\,(dec)}}^{\psi_{3/2}}$. Considering these two lower plots alone, we could have dark matter production from $\psi_{3/2}$ decays followed by annihilations as there is part of the green (black) curves within the allowed green shaded region. However, we recall that this region is below the yellow band for $\chi_{S1}$ and $\chi_{S1}$ would decay after BBN. Since this case is not desirable, sufficient dark matter production can never be achieved for $\psi_{3/2}$ decays.
	
For completeness, we discuss dark matter production in figure \ref{fig:ann_crosssections} for another value of thermal cross section, namely $\left\langle \sigma_{\text{ann}}v_{\text{M\o{}l}}\right\rangle=10^{-10}\textrm{ GeV}^{-2}$. For the lower subfigures, the problem discussed for $\psi_{3/2}$ still stands, i.e., $\chi_{S1}$ cannot decay before BBN while at the same time $\psi_{3/2}$ generates the right relic density $\Omega_\chi h_\textrm{d}^2\simeq 0.12$. For the upper subfigures, the intersection of the blue dashed, green and black lines moves up to $h\simeq7\times10^{-2}$ ($a_\eta=10^{-3}$) and to $h\simeq4\times10^{-1}$ ($a_\eta=10^{-1}$). However, above the green band ($\chi_{S1}$ decays while $\chi$ has not yet frozen out), which means $\Omega_\chi h_\textrm{d}^2\leq 0.12$ from $\chi$ production, with subsequent annihilations, only happens above the green line ans is therefore irrelevant. On the other hand, below the blue dashed line, we have $\Omega_\chi h_\textrm{d}^2\simeq 0.12$ on the black line and $\Omega_\chi h_\textrm{d}^2< 0.12$ below it (and above the lower black line for $a_\eta=10^{-1}$). The region respecting $\Omega_\chi h_\textrm{d}^2\leq 0.12$ and within the green band has been hatched.


Along with non-thermal production of neutralinos, it remains to discuss their thermal production from freezout. The contributions, due to purely thermal neutralino freezout from the $\eta$ plasma, assumes the values for Wino, Bino and Higgsino \cite{Omega_calculation, neutralino_calculation}
\begin{eqnarray}
\Omega_{\textrm{Wino}}^{\textrm{freezout}}h_{\textrm{d}}^{2} & \simeq & 7.03\times10^{-4}~,\\
\Omega_{\textrm{Bino}}^{\textrm{freezout}}h_{\textrm{d}}^{2} & \simeq & 0.0261~,\\
\Omega_{\textrm{Higgsino}}^{\textrm{freezout}}h_{\textrm{d}}^{2} & \simeq & 0.010~.
\end{eqnarray}
Notice then that for $m_{\chi}=100\textrm{ GeV}$, each of the Bino and Higgsino relic densities yields $\sim 10\%$ of the required $\Omega_{\textrm{CDM}}h_{\textrm{d}}^{2}\simeq0.12$. In these cases, if one wants to obtain $\Omega_{\chi}h_{\textrm{d}}^{2}\simeq0.12$ for $m_{\chi}=100\textrm{ GeV}$, one has to consider $\left(h,M\right)$ points slightly off the black and green lines so that $\Omega_{\chi}^{\chi_{S1}}h_{\textrm{d}}^{2}\sim0.9\,\cdotp\Omega_{\textrm{CDM}}h_{\textrm{d}}^{2}$. In the end, we must have $\Omega_{\chi}^{\textrm{freezout}}h_{\textrm{d}}^{2}+\Omega_{\chi}^{\chi_{S1}}h_{\textrm{d}}^{2}\simeq\Omega_{\textrm{CDM}}h_{\textrm{d}}^{2}$.

For the case when $\chi_{S1}$ decays above the green band, that means the thermal freezout must account for all dark matter density, hence $\Omega_{\chi}^{\textrm{freezout}}h_{\textrm{d}}^{2}\simeq0.12$. This can be accomplished considering by, e.g., a Bino LSP, if one considers $m_{\tilde{l}_{R}}\simeq220\textrm{ GeV}$.

%%%%%%%%%%%%%%%%%%%%%%%%%%%%
%%%%%%%%%%%%%%%%%%%%%%%%%%%%
\section{Conclusions} \label{Sec:Conclusions}
%%%%%%%%%%%%%%%%%%%%%%%%%%%%
%%%%%%%%%%%%%%%%%%%%%%%%%%%%

In this work we analyzed the production of dark matter neutralino candidates (Wino, Bino, and Higgsino) within a setup mixing the Minimal Supersymmetric Standard Model (MSSM) with the string/supergravity KL moduli stabilization scenario whose vacuum energy is uplifted with the help of the ISS model as an F-term dynamical supersymmetry breaking sector dual to $\mathcal{N}=1$ SQCD. Before we could probe the possibility of dark matter, we obtained constraints on the parameters $M$ and $h$ of the ISS model imposing that entropy production from the fields $S_1$, $S_2$ and $Q_1$ is negligible. The entropy production constraint enables mainstream baryogenesis mechanisms to work besides easing the analysis on the evolution of the Universe.

We coupled the KL-ISS scenario with the inflaton and MSSM fields. When computing the decay rates associated with interaction terms between the ISS and the MSSM fields, we found that the the three largest contributions originate from decays of the ISS fields to gravitinos via $(Q_1,Q_2,S_2) \rightarrow \psi_{3/2}+\psi_{3/2}$, to $\chi_{S1}$ pairs via $S_1 \rightarrow \bar{\chi}_{S1}+\chi_{S1}$, and to three particles via $Q_1 \rightarrow \bar{\chi}_{S1}+\chi_{S1}+\textrm{Re}Q_2$ and $Q_1 \rightarrow \bar{\chi}_{S1}+\chi_{S1}+\textrm{Im}Q_2$, and $Q_2 \rightarrow \bar{\chi}_{S1}+\chi_{S1}+\textrm{Im}Q_2$.

A detailed study of oscillations from the inflaton $\eta$ and the ISS fields was then performed, where we used $S_1,\,S_2$ and $Q_1$ as the relevant ISS fields for the subsequent analysis on entropy dilution. Furthermore, we discussed the epochs of the decays of the inflaton $\eta$ and these ISS fields. The entropy production upper bounds on $M$ and $h$ are obtained from the entropy densities of the ISS fields and $\eta$ decays such that the former is smaller than the latter --- where the non-relativistic behaviour of the final products from ISS decays was also taken into account. Moreover, we treated the issue of the relativistic degrees of freedom of Im$Q_2$, which is massless, obtaining that its contribution is within the observed error of $N_\textrm{eff}$.

Finally, in section \ref{Sec:DarkMatter}, considering a neutralino dark matter, we provided expressions for dark matter production via direct decays from gravitinos $\psi_{3/2}$ or from the ISS decay product $\chi_{S1}$, or through their decays followed by annihilations. We set these expressions against the constraints for negligible entropy production, as well as constraints on the decay epochs of $\psi_{3/2}$ and $\chi_{S1}$. 

We have obtained that the parameter space of $M$ and $h$ is severely constrained, mainly because of the extremely small $\chi_{S1}$ decay rate (much smaller than the one of $\psi_{3/2}$). Through figures \ref{fig:DM_production} and \ref{fig:ann_crosssections}, it can be noticed that this feature enables before BBN decays for both particles while generating enough DM relic density in a reduced area of the parameter space, i.e., practically the region where $\chi_{S1}$ decays before BBN and after $\chi$ freezout. While the gravitino could generate enough neutralinos, the before BBN constraint for $\chi_{S1}$ forbids this solution. For $\chi_{S1}$ decays before BBN, sufficient DM can be generated, either through direct decays of $\chi_{S1}$ (for both $\left\langle \sigma_{\text{ann}}v_{\text{M\o{}l}}\right\rangle=10^{-7}\textrm{GeV}^{-2}$ and $\left\langle \sigma_{\text{ann}}v_{\text{M\o{}l}}\right\rangle=10^{-10}\textrm{GeV}^{-2}$) or through the subsequent annihilation of neutralinos (only for $\left\langle \sigma_{\text{ann}}v_{\text{M\o{}l}}\right\rangle=10^{-7}\textrm{GeV}^{-2}$). We conclude that for $m_{\chi}=100\textrm{ GeV}$, the standard thermal scenario yields at most $10\%$ of the required DM relic density, while the non-thermal scenario can provide the remaining $90\%$ DM content.

%%%%%%%%%%%%%%%%%%%%%%%%%%%%
%%%%%%%%%%%%%%%%%%%%%%%%%%%%
\begin{center}
\textbf{Acknowledgements} 
\end{center}

Thaisa C. da C. Guio would like to thank financial support from CNPq (Brazil) under grant number 205626/2014-9 and the honors branch of Bonn-Cologne Graduate School of Physics and Astronomy (BCGS). Ernany R. Schmitz would also like to thank financial support from CNPq (Brazil) under grant number 201016/2014-1 and support from the BCGS. The authors would like to thank Manuel Drees for a careful reading of earlier versions of this manuscript, and for his pertinent comments and suggestions. The authors would also like to thank Gl\'auber C. Dorsch for helpful discussions.

%%%%%%%%%%%%%%%%%%%%%%%%%%%%
%%%%%%%%%%%%%%%%%%%%%%%%%%%%

\bibliographystyle{plain}

%%%%%%%%%%%%%%%%%%%%%%%%%%%%
%%%%%%%%%%%%%%%%%%%%%%%%%%%%
\end{document}